\documentclass[prd,nofootinbib,showpacs,unsortedaddress]{revtex4}
\usepackage{natbib}
\usepackage{amssymb,amsbsy,amsmath,amsfonts}
\usepackage{graphicx}
\usepackage{wick}
\usepackage{float}
\usepackage{rotating}
\newcommand{\comments}[1]{}%Never use this command with the package verbatim
\newcommand{\RR}{{\boldsymbol{\mathcal{R}}}}
\newcommand{\rL}{{\rm L}}
\newcommand{\rR}{{\rm R}}

\def\tstrut{\vrule height2.5ex depth0pt width0pt} % used in tables
\begin{document}
\title {
Low-lying even parity meson resonances and spin-flavor symmetry
}

\author{C. Garc{\'\i}a-Recio}
\affiliation{Departamento de F{\'\i}sica
At\'omica, Molecular y Nuclear, Universidad de Granada, E-18071
Granada, Spain}
\author{L.S. Geng}
\affiliation{School of Physics and 
Nuclear Energy Engineering, Beihang University, Beijing 100191, China}
\affiliation{Physik Department, Technische Universit\"at M\"unchen,
  D-85747 Garching, Germany }
\author{J. Nieves}
\affiliation{ Instituto de F{\'\i}sica Corpuscular (IFIC), 
Centro Mixto CSIC-Universidad de Valencia,
Institutos de Investigaci\'on de Paterna, Aptd. 22085, E-46071
Valencia, Spain}

\author{L.L. Salcedo}
\affiliation{Departamento de
F{\'\i}sica At\'omica, Molecular y Nuclear, Universidad de Granada,
E-18071 Granada, Spain}

\begin{abstract}
Based on a spin-flavor extension of chiral symmetry, a novel $s-$wave
meson-meson interaction involving members of the $\rho-$nonet and of the
$\pi-$octet is introduced and its predictions are analyzed. The starting point
is the SU(6) version of the SU(3) flavor Weinberg-Tomozawa Lagrangian.  SU(6)
symmetry breaking terms are then included to account for the physical meson
masses and decay constants in a way that preserves (broken) chiral
symmetry. Next, the $T-$matrix amplitudes are obtained by solving the Bethe
Salpeter equation in coupled-channel and the poles are identified with their
possible Particle Data Group (PDG) counterparts. It is shown that most of the
low-lying even parity PDG meson resonances, specially in the $J^P=0^+$ and
$1^+$ sectors, can be classified according to multiplets of SU(6). The
$f_0(1500)$, $f_1(1420)$ and some $0^+(2^{++})$ resonances cannot be
accommodated within this scheme and thus they would be clear candidates to be
glueballs or hybrids. Finally, we predict the existence of five exotic
resonances ($I \ge 3/2$ and/or $|Y|=2$) with masses in the range 1.4--1.6 GeV,
which would complete the $27_1$, $10_3$, and $10_3^*$ multiplets of
SU(3)$\otimes$SU(2).
\end{abstract}
\date{\today} \maketitle
\section{Introduction}
Chiral perturbation theory (ChPT), a systematic implementation of chiral
symmetry and of its pattern of spontaneous and explicit breaking, provides a
model independent scheme where multitude of low-energy non-perturbative
strong-interaction phenomena can be understood. It has been successfully
applied to study different processes, both in the meson-meson and in the
meson-baryon sectors, involving light ($u$ and $d$) or strange ($s$)
quarks~\cite{Weinberg:1978kz,Gasser:1983yg,Gasser:1984gg,Meissner:1993ah,
  Bernard:1995dp,Pich:1995bw,Ecker:1994gg,Scherer:2002tk,Bernard:2007zu,
  Scherer:2009bt}.

However, by construction, ChPT is only valid at low-energies and it cannot
describe the nature of hadron resonances.  In recent years, it has been shown
that by unitarizing the ChPT amplitudes one can greatly extend the region of
application of ChPT.\footnote{Several frameworks have been proposed to
  unitarize the amplitudes, though the most common and successful are the
  inverse amplitude method and the solution of the Bethe-Salpeter (BS)
  equation. In this latter case, several renormalization schemes have been
  also employed, differing mostly in the treatment of the off-shell
  effects. In general, the different methods give similar results for the
  lowest-lying resonances.} This approach, commonly referred as Unitary Chiral
Perturbation Theory (UChPT), has received much attention and provided many
interesting results~\cite{Truong:1988zp,Dobado:1989qm,Dobado:1992ha,
  Kaiser:1995eg,Dobado:1996ps,Hannah:1997ux,Oset:1997it, Oller:1997ng,
  Oller:1997ti, Nieves:1998hp, Kaiser:1998fi, Oller:1998hw,
  Oller:1998zr, Nieves:1999bx,
  GomezNicola:2000wk, Nieves:2000km, Markushin:2000fa, Oller:2000fj,
  Nieves:2001de,GomezNicola:2001as,Lutz:2001yb, Nieves:2001wt, Hyodo:2002pk,
  GarciaRecio:2002td,
  Kolomeitsev:2003kt,GarciaRecio:2003ks,Lutz:2003fm,Jido:2003cb,
  Nicola:2003zi, Sarkar:2004jh, GarciaRecio:2005hy, Borasoy:2005ie,
  Roca:2005nm, GarciaRecio:2006vx, GomezNicola:2007qj,Toki:2007ab}. In
particular, many meson-meson and meson-baryon resonances and bound states
appear naturally within UChPT. These states are then interpreted as having
``dynamical nature.'' In other words, they are not genuine $q\bar{q}$ or $qqq$
states, but are mainly built out of their meson-meson or meson-baryon
components.  One way to distinguish these two pictures is to study the
dependence on $N_C$ of the resonance masses and widths~\cite{Pelaez:2003dy,
  Pelaez:2006nj,GarciaRecio:2006wb,Hyodo:2007np,Roca:2008kr,Geng:2008ag,
  Nieves:2009ez, Nieves:2009kh}.

Some examples are the low-lying scalar mesons, $f_0(600)$, $f_0(980)$,
$a_0(980)$ and $K^*_0(800)$, which naturally appear as resonant states
of two mesons of the pion octet~\cite{Oller:1997ng,
Oller:1997ti,Kaiser:1998fi,Nieves:1998hp, Nieves:1999bx,
Markushin:2000fa, Dobado:1996ps,Oller:1998hw, GomezNicola:2001as}, or
the low-lying $J^P=1/2^-$ baryonic resonances, $N(1535)$, $N(1650)$,
$\Lambda(1405)$ and $\Lambda(1670)$, which are found after unitarizing
the ChPT amplitudes for the scattering of $\pi$ pseudoscalar octet
mesons off baryons belonging to nucleon
octet~\cite{Kaiser:1995eg,Oset:1997it,Oller:2000fj,Nieves:2001wt,Lutz:2001yb,
GarciaRecio:2002td, GarciaRecio:2003ks, Jido:2003cb, Borasoy:2005ie,
GarciaRecio:2005hy,Hyodo:2002pk}, or the low-lying $J^P=3/2^-$
baryonic resonances (e.g. the $\Lambda(1520)$) found in the
interaction of mesons of the pion octet with baryons of the $\Delta$
decuplet \cite{Kolomeitsev:2003kt,Sarkar:2004jh, Toki:2007ab}, and
finally the low-lying axial vector mesons, $a_1(1260)$, $b_1(1235)$,
$h_1(1170)$, $f_1(1285) $, $K_1(1270)$ that can be described as
resonant states of a $\pi$ octet and a $\rho$ vector nonet
mesons~\cite{Lutz:2003fm,Roca:2005nm}.

These ideas have also been extended to study three-body meson
resonances~\cite{MartinezTorres:2007sr,MartinezTorres:2008gy} and to
systems including a heavy quark, which has allowed to describe also
meson~\cite{Kolomeitsev:2003ac,Hofmann:2003je,Guo:2006fu,Gamermann:2006nm}
and baryon~\cite{Lutz:2003jw, Hofmann:2005sw,
Hofmann:2006qx,Mizutani:2006vq, GarciaRecio:2008dp,Gamermann:2010zz}
charmed resonances. In these latter cases, of course one cannot invoke
chiral dynamics to construct the tree level amplitudes that later on
will be unitarized. Thus for instance in Refs.~\cite{Lutz:2003jw,
Hofmann:2005sw, Hofmann:2006qx,Mizutani:2006vq}, the universal
vector-meson coupling hypothesis is exploited to break the SU(4)
symmetry in a convenient and well-defined manner. This is done by a
$t-$channel exchange of vector mesons between pseudoscalar mesons and
baryons, in such a way that chiral symmetry is preserved in the light
meson sector, while the interaction is still of the Weinberg-Tomozawa
(WT) type~\cite{Weinberg:1966kf,Tomozawa:1966jm}. A serious
limitation of this approach is that it is not consistent with
heavy-quark-spin symmetry, which is a proper QCD symmetry that appears
when the quark masses become much larger than the typical confinement
scale, $\Lambda_{\rm QCD}$. The recent works of
Refs.~\cite{GarciaRecio:2008dp,Gamermann:2010zz} develop a scheme
consistent with both chiral and heavy-quark-spin symmetry by starting
from a larger SU(8) spin-flavor symmetry group, and conveniently
breaking the spin (in the light $u,d,s$ sector) and flavor
symmetries. In both schemes, coupled-channel unitarity plays a major
role.

Coupled-channel unitarity has been also the essential ingredient in other
recent works, where the vector meson-vector
meson~\cite{Molina:2008jw,Geng:2008gx} and vector meson-octet and decuplet
baryon interactions~\cite{Oset:2009vf, Gonzalez:2008pv,Sarkar:2009kx} in the
light sector have been studied. In these cases, the formalism of the hidden
gauge interaction for vector mesons of \cite{Bando:1984ej,Bando:1987br} is
adopted, and it has led to very successful results from a phenomenological
point of view \cite{MartinezTorres:2009uk,Branz:2009cv,Geng:2009iw}.

The approach taken in Refs.~\cite{GarciaRecio:2005hy,GarciaRecio:2006vx},
where SU(6) spin-flavor symmetry is invoked, allows to assign the vector
mesons of the $\rho$ nonet and the pseudoscalar mesons of the $\pi$ octet in
the same ({\bf 35}) SU(6) multiplet, while the baryons of the nucleon octet
and $\Delta$ decuplet are placed in the totally symmetric {\bf 56} SU(6)
representation. The scheme is completely constrained by requiring that its
restriction to the $\pi$-$N$ octet sector will reduce to the SU(3) $s-$wave WT
Lagrangian, leading order (LO) of the chiral expansion in this case. Finally,
the SU(6) symmetry is broken by using physical masses and decay constants. The
corresponding BS amplitudes successfully
reproduce the previous SU(3)-flavor WT results for the lowest-lying $s-$and
$d-$wave odd parity baryon resonances obtained from scattering of the mesons
of the pion octet off baryons of the nucleon octet and delta
decuplet~\cite{GarciaRecio:2006vx}.\footnote{The predictions of this scheme
  for the vector meson nonet-baryon decuplet sector have not been derived
  yet.} The extension of the scheme to the charm
sector~\cite{GarciaRecio:2008dp, Gamermann:2010zz} naturally accommodates
heavy-quark-spin symmetry, as mentioned above, since it encodes spin symmetry
in the charm sector, while this is not the case for those models based on
vector meson exchanges.

In this work, we present the extension of the meson-baryon scheme derived in
Ref.~\cite{GarciaRecio:2005hy} to the meson-meson sector.  The basis of our
approach is rooted in the ideas of Caldi and Pagels
\cite{Caldi:1975tx,Caldi:1976gz}.  These authors identify vector mesons of the
nonet as ``dormant'' Nambu-Goldstone bosons of an extended chiral symmetry
SU(6)$_\rL\otimes$SU(6)$_\rR$. This symmetry is intrinsically an
approximate one and the vector mesons acquire mass through relativistic
corrections. Such scheme naturally solves a number of puzzles involving the
nature and classification of vectors mesons and makes predictions in
remarkable agreement with the experiment \cite{Caldi:1975tx,Caldi:1976gz}. The
low energy theorems derived from partial conservation of the tensor current
have been obtained in \cite{Ibanez:1979vb}. The Skyrmion of the SU(6)$_{\rm
  L}\otimes$SU(6)$_\rR$ has been studied in \cite{Meissner:1986it}. The
validity of the dormant Nambu-Goldstone boson description has been verified in
the lattice \cite{Smit:1980nf}.

Even if, for illustration purposes, SU(3) flavor symmetry is assumed to be
exact and the pseudoscalar mesons are assumed to be massless (these
simplifying assumptions are not essential and can be lifted) the breaking of
SU(6) down to SU(3) (due to the vector mesons masses being nonvanishing)
implies a breaking of SU(6)$_\rL\otimes$SU(6)$_\rR$ down to
SU(3)$_\rL\otimes$SU(3)$_\rR$. One of the contributions of the present
work is the construction of suitable Lagrangian mass terms achieving such
pattern of symmetry breaking. Therefore the approximate spin-flavor symmetry
scheme provides a unified framework to deal with lowest lying mesons,
implementing the required symmetry breaking patterns and in particular
fulfilling low energy theorems derived from chiral symmetry. In addition, when
the present scheme is extended to include heavier quark flavors,
the QCD  heavy-quark spin symmetry can also be naturally accommodated in the
spin-flavor approach through a suitable flavor breaking pattern. 
Although the
predictions so obtained are not so reliable as those derived for pseudoscalar
mesons assuming only the standard chiral symmetry breaking pattern of QCD, the
fact is that vector mesons do exist and they are known to play a relevant role
in hadronic physics.  Inescapably, they will interact among themselves and
with other pseudoscalar mesons and will certainly influence the properties of
the known mesons resonances. However not so many approaches to deal with
vector mesons are available, and the existing ones, e.g., the hidden gauge
approach \cite{Bando:1984ej,Bando:1987br} are also subject to a certain amount
of modeling not directly rooted in QCD. The theoretically founded models to
deal with vector mesons being scarce, we regard the spin-flavor symmetric
scenario (suitably broken) as a reasonable alternative approach and we believe
it makes sense to work out the predictions of such a model. Moreover, we
remarkably find that most of the low-lying even parity PDG meson resonances
can be classified according to multiplets of the spin-flavor symmetry group
SU(6).

We will study the $s-$wave interaction of two members of the {\bf 35}
($\pi-$octet + $\rho-$nonet) SU(6) multiplet by means of an enlarged WT meson
Lagrangian to accommodate vector mesons, which guarantees that chiral symmetry
is recovered when interactions between pseudoscalar Nambu-Goldstone bosons are being
examined. We will pay a special attention to the novel pseudoscalar-vector
($PV$) and vector-vector ($VV$) channels, where we will compare our
predictions with previous recent results ~\cite{Roca:2005nm,
  Molina:2008jw,Geng:2008gx} obtained within the formalism of the hidden gauge
interaction for vector mesons. In the $PV\to PV$ sector, chiral symmetry
constrains the interactions, and our model and that developed in
Ref.~\cite{Roca:2005nm} totally agree at LO in the chiral expansion, despite
their different apparent structure and origin. As a result of this work, we
show that most of the low-lying even parity meson resonances, specially in the
$J^P=0^+$ and $1^+$ sectors, can be classified according to multiplets of the
spin-flavor symmetry group SU(6). This can be seen in Table~\ref{tab:j012},
which summarizes the set of dynamically generated resonances obtained in this
work.  The remaining firmly established positive parity PDG meson states below
2 GeV, that cannot be accommodated within SU(6) multiplets, are clear
candidates to be glueballs or hybrids. This is the case of the $f_0(1500)$,
$f_1(1420)$ and some $0^+(2^{++})$ resonances. On the other hand, we predict
the existence of five exotic resonances ($I \ge 3/2$ and/or $|Y|=2$) with
masses in the region 1.4--1.6 GeV, which would complete the $27_1$ and $10_3$
and $10_3^*$ SU(3)$\otimes$SU(2) multiplets.

\begin{table}
\begin{center}
\caption{Tentative SU(6) classification of the poles found in this work,
  together with their possible PDG counterparts \cite{Amsler:2008zz}. $J^{P},
  Y,I^G$ stand for the spin-parity, hypercharge, isospin and $G-$parity,
  respectively [for non-strange states, charge conjugation is given by
    $G=(-1)^{I}$].  Those resonances marked with $\dagger$ need to be
  confirmed, while a (*) symbol indicates that the resonance does not appear
  in the PDG. Finally, a question mark symbol expresses our reservations on
  the assignment. Mixings between states with the same $J^PI^GY$ quantum
  numbers, but belonging to different SU(6) and/or SU(3) multiplets have not
  been considered. }
\vspace{0.3cm}
\label{tab:j012}
\begin{tabular}{ccc}
\hspace{0cm} $J^P=0^+$  &   & \hspace{-4cm} $J^P=1^+$ \\
\hspace{0cm}\begin{tabular}{c||c|c|c|l|c}
\hline\tstrut
 SU(6)  & SU(3)$\otimes$SU(2)      &$Y$&$I^G$& $\sqrt{s}$ [MeV] &
 PDG~\protect\cite{Amsler:2008zz}  \\
Irrep &Irrep & & & (this work) & \\
\hline
\hline
 {\bf 1}     & $1_1$         &  0  & $0^+$   & $(635,-202)$ & $f_0(600)$ \\\hline
 &       &$\pm 1$& 1/2 & $(830,-170)$ &
 $K^*_0(800)^\dagger$ \\   
          \cline{3-4}\tstrut 
{ ${\bf 35}_s$}             &     $8_1$      &  0  & $1^-$   & $(991,-46)$  & $a_0(980)$ \\
          \cline{3-4}\tstrut 
             &           &  0  & $0^+$   & $(969,0)$ & $f_0(980)$  \\ \hline
    &       &$\pm 2$& 1 & $(1564,-66)$ & (*) \\   
          \cline{3-4}\tstrut 
    &        &$\pm 1$& 3/2 & $(1433,-70)$ &  (*) \\ 
          \cline{3-4}\tstrut 
    &        &$\pm 1$& 1/2 & $(1428,-24)$ & $K^*_0(1430) $ \\         
          \cline{3-4}\tstrut 
    &  $27_1$      &0& $0^+$ & $(1350,-62)$  & $f_0(1370)$ \\
          \cline{3-4}\tstrut 
    &        &0& $1^-$ & $(1442,-5)$  & $a_0(1450)$ \\
          \cline{3-4}\tstrut 
{\bf 189} &        &0& $2^+$ &  $(1419,-54)$ & $X(1420)^\dagger$\\
          \cline{2-6}\tstrut 
    &  &$\pm 1$& 1/2 &  $(1787,-37)$ & $K_0^*(1950)^\dagger$\\
          \cline{3-4}\tstrut 
    &   $8_1$      &0& $1^-$ & $(1760,-12)$ & $a_0(2020)^\dagger$ ?  \\
          \cline{3-4}\tstrut 
    &       &0& $0^+$ & $(1723,-52)$ & $f_0(1710)$\\
          \cline{2-6}\tstrut 
    & $1_1$ &0&  $0^+$ &~~~~ $-$ & \\
\hline
\end{tabular} & & \hspace{-4cm}
\begin{tabular}{c||c|c|c|l|c}
\hline\tstrut
 SU(6)  & SU(3)$\otimes$SU(2)       &$Y$&$I^G$& $\sqrt{s}$ [MeV] &
 PDG~\protect\cite{Amsler:2008zz}  \\
Irrep &Irrep & & & (this work) & \\
\hline
\hline 
 &       &$\pm 1$& 1/2 & $(1188,-64)$ &
 $K_1(1270)$ \\
          \cline{3-4}\tstrut 
            &     $8_3$      &  0  & $1^+$   &$(1234,-57)$
 &  $b_1(1235)$ \\
          \cline{3-4}\tstrut 
    { ${\bf 35}_s$}          &           &  0  & $0^-$   & $(1373,-17)$  &
 $h_1(1380)^\dagger$   \\
\cline{2-6}\tstrut 
&  $1_3$& 0 & $0^-$ & $(1006,-85)$ & $h_1(1170)$
\\ \hline
    &       &$\pm 2$& 0 & $(1608,-114)$ & (*) \\   
          \cline{3-4}\tstrut 
    & $10_3$       &$\pm 1$& 3/2 & $(1499,-127)$ &  (*) \\ 
          \cline{3-4}\tstrut 
    & \&       &$\pm 1$& 1/2 & $(1414,-66)$ & $K_1(1400)$ ? \\         
          \cline{3-4}\tstrut 
    &  $10^*_3$      &0& $1^+$ & $(1642,-139)$ & $b_1(1960)^\dagger$ ?  \\
          \cline{3-4}\tstrut 
 {\bf 189}   &        &0& $1^-$ &  $(1568,-145)$ & $a_1(1640)^\dagger$\\
          \cline{2-6}\tstrut 
    &  &$\pm 1$& 1/2 & $(1250,-31)$ &$K_1(...)$ (*) \\
          \cline{3-4}\tstrut 
    &   $8_3^a$      &0& $1^-$ & $(1021,-251)$ & $a_1(1260)$ \\
          \cline{3-4}\tstrut 
    &       &0& $0^+$ & $(1286,0)$ & $f_1(1285)$\\

          \cline{2-6}\tstrut 
    &  &$\pm 1$& 1/2 & $(1665,-95)$ & $K_1(1650)^\dagger$ ? \\
    &   $8_3^s$      &0& $1^+$ &~~~~ $-$ &  \\
    &       &0& $0^-$ & $(1600,-67)$ & $h_1(1595)^\dagger$\\
\hline
\end{tabular} \\
 & & \\
 & \hspace{-4cm} $J^P=2^+$ & \\
&\hspace{-4cm} \begin{tabular}{c||c|c|c|l|c}
\hline\tstrut
 SU(6)  & SU(3)$\otimes$SU(2)       &$Y$&$I^G$& $\sqrt{s}$ [MeV] &
 PDG~\protect\cite{Amsler:2008zz}  \\
Irrep &Irrep & & & (this work) & \\
\hline
\hline 
{ $\bf 189$} &       &$\pm 1$& 1/2 & $(1708,-156)$ &
 $K_2^*(1430)$ ? \\
          \cline{3-4}\tstrut 
     \&              &     $8_5$      &  0  & $1^-$   & $(1775,-6)$
 &  $a_2(1700)^\dagger$ \\
          \cline{3-4}\tstrut 
{\bf Contact $VV$}         &           &  0  & $0^+$   & $(1783,-19)$
 & $f_2(1640)^\dagger$  ?
   \\
\cline{2-6}\tstrut 
 &  $1_5$& 0 & $0^+$ &$(1289,0)$ & $f_2(1270)$ 
\\ \hline
 &  & $\pm 1$ & 1/2 &~~~~  $-$ & \\
          \cline{3-4}\tstrut
         &     $8_5$    & 0   & $1^-$ & $(1228,0)$ & $a_2(1320)$\\
          \cline{3-4}\tstrut
{\bf Contact $VV$}  &       & 0   & $0^+$ &    ~~~~  $-$ & \\
 \cline{2-6}\tstrut 
&  $1_5$& 0 & $0^+$ &  ~~~~  $-$  &
\\\hline 
\end{tabular}  
\end{tabular}
\end{center}
\end{table}

\comments{
\begin{table}
\begin{center}
\caption{The same as Table~\ref{tab:j0} for $J^P=1^+$. }\vspace{0.3cm}

\end{center}
\end{table}

\begin{table}
\begin{center}
\caption{The same as Table~\ref{tab:j0} for $J^P=2^+$.}\vspace{0.3cm}

\end{center}
\end{table}
}

The extension of the model presented here to the charm sector would
naturally accommodate heavy-quark spin symmetry. On the contrary,
this latter QCD requirement will not be easily met for
models~\cite{Kolomeitsev:2003ac,Hofmann:2003je,Guo:2006fu,
Gamermann:2006nm} based on the formalism of the hidden gauge
interaction for vector mesons, since those would not treat in the same
way pseudoscalar ($D$) and vector ($D^*$) charmed mesons.

%---------------------------

This paper is organized as follows: In Sect.~\ref{sec:ii}, we start from the
chiral Lagrangian for pseudoscalar-pseudoscalar interactions
(Subsect.~\ref{sec:ii.su3}) and extend it to calculate the interaction
vertices between two pseudoscalars, one pseudoscalar and one vector, and
between two vector mesons in terms of SU(6) projectors and the corresponding
eigenvalues (Subsect~\ref{sec:su6}). Also in this subsection, we show how to
obtain three relations connecting these eigenvalues by matching our amplitudes
to the LO ChPT ones for two pseudoscalar scattering and how finally all
eigenvalues can be fixed.  Next in Subsect.~\ref{sec:su6fb}, we discuss the
nature of the SU(6) symmetry breaking terms needed to account for the physical
meson decay constant and masses, without spoiling partial conservation of the
axial current in the light pseudoscalar sector.  Sec~\ref{sec:bs} deals with
the BS equation and with issues related with its
renormalization. In Sect.~\ref{sec:res}, we present results in terms of the
unitarized amplitudes and search for poles on the complex plane. We discuss
the results sector by sector trying to identify the obtained resonances or
bound states with their experimental counterparts~\cite{Amsler:2008zz}, and
compare our results with earlier studies. A brief summary and some conclusions
follow in Sect.~\ref{sec:concl}. In Appendix \ref{app:tables} the various
potential matrices are compiled for the different hypercharge, isospin and
spin sectors. In Appendix \ref{app:eigenvalues} details are given on the
computation of the eigenvalues of various operators.

\section{SU(6) description of vector-pseudoscalar
and vector-vector interactions}
\label{sec:ii}

\subsection{SU(3) and chiral symmetry}
\label{sec:ii.su3}

The lowest-order chiral Lagrangian describing the interaction of pseudoscalar
Nambu-Goldstone bosons is \cite{Gasser:1984gg}
\begin{equation}
 \mathcal{L}=\frac{f^2}{4}\mathrm{Tr}\left(\partial_\mu U^\dagger
 \partial^\mu U+\mathcal{M}(U+U^\dagger-2)\right),
\label{eq:chpt}
\end{equation}
where $f\sim 90$ MeV is the chiral-limit pion decay constant, $U=e^{{\rm
    i}\sqrt{2}\Phi/f}$ is the SU(3) representation of the meson
fields, with
\begin{equation}
 \Phi=\left(\begin{array}{ccc}
             \frac{1}{\sqrt{6}}\eta+\frac{1}{\sqrt{2}}\pi^0 & \pi^+ & K^+\\
              \pi^- & \frac{1}{\sqrt{6}}\eta-\frac{1}{\sqrt{2}}\pi^0& K^0\\
               K^- & \bar{K}^0 & -\sqrt{\frac{2}{3}}\eta 
            \end{array}
\right)
,
\label{eq:su3phi}
\end{equation}
and the matrix
$\mathcal{M}=\mathrm{diag}(m_\pi^2,m_\pi^2,2m_K^2-m_\pi^2)$ is
determined by the pion and kaon meson masses.  Expanding up to four
meson fields, one finds
\begin{equation}
 \mathcal{L^{\rm int}}
=
\frac{1}{12 f^2} \mathrm{Tr}\left( 
[\Phi,    \partial_\mu\Phi] [\Phi, \partial^\mu\Phi]
+\mathcal{M} \Phi^4\right)
.
\label{eq:su3chpt}
\end{equation}
Taking a common mass, $m$, for all the (pseudo) Nambu-Goldstone mesons
and projecting into $s-$wave, the above Lagrangian leads to an
interaction Hamiltonian (on shell)
\begin{equation}\label{eq:su3}
 {\mathcal H}=\frac{3s-4 m^2}{6 f^2} \hat H_1 -\frac{m^2}{3 f^2} \hat H_2.
\end{equation}
where $\sqrt{s}$ is the total energy of the meson pair in the center
of mass system, and $\hat H_1$ and $\hat H_2$ are coupled-channel
matrices; they are $IY$ block diagonal, with $I$ and $Y$ the total
isospin and hypercharge (strangeness) of either the initial or final
meson pair. The normalization can be unambiguously fixed thanks to the
relation of the diagonal matrix elements of ${\mathcal H}(s)$ with the
$s-$wave scattering amplitude, ${\mathcal F}(s)$, the phase shifts
$\delta(s)$ and inelasticities $\eta (s)$,
\begin{equation}
{\mathcal H}_{ii} (s)= -8 \pi \sqrt{s}\, {\mathcal F}_{ii}(s), \qquad
2\, {\rm i}\, p\, {\mathcal F} = \eta \, e^{2\, {\rm i} \delta }-1
,
\end{equation}
where $p$ is the momentum in the center of mass frame of the two mesons.

The operators $\hat H_1$ and $\hat H_2$ are linear combinations of orthogonal
projectors, $P_\mu$, onto the SU(3) $\mu$ representations that appear in the
reduction of the product of representations $8 \otimes 8$. Namely,
\begin{equation}
\hat H_1= -3 P_1-\frac32 P_{8_s}+ P_{27}, 
\quad 
\hat H_2 = 5 P_1+P_{8_s} + P_{27}
.
\label{eq:su3bis}
\end{equation}
Note that only representations which are symmetric under the permutation of
the two octets appear. This is a consequence of $s-$wave Bose statistics, once
we have assigned a common mass for all pseudoscalar mesons.
\comments{ The basis of states
in which the projectors are defined is also symmetrized under the exchange of
the two mesons,
\begin{equation}
|M_1M_2;YIJ\rangle_{\rm sym}= \frac{\sum_{\mu_{\rm sym}} P_{\mu_{\rm sym}}
|M_1M_2;YIJ\rangle}{||\sum_{\mu_{\rm sym}} P_{\mu_{\rm sym}}
|M_1M_2;YIJ\rangle||\,}, \qquad \mu_{\rm sym}= 1, 8_s, 27
.
\label{eq:states-sym}
\end{equation}
}

On the other hand, by imposing just SU(3) flavor symmetry, the
interaction Hamiltonian would be of the form
\begin{equation}
{\mathcal H}(s) = \sum_{\mu} F_\mu(s) P_\mu
\end{equation}
with the SU(3) representation $\mu$ running over the $1,8_s$ and $27$
irreducible symmetric representations that appear in the reduction of $8
\otimes 8$, and $F_\mu$ arbitrary functions of the Mandelstam variable
$s$. The approximate chiral symmetry of QCD, which is much more restrictive
than just flavor symmetry, and its pattern of spontaneous and explicit
symmetry breaking fixes this enormous freedom and allows to determine the
chiral expansion of the functions $F_\mu(s)$.  At LO, the functions $F_\mu$
can be easily read off from Eqs.~(\ref{eq:su3}) and (\ref{eq:su3bis}).

The first contribution in $\mathcal{L^{\rm int}}$ of
Eq.~(\ref{eq:su3chpt}) is the WT term in this $\pi\pi$ case.  There is a WT
term for the interaction of Nambu-Goldstone bosons off any target. Its
form follows entirely from chiral symmetry (and its pattern of
symmetry breaking) \cite{Weinberg:1966kf,Tomozawa:1966jm} and fully
accounts for the interaction near threshold. Specifically, assuming
exact chiral symmetry (and so massless Nambu-Goldstone bosons) and for 
$s$-wave, ${\mathcal H}(s)$ vanishes at threshold and moreover
\begin{equation}
\frac{d{\mathcal H}(s)}{ds}\Big|_{\rm threshold}
=
\xi \frac{1}{2f^2}\hat H_{\rm WT}
\label{eq:9}
\end{equation}
where $\xi$ is the symmetry factor, namely, $1/2$ if the target is
another Nambu-Goldstone boson and $1$ if it is not, and
\begin{equation}
\hat H_{\rm WT} = \sum_\mu \lambda_\mu P_\mu
,
\end{equation}
where $\mu$ runs over the allowed SU(3) representations.  Note that
${\mathcal H}(s)$ acts on different spaces depending on the target, e.g.,
$(8\otimes 8)_{\rm sym}$ for $\pi\pi$, $8\otimes8$ for $\pi\rho$, and
$8\otimes 1$ for $\pi\omega_1$.  For two flavors the WT interaction
comes as the scalar product of the Nambu-Goldstone boson and target
isospin operators \cite{Birse:1996hd} (and so it depends only on the
isospin target). For any number $N_F$ of (massless) flavors one has
instead $\sum_{a=1}^{N_F^2-1}J^a_{\rm NG}J^a_{\rm target}$ and this
fixes the eigenvalues (see e.g. \cite{Hyodo:2006kg})
\begin{equation}
\lambda_\mu= C_2(\mu)- C_2(\mu_{\rm NG})-C_2(\mu_{\rm target}),
\label{eq:11}
\end{equation}
where $C_2(\mu)$ refers to the value of the quadratic Casimir of the
irrep $\mu$ in SU$(N_F)$ (with normalization $C_2({\rm adj})= N_F$),
$\mu_{\rm NG}$ is the adjoint representation.

This gives for $\pi\pi$ scattering the eigenvalues quoted in (\ref{eq:su3bis})
for $N_F=3$, and new ones for $\pi\rho$ and $\pi\omega_1$ ($\omega_1$ refers
to the SU(3) singlet):
\begin{equation}
\lambda^\pi_1=\lambda^\rho_1=-6,\quad
\lambda^\pi_{8_s}=\lambda^\rho_{8_s}=\lambda^\rho_{8_a}=-3,\quad
\lambda^\pi_{27}= \lambda^\rho_{27}=2,\quad 
\lambda^\rho_{10}=\lambda^\rho_{10^*}=0,\quad
\lambda^{\omega_1}_{8}=0\,.
\label{eq:12}
\end{equation}
Exact SU(3) symmetry has been assumed throughout in this discussion, so $\pi$
refers to the full $\pi$ octet, and so on. Note that no configuration mixing
(e.g. $|\pi\rho;8_s\rangle \to |\pi\rho;8_a\rangle$) takes place within the WT
interaction. These results will be used next.

\subsection{Spin-flavor and chiral symmetries}
\label{sec:su6}

With the inclusion of spin there are 36 quark-antiquark
($q\bar{q}$) states, and the SU(6) group representation reduction
(denoting the SU(6) multiplets by their dimensionality and an SU(3)
multiplet $\mu$ of spin $J$ by
$\mu_{2J+1}$) reads
\begin{equation}
{\bf 6}\otimes {\bf 6^*} = {\bf 35} \oplus {\bf 1} = 
\underbrace{8_1 \oplus 8_3 \oplus 1_3}_{\bf 35} \oplus \underbrace{1_1}_{\bf 1}
.
\end{equation}
The lowest bound $q\bar{q}$ state is expected to be an $s-$state and the
relative parity of a fermion-antifermion pair being odd, the octet of
pseudoscalar ($K, \pi,\eta, {\bar K}$) and the nonet of vector ($K^*,
\rho,\omega, {\bar K}^{*}, \phi$) mesons are commonly placed in the {\bf 35}
representation \cite{Gursey:1992dc,Pais:1964,Sakita:1964qq}. 

Strong interaction conserves total spin ($J$), hypercharge ($Y$), and isospin
$(I)$ (assuming equal masses for the up and down quarks).  Furthermore, since
we consider only $s-$wave states, the total spin of the meson-meson states is
simply the sum of their individual spins.  Therefore, on account of the SU(6)
group reduction
\begin{equation}
{\bf 35}\otimes {\bf 35}={\bf 1}\oplus {\bf 35}_s\oplus {\bf 35}_a\oplus
{\bf 189}\oplus {\bf 280}\oplus {\bf 280}^* \oplus {\bf 405}\, ,
\label{eq:13}
\end{equation}
a meson-meson state written in terms of the SU(6) basis takes the form
\begin{equation}\label{eq:mm}
 |M_1M_2;YIJ\rangle=\sum_{\mu,\alpha,\RR}
\left(\begin{array}{cc|c}
\mu_{M_1} &\mu_{M_2} &\mu\\ I_{M_1} Y_{M_1} &I_{M_2} Y_{M_2} & IY
\end{array}\right)
%\times
\left(\begin{array}{cc|c}
{\bf 35} & {\bf 35} & \RR\\ \mu_{M_1}J_{M_1} &\mu_{M_2}J_{M_2}&
        \mu J\alpha
       \end{array}
\right)
|\RR;\mu^\alpha_{2J+1} I Y\rangle,
\end{equation}
where $Y=Y_{M_1}+Y_{M_2}$, $|I_{M_1}-I_{M_2}|\le I\le
I_{M_1}+I_{M_2}$, $|J_{M_1}-J_{M_2}|\le J\le J_{M_1}+J_{M_2}$, $\mu$
and $\RR$ denote generic SU(3) and SU(6) representations, respectively.
$I_{M_{1,2}}$, $Y_{M_{1,2}}$, $J_{M_{1,2}}$ are the isospin,
hypercharge, and spin of the two mesons. In the above equation,
$\RR={\bf 1}, {\bf 189}, {\bf 35}_s, {\bf 405}, {\bf 35}_a, {\bf 280}
$ and ${\bf 280}^*$,
$\mu=1,8_s,8_a,27,10,10^*$, and $\alpha$ accounts for the
multiplicity of each of the $\mu_{2J+1}$ SU(3) multiplets of spin
$J$.  In Eq.~(\ref{eq:mm}), the two coefficients multiplying each
element of the SU(6) basis, $|\RR,\mu^{\alpha}_{2J+1}IY\rangle$,
are the SU(3) isoscalar factors~\cite{deSwart:1963gc} and the SU(6)
Clebsch-Gordan coefficients~\cite{Cook65}, respectively.

At this point one can ask how does SU(6) symmetry go along with chiral
symmetry. Certainly, because chiral symmetry must be present in any reliable
approach this is a central point in this work. To clarify this issue we will
consider the following exercise, namely, whether it is possible for an SU(6)
invariant interaction to reproduce the low energy theorems quoted in
Eq.~(\ref{eq:9}) with the correct eigenvalues in Eq.~(\ref{eq:12}). (Again we
assume exact chiral symmetry and $s$-wave.) As it turns out, this is indeed
possible. Such solutions correspond to operators ${\mathcal H}^{\rm SU(6)}(s)$
acting on the spin-flavor space ${\bf 35}\otimes {\bf 35}$ of the form
\begin{equation}
{\mathcal H}^{\rm SU(6)}(s)
= 
\frac{1}{2f^2}\sum_\RR F_\RR(s)\lambda_\RR P_\RR 
\label{eq:16}
\end{equation}
where $\RR$ runs over the seven SU(6) irreps in Eq.~(\ref{eq:13}) and $P_\RR$
are the corresponding projectors.  The functions $F_\RR(s)$ vanish at
threshold and are normalized by the condition $dF_\RR(s)/ds|_{\rm
  threshold}=1$, e.g. $F_\RR(s)=s-\frac{1}{2}\sum_{i=1}^4 q_i^2$. Therefore,
\begin{equation}
\frac{d{\mathcal H}^{\rm SU(6)}(s)}{ds}\Big|_{\rm threshold}
=
\frac{1}{2f^2}\hat H^{\rm SU(6)}_{\rm WT}
\label{eq:17}
\end{equation}
with
\begin{equation}
\hat H^{\rm SU(6)}_{\rm WT} = \sum_\RR \lambda_\RR P_\RR \,
.
\end{equation}
Finally, the eigenvalues $\lambda_\RR$
reproducing those in (\ref{eq:12}) are\footnote{For $\pi M$ scattering
the relation between SU(3) and SU(6) eigenvalues is
\begin{equation}
\lambda^M_{\mu_{2J+1}\alpha} =
2\sum_{\RR}
\lambda_\RR \left(\begin{array}{cc|c}
{\bf 35} & {\bf 35} & \RR\\ 8_1 & \mu^M_{2J+1}&
        \mu_{2J+1}\alpha
       \end{array}
\right)^2
.
\label{eq:19}
\end{equation}
}
\begin{equation}
\lambda_{{\bf 1}}=-12,\quad
\lambda_{{\bf 35}_s}=\lambda_{{\bf 35}_a}=-6,\quad
\lambda_{{\bf 189}}=-2,\quad
\lambda_{{\bf 405}}=2,\quad
\lambda_{{\bf 280}}=\lambda_{{\bf 280}^*}=0.
\label{eq:18}
\end{equation}

Several comments are pertinent here. i) The functions $F_\RR(s)$ depend on the
concrete model.  Chiral symmetry fixes the derivative of ${\mathcal H}^{\rm
  SU(6)}(s)$ with respect to $s$ at threshold. (A detailed model is developed
below.)  ii) The eigenvalues $\lambda_\RR$ are unique and are such that $\hat
H^{\rm SU(6)}_{\rm WT}$, when restricted to the $\pi\pi$, $\pi\rho$ and
$\pi\omega_1$ subspaces yield the correct SU(3) eigenvalues of subsection
\ref{sec:ii.su3}. iii) The projectors on antisymmetric representations vanish
on $PP$ states. However, for the more general case involving vector mesons,
both symmetric and antisymmetric representations (e.g. ${\bf 35}_s$ and ${\bf
  35}_a$) are required even in $s$-wave. Although the $\pi$ octet and the
$\rho$ nonet fall in the same SU(6) representation they are kinematically
distinguishable through their mass. To give mass to the vector mesons
certainly requires breaking SU(6) in the Lagrangian (not only through mass
terms but also by interaction terms, due to chiral symmetry). This simply
means that ${\mathcal H}^{\rm SU(6)}(s)$ is not the full ${\mathcal H}(s)$
acting on the space ${\bf 35}\otimes{\bf 35}$. Besides ${\mathcal H}^{\rm
  SU(6)}(s)$ there are further terms, $\delta{\mathcal H}(s)$, which do not
have a contribution to ${\mathcal H}(s_{\rm threshold})$ nor $d{\mathcal
  H}(s)/ds|_{\rm threshold}$ when they are restricted to the subspaces $PP\to
PP$ or $PV\to PV$. (Once again we refer to the model below which fulfills
these requirements.)  iv) $\hat H^{\rm SU(6)}_{\rm WT}$ can be regarded as an
extension from WT in flavor SU(3) to a WT-like term in spin-flavor SU(6). The
eigenvalues $\lambda_\RR$ in Eq.~(\ref{eq:18}) obey the general WT rule in
Eq.~(\ref{eq:11}) applied to SU(6) instead of SU(3). Actually there is an
extra factor of two in Eq.~(\ref{eq:17}) since the symmetry factor is $\xi=1/2$
for ${\bf 35}\times {\bf 35}$. However this is not related to the validity of
Eq.~(\ref{eq:11}) in the SU(6) extended version of WT but to the fact that
$f_6= f/\sqrt{2}$ applies instead of $f$ in this extended version. (The same
factor 2 appears in Eq.~(\ref{eq:19}).) No such factors appear when SU($N_F$) is
extended to SU($N_F^\prime$) (a larger number or flavors). This is because the
embedding of SU($N_F$) into SU($N_F^\prime$) is different from the embedding
of SU($N_F$) into spin-flavor SU($2N_F$). (See below.)  As will be obvious
from what follows the same exercise can be repeated, successfully, for any
number of flavors and not only for meson-meson scattering but also for
meson-baryon \cite{GarciaRecio:2005hy,GarciaRecio:2006wb}.

The previous discussion suggests that chiral symmetry, SU(3)$_{\rm
  L}\otimes$SU(3)$_\rR$, is compatible with spin-flavor symmetry,
  SU(6). (Note that 10 couplings, $\lambda_\mu$, have been reproduced
  using only 7 unknowns, $\lambda_\RR$, and a similar
  overdetermination exists for more flavors of for meson-baryon.) In
  fact such compatibility was exposed by Caldi and Pagels in
  \cite{Caldi:1975tx,Caldi:1976gz} by the simple method of extending
  SU(3)$_\rL$ to SU(6)$_\rL$ and SU(3)$_\rR$ to
  SU(6)$_\rR$ where SU(6) refers to spin-flavor. This produces a
  larger symmetry group, SU(6)$_\rL\otimes$SU(6)$_\rR$, which
  includes chiral and spin-flavor groups as subgroups. Specifically,
  the usual spin-flavor SU(6) corresponds to the subgroup of diagonal
  transformations (i.e., the same SU(6) transformation in L and R
  sectors) similar to SU(3)$_{\rm V}$ (flavor group) in SU(3)$_{\rm
  L}\otimes$SU(3)$_\rR$.

The spin-flavor extended chiral group SU(6)$_{\rm
L}\otimes$SU(6)$_\rR$ is a realization of the
Feynman--Gell-Mann--Zweig algebra \cite{Feynman:1964fk} and was
introduced in \cite{Caldi:1975tx,Caldi:1976gz} precisely to solve an
apparent inconsistency. Namely, on the one hand the phenomenological
successful spin-flavor symmetry in the quark model puts $\pi$ and
$\rho$ in the same SU(6) multiplet. On the other, the pion is a
collective state, the Nambu-Goldstone boson from spontaneous breaking
of chiral symmetry. In the scenario of
\cite{Caldi:1975tx,Caldi:1976gz}, one would find that in an exactly
SU(6) symmetric world the $\pi$ octet and the $\rho$ nonet are the
Nambu-Goldstone bosons of the spontaneous breaking of SU(6)$_{\rm
L}\otimes$SU(6)$_\rR$ down to SU(6). Spin-flavor symmetry is an
approximated one and so the $\rho$ nonet mesons are regarded as ``dormant''
Nambu-Goldstone bosons.  As it is known, spin-flavor symmetries cannot
be exact as they cannot be accommodated with full Poincare invariance
\cite{Coleman:1967ad}. Still one can consider a static limit enjoying
SU(6)$_\rL\otimes$SU(6)$_\rR$ symmetry. In the Caldi-Pagels
scenario, relativistic (and so SU(6) breaking) corrections give mass
to the vector mesons while pions are still protected by the usual
SU(3) chiral symmetry.

The scenario just described solves a number of puzzles involving vector mesons
while maintaining vector meson dominance, KSFR relations and so on
\cite{Caldi:1975tx,Caldi:1976gz}. Here we comment only on two issues, namely,
the consequences regarding the chiral and Lorentz transformations of vector
mesons. Because the pion falls in the $(3,3^*)+(3^*,3)$ representation of the
chiral group, spin-flavor symmetry requires the $\rho$ to fall in the same
representation (and both in $(6,6^*)+(6^*,6)$ of SU(6)$_\rL\otimes$SU(6)$_{\rm
  R}$). This is different from vector and axial currents which transform
instead as $(8,1)+(1,8)$ under the chiral group. At first the fact that the
$\rho$ meson and the vector current transform differently seems to be in
conflict with vector meson dominance. As shown in
\cite{Caldi:1975tx,Caldi:1976gz} this not so, due to the spontaneous breaking
of chiral symmetry, for the same reason that PCAC relates pion and axial
current, also in different chiral representations.

Related to the chiral representation is the nature of vectors mesons under
Lorentz transformations. This is most easily exposed by coupling the meson
fields to quark bilinears (alternatively the quark bilinear can be regarded as
a representation or interpolating field of the meson, as in
Nambu--Jona-Lasinio models). Let us for this discussion consider just two
flavors ($N_F=2$) and use a linear sigma model representation (as opposed to
the non-linear one) as there it is simpler to expose the chiral transformation
properties of the fields.  The pion and $\sigma$ mesons couple to $\bar q
i\gamma_5 \vec\tau q$ and $\bar q q$. Of course, this just of the form $\bar
q_\rL {\mathcal M}q_\rR + \bar q_{\rm R} {\mathcal M}^\dagger q_\rL$
corresponding to the chiral representation $(1/2,1/2)$. The coupling can be
extended to include vector mesons while preserving spin-flavor
\begin{equation}
{\mathcal M} = \sigma + i\pi_a \tau_a 
+ i \rho_{ai} \tau_a \sigma_i 
+ \cdots .
\end{equation}
(Let us remark that these are the linear sigma model mesons fields and
will be used only in this subsection. Elsewhere in this paper the
non-linear meson fields are used. Also note that this ${\mathcal M}$
is unrelated to the mass term in Eq.~(\ref{eq:chpt}).) The dots
represent further meson fields to complete a general $2N_F\times 2N_F$
complex matrix. The space spanned by such matrices ${\mathcal M}$
carries a representation of the group ${\rm SU}(2N_F)_\rL\otimes {\rm
SU}(2N_F)_\rR$, acting as ${\mathcal M}\to\Omega_\rL^\dagger{\mathcal
M} \Omega_\rR$ with $\Omega_{\rm L,R}\in$SU($2N_F$). This
SU($2N_F$)$\supset$SU($N_F$)$\otimes$SU(2) is generated by $\tau_a$,
$\sigma_i$ and $\tau_a\sigma_i$. In particular, the $\sigma$ and $\pi$
fields mix under transformations generated by $\tau_a$ with
$\Omega_\rL\not=\Omega_\rR$. These are the usual chiral
transformations. On the other hand $\pi$ mixes with $\rho$ (and other
mesons) under spin-flavor transformations ($\Omega_\rL=\Omega_{\rm
R}$).  Using the chiral representation of the Dirac gammas one
immediately obtains the coupling
\begin{equation}
\bar q_\rL {\mathcal M}q_\rR + \bar q_\rR {\mathcal M}^\dagger
q_\rL
=
\bar{q}(\sigma +i\pi_a \tau_a\gamma_5
+ \rho_{ai} \tau_a \sigma^{0i} 
+ \cdots
)q
\,.
\label{eq:22}
\end{equation}
The LR+RL structure (i.e. $(3,3^*)+(3^*,3)$) requires quark bilinears
constructed with $1$, $\gamma_5$ or $\sigma_{\mu\nu}$ which commute
with $\gamma_5$, while $\gamma_\mu$ or $\gamma_\mu \gamma_5$ produce
LL+RR (i.e. $(1,8)+(8,1)$). This implies that the spin-flavor approach
favors an antisymmetric tensor representation of vector mesons
\cite{Gasser:1983yg}. The antisymmetric tensor $V_{\mu\nu}$ contains 3
spurious degrees of freedom and one can choose to take $V_{0i}$ as the dynamical fields \cite{Ecker:1988te}. This
is the choice in Eq.~(\ref{eq:22}) with $\rho_{ai}\sim\bar{q}\tau_a
\sigma^{0i}q$. The description of vector mesons using antisymmetric
tensors has been shown to be consistent with all expected properties
of vector mesons \cite{Ecker:1988te,Ecker:1989yg,Birse:1996hd}.  (See
e.g. \cite{Jaminon:1998de} for the use of $\bar{q}\sigma_{\mu\nu}q$ as
interpolating field of the vector meson in the context of
Nambu--Jona-Lasinio models.)

Conversely, the quark bilinear construction naturally favors a
$(3,3^*)+(3^*,3)$ representation instead of $(1,8)+(8,1)$ for vector
mesons if they are considered as antisymmetric tensors. In principle,
one would expect that the chiral representation under which the meson
transforms would reflect itself on the observable properties of the
meson. However, this is not at all obvious. In the context of
effective chiral Lagrangians for mesons a very convenient treatment is
that based on the non-linear realization of chiral symmetry
\cite{Weinberg:1968de}. (Let us remark that only the linear
realization is used in this work.)  In this approach a field $u$ is
constructed out of $\mathcal M$ such that under chiral transformations
$u\to \Omega^\dagger_\rL u h = h^\dagger u \Omega_\rR$ (where $u$ and
$h$ are unitary matrices) \cite{Birse:1996hd}. A field of the type LL
such as the chiral current $V_\rL^\mu$ transforms as
$\Omega_\rL^\dagger V_\rL^\mu \Omega_\rL$, and so belongs to the
chiral representation (8,1). This field is represented in the
non-linear realization by the new field $\tilde{V}_\rL^\mu= u^\dagger
V_\rL^\mu u$ which transforms instead as $h^\dagger \tilde{V}_\rL^\mu
h$. Likewise, the field $\mathcal M$, of the type LR, will be
represented by $\tilde{\mathcal M}= u^\dagger \mathcal M
u^\dagger$. This new field transforms as $h^\dagger \tilde{\mathcal M}
h$, that is, exactly in the same way as the chiral currents or the
vector or axial currents, etc. That was precisely the point of the
non-linear realization, namely, all fields in the same representation
with respect to SU$(N_F)_V$ will be represented by fields transforming
in the same way under general chiral transformations, regardless of
their detailed chiral representation.  Therefore, such detailed chiral
representation does not reflect on the properties of the meson, at
least to the extend that effective chiral Lagrangians are sufficient
to describe them.  This should not be surprising as it was already
noted before that, e.g., the WT term is only sensible to the isospin
(or more generally, flavor) of the target.

\subsection{The model: SU(6) invariant part}

In view of the previous remarks, we introduce now a model for meson
interaction, including the $\pi$ octet and the $\rho$ nonet, with
simultaneous chiral symmetry and spin-flavor symmetry, suitably broken.

The natural SU(6) extension of Eq.~(\ref{eq:chpt}) from SU(3) to SU(6) is
\begin{equation}
 \mathcal{L}_\mathrm{SU(6)}=\frac{f^2_6}{4}\mathrm{Tr}\left(\partial_\mu U^\dagger_6
 \partial^\mu U_6+\mathcal{M}_6(U_6+U_6^\dagger-2)\right), \qquad U_6=e^{{\rm
    i}\sqrt{2}\Phi_6/f_6} 
.
\label{eq:lsu6}
\end{equation}
$U_6$ is a unitary $6\times 6$ matrix that transforms under the linear
realization of SU(6)$_\rL\otimes$SU(6)$_{\rm R}$.  The Hermitian
matrix $\Phi_6$ is the meson field, in the {\bf 35} irreducible
representation of SU(6), and $f_6= f/\sqrt{2}$, as shown in Appendix B
of Ref.~\cite{GarciaRecio:2006wb}.  The first term in
$\mathcal{L}_\mathrm{SU(6)}$ preserves both chiral and spin-flavor
symmetry. The second term breaks chiral symmetry and possibly flavor
symmetry. This is not the most general breaking and this issue will be
discussed in the next subsection. For the time being this term will be
kept for illustration purposes with $\mathcal{M}_6=m_6 I_{6\times 6}$,
with $m_6$ a common mass for all mesons belonging to the SU(6) {\bf
35} irreducible representation.

Expanding the previous Lagrangian up to ${\mathcal O}(\Phi_6^4)$ gives
the interaction Lagrangian,
\begin{equation}
 \mathcal{L}_\mathrm{SU(6)}^{\rm int}= 
\frac{1}{12 f^2_6}\mathrm{Tr}\left ( [\Phi_6,
   \partial_\mu\Phi_6] [\Phi_6, \partial^\mu\Phi_6]
+m_6 \Phi_6^4\right) .
\label{eq:su6-ampl}
\end{equation}
The restriction of this Lagrangian to the SU(3) pseudoscalar $8
\otimes 8$ sector reproduces that given in Eq.~(\ref{eq:su3chpt})
(with a common mass for all pseudoscalars).  The kinetic term in
Eq.~(\ref{eq:su6-ampl}) amounts to a coupling of the type $\left[({\bf
35}\otimes {\bf 35})_{{\bf 35}_a}\otimes({\bf 35}\otimes {\bf
35})_{{\bf 35}_a}\right]_{\bf 1}$ in the $t-$channel. That is, each
meson pair, respecting Bose statistics, is coupled to the
antisymmetric SU(6) adjoint representation (${\bf 35}_a$) and the two
resulting ${\bf 35}_a$'s couple into the singlet one to built up an
SU(6) invariant interaction. This mechanism is completely analogous to
that in Eq.~(\ref{eq:su3chpt}) for SU(3) and so it is a natural
extension of the WT chiral Lagrangian.

$\Phi_6$ is a dimension six matrix made of full meson fields, which
depend on the space-time coordinates.  A suitable choice for the
$\Phi_6$ field is\footnote{Matrices, $A^i_j$, in the dimension 6 space
are constructed as a direct product of flavor and spin matrices. Thus,
an SU(6) index $i$, should be understood as $i \equiv
(\alpha,\sigma)$, with $\alpha= 1,2,3$ and $\sigma=1,2$ running over
the (fundamental) flavor and spin quark degrees of freedom,
respectively.}
\begin{equation}
\Phi_6 = \underbrace{P_a \frac{\lambda_a}{\sqrt{2}}\otimes
  \frac{I_{2\times 2}}{\sqrt{2}}}_{{\displaystyle {\Phi_P}}} + \underbrace{R_{ak}
\frac{\lambda_a}{\sqrt{2}} \otimes \frac{\sigma_k}{\sqrt{2}} + W_k
\frac{\lambda_0}{\sqrt{2}} \otimes
  \frac{\sigma_k}{\sqrt{2}}}_{\displaystyle{\Phi_V}}, \quad a=1,\ldots,8,\quad
  k=1,2,3 
\end{equation}
with $\lambda_a$ the Gell-Mann and $\vec{\sigma}$ the Pauli spin
matrices, respectively, and $\lambda_0= \sqrt{2/3} \,I_{3\times 3}$
($I_{n\times n}$ denotes the identity matrix in the $n$ dimensional
space). $P_a$ are the $\pi, K, \eta$ fields, while  $R_{ak}$ and $W_k$
stand for the $\rho-$vector nonet fields, considering explicitly the
spin degrees of freedom.
The annihilation part of the meson matrix $[\Phi_6]^i_j$ is determined by the
operators $M^i_{\ j}$.  Regarding $M$ as a matrix with respect to $i$ and $j$,
the convention is that the upper/lower index acts as the first/second index of
the matrix. $M$ is traceless and transforms under SU(6) in the same way as the
quark operators
\begin{equation}
 q^i{\bar q}_j - \frac{1}{2N_F} q^m{\bar q}_m \delta^i_j,\quad
i,j=1,\ldots 2N_F,
\label{eq:defm}
\end{equation}
where $N_F$ is the number of flavors, three in this work.

We have denoted the contravariant and covariant spin-flavor quark  and
antiquark components
\begin{equation}
q^i = \left(\begin{array}{c}
u\uparrow \cr d\uparrow \cr s\uparrow \cr u\downarrow \cr d\downarrow \cr
s\downarrow 
\end{array}\right),
\qquad 
{\bar q}_i = \left(
{\bar u}\downarrow,-{\bar
  d}\downarrow,-{\bar s}\downarrow, -{\bar u}\uparrow, {\bar
  d}\uparrow, {\bar s}\uparrow  
\right)
\end{equation}
where $q^i$ (${\bar q}_i$) annihilates\footnote{Our convention is
such that $\left(\begin{array}{c}{\bar d}\cr {\bar
u}\end{array}\right)$ is a standard basis of SU(2), that is, ${\bar d}
= |1/2,1/2\rangle $ and ${\bar u} = |1/2,-1/2\rangle $. Thus, ${\bar
u}, {\bar d} , {\bar s}$ is a standard basis of the $3^*$
representation of SU(3) with the Swart's
convention~\cite{deSwart:1963gc}.} a quark (antiquark) with the
spin-flavor $i$. For instance ${\bar u}\downarrow$ annihilates an
antiquark with flavor ${\bar u}$ and $S_z=-1/2$.  The corresponding
Wick's contractions of these operators read
\begin{eqnarray}
\underwick{1}{<1M^k_l >1M^{\dagger i}_j}&=&
\delta^k_j\delta^i_l-\frac{1}{2N_F} \delta^i_j\delta^k_l \label{eq:wick}
\end{eqnarray}

\begin{figure}[b]
%\vspace{-3cm}
\centerline{\includegraphics[height=3.cm]{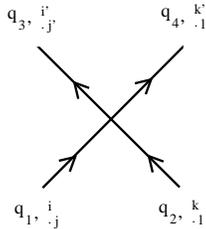}}
%\vspace{-15cm}
\caption{%\footnotesize
Diagrammatic representation of four meson  scattering in SU(6).    }
\label{fig:mm}
\end{figure}

For the process depicted in Fig.~\ref{fig:mm}, the Lagrangian of
Eq.~(\ref{eq:su6-ampl}) provides the following amplitudes (${\mathcal H}^{\rm
  SU(6)}$)
\begin{eqnarray}
{\mathcal H}^{\rm SU(6)} &=&
{\mathcal H}^{\rm SU(6)}_+ + {\mathcal H}^{\rm SU(6)}_-
\end{eqnarray}
with
\begin{eqnarray}
{\mathcal H}^{\rm SU(6)}_+ &=& 
\frac{1}{12f^2_6} \left(3s-\sum_{i=1}^4 q_i^2\right)\langle 0 |
M^{i^\prime}_{j^\prime}M^{k^\prime}_{l^\prime} \hat {\mathcal G}_+ M^{\dagger i}_{j}M^{\dagger
  k}_{l} | 0 \rangle -\frac{m^2_6}{12f^2_6} \langle 0 |
M^{i^\prime}_{j^\prime}M^{k^\prime}_{l^\prime} \hat {\mathcal G}_{\mathcal M} M^{\dagger i}_{j}M^{\dagger
  k}_{l} | 0 \rangle
,
\\
{\mathcal H}^{\rm SU(6)}_- &=& \frac{u-t}{4f^2_6} 
\left ( {\mathcal G}_d - {\mathcal G}_c \right)
,
\end{eqnarray}
where
\begin{equation}
\hat {\mathcal G}_+ \, = \,\, 
:\frac{1}{2} {\rm Tr}\left([M^\dagger,M]^2\right): \,, 
 \qquad
\hat {\mathcal G}_{\mathcal M} \, = \,\, 
: \mathrm{Tr}\left((M+M^\dagger\right)^4): \,,
\end{equation}
and
\begin{eqnarray}
\langle 0 |
M^{i^\prime}_{j^\prime}M^{k^\prime}_{l^\prime} \hat {\mathcal G}_+ 
M^{\dagger i}_{j} M^{\dagger k}_{l} | 0 \rangle 
&=& 
%\overbrace{
\underwick{2112}{<1M^{i^\prime}_{j^\prime} <2M^{k^\prime}_{l^\prime} {\rm Tr}\big(
[>2M^\dagger, <3M] [>1M^\dagger, <4M]\big) >4M^{\dagger i}_{j}>3M^{\dagger
  k}_{l}}
%}^{{\mathcal G}_d} 
\nonumber \\ &&
+ 
%\overbrace{
\underwick{2112}{<1M^{i^\prime}_{j^\prime}
<2M^{k^\prime}_{l^\prime} {\rm Tr}\big(
[>2M^\dagger, <3M] [>1M^\dagger, <4M]\big) >3M^{\dagger i}_{j}>4M^{\dagger
  k}_{l}}
%}^{{\mathcal G}_c} 
\\
&\equiv& {\mathcal G}_d + {\mathcal G}_c 
\,.
\end{eqnarray}
In these expressions $s=(q_1+q_2)^2$, $t=(q_1-q_3)^2$, $u=(q_1-q_4)^2$,
$|0\rangle$ is the hadron vacuum state and $:\cdots:$ denotes the normal
product.

For a fully SU(6) symmetric theory and because of Bose statistics, the
interaction must be symmetric under the simultaneous exchange $(i,j)
\leftrightarrow (k,l)$ and $q_1\leftrightarrow q_2 $ or $(i^\prime,j^\prime)
\leftrightarrow (k^\prime,l^\prime)$ and $q_3\leftrightarrow q_4$. This can be
realized in two different manners: i) being both symmetric in flavor and
momentum spaces, or ii) being both antisymmetric in flavor and momentum
spaces. This corresponds to the decomposition ${\mathcal H}^{\rm
  SU(6)}={\mathcal H}^{\rm SU(6)}_+ + {\mathcal H}^{\rm SU(6)}_-$. The first
of the amplitudes turns out to be purely $s-$wave, while ${\mathcal H}^{\rm
  SU(6)}_-$ describes $p-$wave scattering when mesons are degenerate in mass.

In terms of SU(6) projectors, the above
amplitudes read (see Appendix \ref{app:eigenvalues} for details)
\begin{eqnarray}
{\mathcal H}^{\rm SU(6)}_+ &=& \frac{1}{6f^2} \left(3s-\sum_{i=1}^4
q_i^2\right) \left (-12  P_{\bf 1}-6 P_{{\bf 35}_s}-2P_{{\bf 189}}+2P_{{\bf 405}} \right)
\nonumber \\
&& - \frac{ m^2_6}{3f^2} \left (23  P_{\bf 1}+10 P_{{\bf 35}_s}-2P_{{\bf
    189}}+2P_{{\bf 405}}
\right) 
,
\label{eq:hsu6}
 \\
{\mathcal H}^{\rm SU(6)}_- &=& 3\frac{u-t}{f^2}
P_{{\bf 35}_a}
.
\label{eq:hsu6s}
\end{eqnarray}
%------------------------------------------------------
%
We will not discuss in this work the $p-$wave part, and we will focus
here on the $s-$wave amplitude. However, ${\mathcal H}^{\rm SU(6)}_-$ will
lead to a non-vanishing $s-$wave contribution for pseudoscalar-vector
meson scattering when SU(6) symmetry breaking mass terms are
considered, since in that case $(u-t)$ provides a non zero projection
into $s-$wave. We return to this point below.

\subsection{SU(6) spin-flavor symmetry breaking effects}

\label{sec:su6fb}

The SU(6) spin-flavor symmetry is severely broken in nature. Certainly it is
mandatory to take into account mass breaking effects by using different
pseudoscalar and vector mesons masses. However, this cannot be done by just
using these masses in the kinematics of the amplitudes derived in the previous
subsection as this would lead to flagrant violations of the soft pion theorems
in the $PV\to PV$ sector due to the large vector meson masses. Instead, the
proper mass terms have to be added to the Lagrangian to give different mass to
pseudoscalars and vectors mesons {\em while preserving, or softly breaking,
  chiral symmetry}. In addition, SU(2)$_{\rm spin}$ invariance must also be
maintained since in the $s$-wave sector it is equivalent to angular momentum
conservation.

To this end, we consider the following mass term (which replaces that in
Eq.~(\ref{eq:lsu6}))
\begin{eqnarray}
\mathcal{L}_\mathrm{SU(6)}^{(m)}
&=& 
\frac{f^2_6}{4} \mathrm{Tr}\left(
{\mathcal M}(U_6+U_6^\dagger -2)\right)
+\frac{f^2_6}{32} 
\mathrm{Tr} \left( 
{\mathcal M}^\prime 
(
\vec{\sigma}\,U_6 \,\vec{\sigma} \, U^\dagger_6
+
\vec{\sigma}\,U_6^\dagger \,\vec{\sigma} \,
U_6
-6
)
\right)
.
\label{eq:lm}
\end{eqnarray}

Here the matrix ${\mathcal M}$ acts only in flavor space and is to be
understood as ${\mathcal M}\otimes I_{2\times 2}$, and similarly for
${\mathcal M}^\prime$, so that SU(2)$_{\rm spin}$ invariance is
preserved. Besides, these matrices should be diagonal in the isospin
basis of Eq.~(\ref{eq:su3phi}) so that charge is conserved. Also,
$\vec{\sigma}$ stands for $I_{3\times 3}\otimes \vec{\sigma}$.

The first term in $\mathcal{L}_\mathrm{SU(6)}^{(m)}$ is fairly standard. It
preserves spin-flavor symmetry when ${\mathcal M}$ is proportional to the
identity matrix and introduces a soft breaking of chiral symmetry when
${\mathcal M}$ is small. As it is shown below this term gives the same mass to
pseudoscalar and vector mesons multiplets. Note that terms of this type are
sufficient to give different mass to pseudoscalars (e.g. $\pi$ and $K$) when
SU($N_F)$ is embedded into SU($N_F^\prime$) (a larger number of flavors). They
are not sufficient however to tailor different $P$ and $V$ masses when
SU($N_F$) is embedded into SU($2N_F$) (spin-flavor).

The second term in $\mathcal{L}_\mathrm{SU(6)}^{(m)}$ only gives mass
to the vector mesons: indeed, if one would retain in $U_6$ only the
pseudoscalar mesons, $U_6$ would cancel with $U_6^\dagger$ (since
these matrices would commute with $\vec\sigma$) resulting in a
cancellation of the whole term. This implies that this term does not
contain contributions of the form $PP$ (pseudoscalar mass terms) nor
$PPPP$ (purely pseudoscalar interaction). In addition, when ${\mathcal
M}^\prime$ is proportional to the identity matrix (i.e., exact flavor
symmetry) chiral symmetry is also exactly maintained, because the
chiral rotations of $U_6$ commute with $\vec\sigma$.  This guarantees
that this term will produce the correct $PV\to PV$ contributions to
ensure the fulfillment of soft pion WT
theorem~\cite{Weinberg:1966kf,Tomozawa:1966jm} even when the vector
mesons masses are not themselves small.\footnote{Under a chiral
transformation, the vector meson mass term ($m_V^2
\mathrm{Tr}(\Phi_V^2))$) will give rise to a $PVV$ term which can only
get canceled by the corresponding variation of the $PPVV$ contact term
(recall $\delta P \sim {\mathcal O}(1)$ and $\delta V \sim {\mathcal
O}(PV, VV)$) to ensure that the whole Lagrangian is invariant. Thus
the contact $PPVV$ piece is fixed by chiral symmetry.}

Expanding to order $\Phi_6^2$ to isolate the genuine mass terms involved, we
find
\begin{eqnarray}
\mathrm{Tr}\left({\mathcal M}(U_6+U_6^\dagger -2)\right) 
&=& 
-\frac{2}{f^2_6}
     \mathrm{Tr}\left({\mathcal M}\,\Phi_6^2\right)+
     {\mathcal O}(\Phi_6^4) = -\frac{2}{f^2_6}
     \mathrm{Tr}\left({\mathcal M}(\Phi_P^2+\Phi_V^2)\right)+
     {\mathcal O}(\Phi_6^4)
,
\nonumber  \\
\mathrm{Tr}
\left( {\mathcal M}^\prime(\vec{\sigma}\,U_6 \,\vec{\sigma} \, U^\dagger_6-3) \right) 
&=& 
\frac{1}{f^2_6}
     \mathrm{Tr}\left({\mathcal M}^\prime\,
[\vec{\sigma},\Phi_6]\,[\vec{\sigma},\Phi_6]
\right) +
     {\mathcal O}(\Phi_6^3)
=
-\frac{8}{f^2_6} \mathrm{Tr}\left({\mathcal M}^\prime
\,\Phi_V^2\right) +
     {\mathcal O}(\Phi_6^3)
.
\end{eqnarray}
Therefore
\begin{eqnarray}
\mathcal{L}_\mathrm{SU(6)}^{(m)} &=& - 
\frac{1}{2} \mathrm{Tr}\left({\mathcal M}\,\Phi_P^2\right) - 
\frac{1}{2} \mathrm{Tr}\left(
({\mathcal M}+{\mathcal M}^\prime)
\Phi_V^2\right) + {\mathcal O}(\Phi_6^4)
.
\end{eqnarray}

As advertised, $\mathcal{M}$ is the only source of mass for the pseudoscalars
and so $\mathcal{M}=\mathrm{diag}(m_\pi^2,m_\pi^2,2m_K^2-m_\pi^2)\,I_{2\times
  2}$ is the usual SU(3) mass breaking matrix. On the other hand, vector
mesons pick up a contribution to their mass from both $\mathcal{M}$ and
$\mathcal{M}^\prime$.

For simplicity, in this exploratory work we will neglect the chiral breaking
mass term ($\mathcal{M}=0$) and take a common mass, $m_V$, for all vector
mesons ($\mathcal{M}^\prime= m_V^2 I_{3\times 3}\,I_{2\times 2}$). We use a
vector meson nonet averaged mass value $m_V= 856$ MeV.  Let us stress that the
simplifying choice $\mathcal{M}=0$, $\mathcal{M}^\prime=m_V^2$, refers only to
the interaction terms derived from the Lagrangian
$\mathcal{L}_\mathrm{SU(6)}^{(m)}$. For the evaluation of the kinematical
thresholds of different channels we use physical meson masses.

With the abovementioned choice, picking up the terms of ${\mathcal
  O}(\Phi_6^4)$ in $\mathcal{L}_\mathrm{SU(6)}^{(m)}$ yields the following
four-meson interaction terms
\begin{eqnarray}
\mathcal{L}_\mathrm{SU(6)}^{(m;\, {\rm int} )} &=&
\frac{m_V^2}{8f^2}\mathrm{Tr}\left( 
\Phi_6^4 
+ \vec{\sigma}\,\Phi_6^2 \,\vec{\sigma}  \Phi_6^2  
-\frac{4}{3} \vec{\sigma}\, \Phi_6 \,\vec{\sigma}\, \Phi_6^3 
\right)
.
\label{eq:lmint}
\end{eqnarray}

As noted above, this Lagrangian contains only $PPVV$ and $VVVV$ interactions
and no $PPPP$ ones. In addition, the $PPVV$ terms are consistent with soft
pion (or soft Nambu-Goldstone boson) theorems.

Altogether, the amplitude, ${\mathcal H}$, of the process
 depicted in Fig.~\ref{fig:mm} after  projecting into $s-$wave and for
 massless pseudoscalar bosons and equal mass vector mesons, takes the form
\begin{eqnarray}
{\mathcal H} &=& 
\left ({\mathcal H}^{\rm SU(6)}_+ + \delta {\mathcal H}^{\rm SU(6)}_+ 
+ {\mathcal H}^{\rm SU(6)}_-\right)_{\mathcal{M}=0,\mathcal{M}^\prime=m_V^2}
 = \frac{1}{6f^2} \left(3s-\sum_{i=1}^4
q_i^2\right) {\mathcal D}_{\rm kin} 
+
\frac{m^2_V}{8f^2} {\mathcal D}_m
+  
\frac{1}{2f^2} \frac{m_V^4}{ s} {\mathcal D}_a
.
\label{eq:vsu6}
\end{eqnarray}
Here
${\mathcal D}_{\rm kin}=-12 P_{\bf 1}-6 P_{{\bf 35}_s}-2P_{{\bf 189}}+2P_{{\bf
    405}}$. It accounts for the first (kinetic) term in Eq.~(\ref{eq:hsu6}),
which reduces to the chirally invariant interaction proportional to $\hat H_1$
in Eq.~(\ref{eq:su3}).
${\mathcal D}_m$ is a matrix in spin-flavor space determined by the interaction
$\mathcal{L}_\mathrm{SU(6)}^{(m;\, {\rm int} )}$ of Eq.~(\ref{eq:lmint}). This
matrix is identically zero in the $PP\to PP$ subspace and it cannot be
expressed as a sum of SU(6) projectors, since
$\mathcal{L}_\mathrm{SU(6)}^{(m;\, {\rm int} )}$ breaks spin-flavor symmetry,
though it is of course $YIJ$ block diagonal.

The last matrix, ${\mathcal D}_a$, is just $6 P_{{\bf 35}_a}$ in the $PV\to PV$
sector and zero otherwise.  ${\mathcal H}^{\rm SU(6)}_-$ of Eq.~
(\ref{eq:hsu6s}) has now a nonvanishing contribution and once again,
neglecting these terms would introduce a large violation of chiral symmetry,
proportional to the vector meson mass. This comes about because $(u-t)$ gives
rise to non-zero $s-$wave contributions for $PV\to PV$ scattering, once
pseudoscalar-vector mass breaking is taken into account.  Indeed, in the limit
of massless Nambu-Goldstone bosons and mass degenerated vector mesons, one finds for
the $PV$ sector (assuming that the legs 1 and 3 in Fig.~\ref{fig:mm} are of
type $P$ and 2 and 4 of type $V$)
\begin{equation}
\langle u-t\rangle_{l=0} = \frac{m^4_V}{s}
\qquad (PV\to PV )
.
\label{eq:33}
\end{equation}
The same average vanishes for $PP$ or $VV$ sectors, since there Bose symmetry
still applies.

Regarding the fulfillment of the relations
Eqs.~(\ref{eq:16}-\ref{eq:18}) in subsection \ref{sec:su6}, we can see that 
$\mathcal H$ of Eq.~(\ref{eq:vsu6}) reduces to
\begin{equation}
{\mathcal H}_{PP\to PP}= \frac{s}{2f^2} {\mathcal D}_{\rm kin}
\end{equation}
in the $PP$ sector. On the other hand, in the $PV$ sector, 
the relation
\begin{equation}
\frac{1}{3}{\mathcal D}_{\rm kin}+\frac{1}{4}{\mathcal D}_m+{\mathcal
    D}_a = 0 \qquad (\mbox{\rm $PV\to PV$ sector}),
\end{equation}
guarantees that 
$\mathcal H(s)$  vanishes at threshold in this sector, and moreover
\begin{equation}
{\mathcal H}_{PV\to PV}= \frac{(s-m_V^2)}{2f^2} \left(
{\mathcal D}_{\rm kin}
-\frac{m_V^2}{s}{\mathcal D}_a\right) .
\end{equation}
These expressions fulfill the relations Eqs.~(\ref{eq:16}-\ref{eq:18}) by
taking $F_\RR(s)=s-\frac{1}{2}\sum_{i=1}^4 q_i^2$ for the symmetric
representations and $F_\RR(s)=(s-\frac{1}{2}\sum_{i=1}^4 q_i^2)m_V^2/s$
for the antisymmetric ones.

We have also considered spin-flavor symmetry breaking effects due to
the difference between pseudoscalar and vector meson decay
constants. The pseudoscalar meson decay constants, $f_P$, are defined
by
\begin{equation}
\langle 0 | {\bar q}_1 \gamma^\mu \gamma_5 q_2 (0) | P(p)\rangle =
-{\rm i} \sqrt{2} f_P \, p^\mu
\label{eq:fp}
\end{equation}
and vector meson decay constants, $f_V$, by
\begin{equation}
\langle 0 | {\bar q}_1 \gamma^\mu  q_2 (0) | V(p,\epsilon)\rangle
= \sqrt{2} m_{V}f_{V}\epsilon^\mu ,
\label{eq:fv}
\end{equation}
where ${\bar q}_1$, $q_2$ are the quark fields, $\epsilon_\mu$ is the
polarization vector of the meson, and $m_V$ its mass. With the above
definitions, in the limit where either the quark or the antiquark that forms
the meson becomes infinitely heavy and thus spin symmetry turns out to be
exact, QCD predicts $f_P=f_V$~\cite{Manohar:2000dt}.  This guarantees that the
normalizations of the coupling constants in Eqs.~(\ref{eq:fp})
and~(\ref{eq:fv}) are consistent. For light mesons there exist sizable
corrections to the heavy quark symmetry-type relation $f_P=f_{V}$.  For
instance, the ratio $f_\rho/f_\pi$ is of the order of 1.7. To take this into
account, in Eq.~(\ref{eq:vsu6}) we apply the prescription
\begin{equation}
\frac{1}{f^2} \to  \frac{1}{\left(f_1 f_2 f_3 f_4\right)^{1/2}} 
\end{equation}
where the labels $1,2,3,4$ refer the four interacting mesons.

The meson decay constants (taken from Ref.~\cite{GarciaRecio:2008dp})
and masses used throughout this work to compute the kinematical
thresholds and loop functions are compiled in Table~\ref{tab:cts},
while the coupled-channel matrices ${\mathcal D}_{\rm kin}$,
${\mathcal D}_{m}$ and ${\mathcal D}_a$ can be found in
Tables~\ref{tab:initial}--\ref{tab:final} of Appendix
\ref{app:tables}. We assume an ideal mixing in the vector meson
sector, namely, $\omega=\sqrt{\frac{2}{3}}\omega_1+\frac{1}{\sqrt
3}\omega_8$ and $\phi=\sqrt{\frac{2}{3}}\omega_8-\frac{1}{\sqrt
3}\omega_1$. The conventions of \cite{deSwart:1963gc} are used
throughout. Note that for the $Y=0$ channels, $G-$parity is
conserved,\footnote{Recall that the $G-$parity operation can be
defined through its action on an $YII_3$ eigenstate as $G|YII_3\rangle
= \chi (-1)^{Y/2+I} |-YII_3\rangle$, with $\chi$ the charge
conjugation of a neutral non-strange member of the SU(3) family. } and
that all $Y=0$ states have well-defined $G-$parity except the $\bar
K^* K $ and $ K^* \bar K $ states, but the combinations $\left(\bar K
K^* \pm K \bar K^* \right)/\sqrt 2$ are actually $G-$parity
eigenstates with eigenvalues $\pm 1$. These states will be denoted
$(\bar{K}K^*)_S$ and $(\bar{K}K^*)_A$, respectively.

A final remark is in order here. The new model introduced in this work
is given by $\mathcal{L}_\mathrm{SU(6)}^{\rm kin} +
\mathcal{L}_\mathrm{SU(6)}^{(m)}$ (namely, the first term in
Eq.~(\ref{eq:lsu6}) and that in Eq.~(\ref{eq:lm})).  It implements the
approximate spin-flavor chiral symmetry (as opposed to the standard
flavor chiral symmetry) advocated by Caldi and Pagels
\cite{Caldi:1975tx,Caldi:1976gz}. In their approach vector mesons are
identified as dormant Nambu-Goldstone
bosons. $\mathcal{L}_\mathrm{SU(6)}^{\rm kin}$ displays such extended
chiral symmetry, while $\mathcal{L}_\mathrm{SU(6)}^{(m)}$ models the
pattern of symmetry breaking.  Regarding this latter term, it should
be noted that there is a large ambiguity in choosing it. Being a
contact term, it cannot contain $PPPP$ contributions, due to chiral
symmetry, and for the same reason the terms $PPVV$ are also fixed, as
already noted. However, $VVVV$ terms are not so constrained. One can
easily propose alternative forms for
$\mathcal{L}_\mathrm{SU(6)}^{(m)}$ which would still be acceptable
from general requirements but would yield different $VVVV$
interactions. For instance, any term of the form $\mathrm{Tr}\left(
{\mathcal M}\, \vec{\sigma} U_6 \vec{\sigma} U^\dagger_6 \vec{\sigma}
U_6 \vec{\sigma} U_6^\dagger \cdots \right)$, with the indices of the
$\vec{\sigma}$ matrices contracted in any order, could be present in
$\mathcal{L}_\mathrm{SU(6)}^{(m)}$. Our choice in Eq.~(\ref{eq:lm}) is
just the simplest or minimal one.\footnote{As it turns out, the same
term has been proposed by Caldi \cite{Caldi:1984xr} as a Lorentz
symmetry restoration correction.} Of course, such minimal choices are
also present in any other model, often tied to some expansion
parameter. We have not yet identified a hierarchy to choose among the
various available operators. Ultimately, the ambiguity should be fixed
by requiring consistency with the asymptotic behavior of
QCD~\cite{Ecker:1989yg}. In what follows we will present results
obtained with the interaction ${\mathcal H}$ given in
Eq.~(\ref{eq:vsu6}).

\section {BS meson-meson scattering amplitude}

\label{sec:bs}

To describe the dynamics of resonances one needs to have exact elastic
unitarity in coupled-channel. For that purpose, we solve the coupled-channel
BS equation and use the SU(6) broken potential defined above to construct its
interaction kernel. In this way in any $YIJ$ sector, the solution for the
coupled-channel $s-$wave scattering amplitude, $T^{YIJ}$, satisfies exact
unitarity in coupled-channel. In the so called {\it on-shell}
scheme~\cite{Oller:1998zr,Nieves:1999bx, Oller:2000fj, GarciaRecio:2003ks}, $T^{YIJ}$ is given by
\begin{equation}
T^{YIJ}(s)=\frac{1}{1-V^{YIJ}(s)\,G^{YIJ}(s)}V^{YIJ}(s)
.
\label{eq:t-matrix}
\end{equation}
$V^{YIJ}(s)$ (a matrix in coupled-channel space) stands for the projection of
the scattering amplitude, ${\mathcal H}$, in the $YIJ$ sector. $G^{YIJ}(s)$ is
the loop function and is diagonal in the coupled-channel space. Suppressing
the indices, it is written for each channel as
\begin{equation}
 G(s)=i\int\frac{d^4q}{(2\pi)^4}\frac{1}{q^2-m^2_1}\frac{1}{(P-q)^2-m^2_2}
 \end{equation}
where $m_1$ and $m_2$ are the masses of the mesons corresponding to
the channel, for which we take physical values, and $P^\mu$ is the
total four momentum ($P^2=s$). The loop function involves a
logarithmic ultraviolet divergence which needs to be dealt with.
Extracting a suitable infinite constant, one can write
\begin{equation}
G(s)=\bar G(s) + G((m_1+m_2)^2)
.
\end{equation}
The finite function $\bar G(s)$ can be found in Eq.~(A9) of
Ref.~\cite{Nieves:2001wt}, and it displays the unitarity right-hand cut of
the amplitude. On the other hand, the constant $G((m_1+m_2)^2)$ contains the
logarithmic divergence. After renormalizing using the dimensional
regularization scheme, one finds
\begin{equation}
G(s=(m_1+m_2)^2) = \frac{1}{16\pi^2} \left ( a(\mu) +
\frac{1}{m_1+m_2} \left \{ m_1 \ln \frac{m^2_1}{\mu^2} + m_2 \ln
\frac{m^2_2}{\mu^2} \right\}\right) \label{eq:rs}
\end{equation}
where $\mu$ is the scale of the dimensional regularization. Changes in
the scale are reabsorbed in the subtraction constant $a(\mu)$, so that
the results remain scale independent.

We fix the Renormalization Scheme (RS) used in this work as follows. We adopt
a reasonable scale $\mu = 1\,{\rm GeV}$ and we allow $a(\mu)$ to vary
around the value $-2$ to best describe the known phenomenology in each $YIJ$
sector.\footnote{One can instead use an ultraviolet hard cutoff $\Lambda$ to
  renormalize the loop function. The relation between the subtraction constant
  $a(\mu)$, at the scale $\mu$, and $\Lambda$ is
\begin{equation}
a(\mu)= - \frac{2}{m_1+m_2} \left \{ m_1\, \ln \left[
\frac{\Lambda+\sqrt{\Lambda^2 +m_1^2}}{\mu}\right] + m_2 \,\ln \left[
\frac{\Lambda+\sqrt{\Lambda^2 +m_2^2}}{\mu} \right]\right\}
\end{equation}
For $\mu=0.7-1$ GeV, and assuming a cutoff of the same order of magnitude,
$-2$ turns out to be a natural choice for the subtraction constant
$a(\mu)$.} Results, of course, have some dependence on the adopted
RS, as
they also depend on the assumed SU(6) breaking pattern of the
couplings ($1/f^2\rightarrow 1/\left (f_1 f_2 f_3
f_4\right)^{1/2}$).  Indeed, both choices are not independent from
each other. That is the reason why we do not mind to scale, for
instance, the $\pi \rho \to \pi \rho$ channel by $1/(f_\pi f_\rho) $
instead of by $1/f^2_\pi$, as one will naturally expect from chiral
symmetry~\cite{Weinberg:1968de}, since a change in the renormalization
scale or in the subtraction constant for this channel would easily
cover the differences among these two choices for the couplings.

Since $f_V$ is significantly higher than $f_P$, the adopted breaking
pattern for the couplings guarantees that low-lying $J^P=0^+$
resonances, such as the $f_0(980)$ or the $f_0(600)$, described previously
by unitarizing pseudoscalar-pseudoscalar meson
amplitudes~\cite{Oller:1997ng, Oller:1997ti, Oller:1998hw,
GomezNicola:2001as} are not much affected by the inclusion of
vector-vector meson channels. It will be shown below that the adopted RS
successfully describes the main features of these positive parity
scalar resonances.

Other on-shell renormalization schemes can be also adopted. For instance, one
can take a certain scale, $\mu$, such that $G(\mu^2)=0$ and
the $T^{YIJ}$-amplitude reduces to the two-particle irreducible amplitude
$V^{YIJ}$, i.e., $T^{YIJ}(\mu^2)=V^{YIJ}(\mu^2)$. This fixes the value of the
subtraction constant $G((m_1+m_2)^2)$. This approach has been adopted in
\cite{GarciaRecio:2003ks, GarciaRecio:2005hy, GarciaRecio:2006vx, Toki:2007ab,
  GarciaRecio:2008dp} for meson-baryon $s-$wave scattering.  The use of one RS
or another is part of the uncertainties of the present approach, though, they
are smaller than those associated to our incomplete knowledge of the
two-particle irreducible amplitude $V^{YIJ}$.  We do not expect large
differences in the gross features of the picture that emerges, though the
exact position of the poles can of course be affected by modifying the RS.  In
the present work, the use of the RS based on dimensional regularization, as
outlined above, is preferable, because the same RS has been adopted in previous
studies of vector meson-vector meson ($VV$) and pseudoscalar meson-vector
meson ($PV$) scattering within the hidden gauge unitary approach
\cite{Roca:2005nm, Molina:2008jw, Geng:2008gx}. This makes it easier to
compare our results with those obtained in these references.

\begin{table}[t]
      \renewcommand{\arraystretch}{2} \setlength{\tabcolsep}{0.4cm}
     \caption{Values for the meson masses and decay constants used in
     the numerical calculations. All units are in MeV. Besides, we use
     $m_V=856\,{\rm MeV}$ as parameter of the Lagrangian.}
\begin{tabular}{cc|cc}
\hline\hline
 $m_\pi$ & 138.0 & $f_\pi$ & 92.4  \\
 $m_K$ & 495.7 &  $f_K$ & 113.0\\
  $m_\eta$ & 547.5 & $f_\eta$ & $1.2\times f_\pi$\\
   $m_\rho$ & 775.5 & $f_\rho$ & 153\\
   $m_{K^*}$ & 893.8 & $f_{K^*}$ & 153\\
   $m_\omega$ & 782.7 & $f_\phi$ & 163\\
   $m_\phi$ & 1019.5 & $f_\omega$ & $f_\rho$\\
\hline\hline
\end{tabular}
\label{tab:cts}
\end{table}

\section{Results and discussion}

\label{sec:res}
In this section, we show the results obtained using the
 approach described above and compare them with those obtained earlier
within different schemes, and to data when possible.

The mass and widths of the dynamically generated resonances in each
$YIJ$ sector are determined from the positions of the poles, $s_R$, in
the Second Riemann Sheet (SRS) of the corresponding scattering
amplitudes, namely $s_R= M^2_R-{\rm i}\ M_R \Gamma_R$. For narrow
resonances ($\Gamma_R \ll M_R$), $\sqrt{s_R} \sim M_R - {\rm
i}\Gamma_R/2 $ constitutes a good approximation. In some cases, we
also find real poles in the First Riemann Sheet (FRS) of the
amplitudes which correspond to bound states.

The coupling constants of each resonance to the various meson-meson states
are obtained from the residues at the pole, by matching the amplitudes to
the expression
\begin{equation}
T^{YIJ}_{ij}(s)=\frac{g_i  g_j }{(s-s_R)} \ , \label{eq:pole}
\end{equation}
for energy values $s$ close to the pole. The couplings, $g_i$,  
are complex in general.

Since our starting point is the chiral dynamics governing the interaction
among Nambu-Goldstone bosons, low energy results should be similar to those
previously obtained by unitarizing one loop ChPT
amplitudes~\cite{Oller:1997ng,Oller:1997ti,GomezNicola:2001as}. Because of the
inclusion of vector meson degrees of freedom the scalar sector has an enlarged
coupled-channel space in our case. However, we expect small effects from these
new degrees of freedom on the low-lying scalar resonances, since vector
meson-vector meson thresholds are relatively far away from the low-energy
region, where the pseudoscalar-pseudoscalar interaction dominates.

To facilitate the discussion of our results, let us point out the main
differences between the approach advocated in the present work and the
approaches followed in Ref.~\cite{Roca:2005nm} for the pseudoscalar-vector
sector, and in Refs.~\cite{Molina:2008jw,Geng:2008gx} for the vector-vector
one. These latter works are based\footnote{Strictly speaking, the study of
  axial-vector resonances carried out in Ref.~\cite{Roca:2005nm} does not use
  the hidden gauge formalism. There, a contact WT type Lagrangian is
  employed. However, the tree level amplitudes so obtained coincide with those
  deduced within the hidden gauge formalism, neglecting $q^2/m_V^2$ in the
  $t-$exchange contributions~\cite{Nagahiro:2008cv} and considering only the
  propagation of the time component of the virtual vector mesons.} on the
formalism of the hidden gauge interaction for vector
mesons~\cite{Bando:1984ej, Bando:1987br}.  The main differences are:
\begin{enumerate}

\item Previous works~\cite{Roca:2005nm,Molina:2008jw,Geng:2008gx}
  treat separately pseudoscalar-pseudoscalar, pseudoscalar-vector
  and vector-vector meson sectors. However, for instance, vector-vector
  channels could modify the properties of some vector-axial
  resonances, generated in Ref.~\cite{Roca:2005nm}, where only
  pseudoscalar-vector meson interactions are considered. Within the
  formalism of the hidden gauge interaction for vector mesons there 
  exist no  $s-$wave $PV\to VV$ transition potentials at tree level, and
  thus it is difficult to overcome this limitation in that scheme.

\item {\it Pseudoscalar-vector channels~\cite{Roca:2005nm}}: Though,
 in a first view, the two-particle irreducible amplitude ($V^{YIJ}$)
 employed here and that used in ~\cite{Roca:2005nm} might look quite
 different, this is not really the case and they just differ at order
 ${\mathcal O}(m^2, \vec{k}^{\,2})$ (with $m$ and $k^\mu $, the mass and
 the momentum of the Nambu-Goldstone boson) in the chiral expansion, which
 is not fixed by the LO WT
 theorem~\cite{Weinberg:1966kf,Tomozawa:1966jm} that both approaches
 satisfy. Thus, in both schemes, the {\it potentials} $V^{YIJ}$ totally
 agree at LO ${\mathcal O}(k^\mu)$ and take the common value 
\begin{equation}
V^{YIJ} = C^{YIJ} \, m_V k^0/ f^2 \label{eq:lowtpv}
\end{equation}
where the $C^{YIJ}$  coupled-channel matrices are given in
 ~\cite{Roca:2005nm}. The $PV \to PV$ amplitudes vanish in the
 soft Nambu-Goldstone boson limit $k^0 \to 0$, as required by the LO WT
 theorem (see discussion in \cite{Birse:1996hd} for some more
 details). 

 As a consequence, and apart from the influence of the vector
 meson-vector meson channels (see point below), of the use here of massless
 Nambu-Goldstone bosons and physical decay constants in the computation of
 $V^{YIJ}$, we expect a rather good agreement with the results of
 Ref.~\cite{Roca:2005nm} for the lowest lying axial resonances, which
 will not be much affected by higher orders of the chiral expansion.

\item {\it Vector-vector channels~\cite{Molina:2008jw,Geng:2008gx}}: In
  Refs.~\cite{Molina:2008jw,Geng:2008gx}, contact, box and $t-$ and
  $u-$exchange contributions were considered, within a scheme based on the
  hidden gauge interaction for vector mesons; the exchange and contact terms
  being the dominant mechanisms. The exchange mechanism is closely related to
  the kinetic interaction derived within our SU(6) symmetric scheme
  (${\mathcal D_{\rm kin}}$). Indeed, one finds that by symmetrizing the
  interaction in the $\rho\rho$ channel in Table I of
  Ref.~\cite{Molina:2008jw} and adding a factor 4/3, our SU(6)
  symmetric $\rho\rho$ interaction is reproduced ($\mathcal{D}_{\rm kin}$ can
  be looked up in Appendix~\ref{app:tables}).  Note that in the $\rho\rho$
  channel SU(6) symmetry implies having symmetric interactions under the
  exchange $I \leftrightarrow J$.

  As we commented above, the kinetic interaction of our model is of the form
  $\left[({\bf 35}\otimes {\bf 35})_{{\bf 35}_a}\otimes({\bf 35}\otimes {\bf
      35})_{{\bf 35}_a}\right]_{\bf 1}$ in the $t-$channel. This can be
  regarded as the zero-range $t-$channel exchange of a full ${\bf 35}$
  irreducible representation, carried by an octet of spin 0 and a nonet of
  spin 1 mesons of even parity. In Refs.~\cite{Molina:2008jw,Geng:2008gx}
  these kinetic terms are originated by the $t-$exchange of the time component
  of vector mesons, which has certain resemblance with our zero-range exchange
  of $0^+$ mesons. Parity and angular momentum conservation also allow the
  exchange of $1^+$ and $2^+$ mesons. The latter exchange is missing in both
  approaches, and the former one is included within our scheme, as required by
  SU(6) symmetry, while it is not present in the hidden gauge formalism
  adopted in Refs.~\cite{Molina:2008jw,Geng:2008gx}. We do not see a priori any
  compelling reason to favor any of the two approaches.
 
  The contact terms in both approaches seem to be totally unrelated. We remind
  here the ambiguities mentioned above associated to this term and that
  presumably its actual nature can only be fixed by the asymptotic behavior of
  QCD.

\item We use $f_V \ne f_P$ for those channels which involve vector mesons,
  while a universal $1/f^2$ coupling is assumed for all channels in the
  previous works. As commented above, this is somehow related with the RS.

\end{enumerate}

In what follows, we show results for the different $YIJ$
sectors, considering only nonnegative hypercharge values.

\begin{table}[htpb]
      \renewcommand{\arraystretch}{2}
     \setlength{\tabcolsep}{0.1cm}
     \caption{Pole positions and modulus of the couplings $|g|$ (MeV units) in
       the $(Y,I,J)=(0,0,0)$ sector. $I^G(J^{PC})=0^+(0^{++})$. The
       subtraction constant has been set to its default value, $a=-2$.
       Possible PDG counterparts: $f_0(600)$, $f_0(980)$, $f_0(1370)$,
     and $f_0(1710)$. } 
\label{tab:000}
\vspace{0.5cm}
\begin{tabular}{c|cccccccc}
\hline\hline
  $\sqrt{s_R}$& $\pi\pi$ &  $\bar{K}K$ & $\eta\eta$ & $\rho\rho$ & $\omega\omega$ & $\omega\phi$  & $\bar{K}^* K^*$ &  $\phi\phi$ \\\hline 
$(635,-202)$   & 3516 & 432 & 333 & 7592 & 7909 & 117 & 7306 & 1850\\

$(969,0) $ &  28 &  2983 & 2477 & 3393 &  2401 & 1627 & 4305 & 3831\\

$(1350,-62)$ &  553 & 3257 & 840 &  1336 & 2841 & 7074 & 10697 & 10647\\
$(1723,-52)$ & 43 &  853 & 3154 &  318 & 408 & 3400 &   2470 & 13698\\
\hline\hline
\end{tabular}
\end{table}
\begin{table}[tbh]
      \renewcommand{\arraystretch}{2}
     \setlength{\tabcolsep}{0.1cm}
     \caption{Pole positions and the modulus of the couplings (MeV units) in
       the $(Y,I,J)=(0,0,0)$
       sector when only block diagonal pseudoscalar-pseudoscalar and
       vector-vector meson interactions are used.   $I^G(J^{PC})=0^+(0^{++})$.
       Possible PDG counterparts: $f_0(600)$, $f_0(980)$, $f_0(1370)$
       and 
$f_0(1710)$.} \label{tab:000bis}
\vspace{0.2cm}
\begin{tabular}{c|cccccccc}
\hline\hline
  $\sqrt{s_R}$ & $\pi\pi$ &  $\bar{K}K$ & $\eta\eta$ & $\rho\rho$ & $\omega\omega$ & $\omega\phi$  & $\bar{K}^* K^*$ &  $\phi\phi$ \\\hline
     $(485,-156)$ &   2807 &  600 &    86&\\
     $(990,-6)$ &     862 &  2746 &  2146\\
    $(1217,0)$ &      &  &  &     3637 &  3111 &  1378 &  12465 &   13130\\
    $(1981,-110)$&    &  &  &      848 &  1896 &  4984 &  3708 &  10747\\
\hline\hline
\end{tabular}
\end{table}

\subsection{Hypercharge 0, isospin 0 and spin 0}

\begin{figure}[tbh]
%\vspace{-3cm}
\centerline{\hspace{-2cm}\includegraphics[height=3.cm]{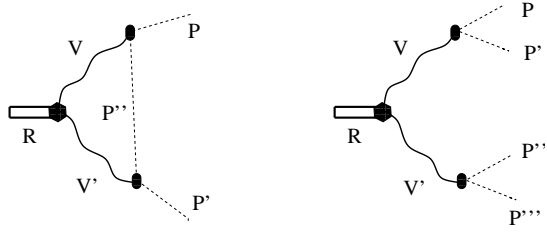}}
%\vspace{-15cm}
\caption{%\footnotesize 
Resonance ($R$) decay to two (left) or four
    (right) pseudoscalar mesons ($P$, $P^\prime$, $P^{\prime\prime}$, $P^{\prime\prime\prime}$) through its
    $s-$wave (hexagon) coupling to two vector mesons ($V$,$V^\prime$) and the
    $p-$wave coupling (ovals) of these latter mesons to two
    pseudoscalar mesons.}
\label{fig:box}
\end{figure}

There are eight coupled channels, i.e., $\pi\pi$, $\bar{K}K$,
$\eta\eta$, $\rho\rho$, $\omega\omega$, $\omega\phi$, $\bar{K}^* K^*$
and $\phi\phi$. In all cases the $G-$parity is positive. Four poles
are found on the complex plane of the SRS. These are compiled in Table
\ref{tab:000}, where the modulus of the couplings to the different
channels (see Eq.~(\ref{eq:pole})) are also given.  The lowest two
poles can be easily identified with the $f_0(600)$ and $f_0(980)$
resonances. There are some differences with other
works~\cite{Oller:1998hw,GomezNicola:2001as} mainly because we have
neglected the pseudoscalar meson mass terms and have incorporated
vector meson-vector meson channels. On the other hand, the
identification of the other two poles is not so direct, though it is
tempting to associate them to the $f_0(1370)$, and $f_0(1710)$
resonances. Thus, in our model the $f_0(1370)$ resonance has a
sizeable coupling to the $\rho\rho$ channel which would lead to a four
pion decay mode. For the decay of the resonance, the $\rho\rho$
channel is more relevant than the other ones (for instance the
$\omega\omega$ or $\bar K^* K^*$), thanks to the large width of the
$\rho-$meson, which enhances the decay of the resonance to the decay
products of the $\rho\rho$ pair. Indeed, the width of these $f_0$
resonances will be enhanced when new mechanisms constructed out of
$VPP$ $p-$wave couplings are considered (see for example
Fig.~\ref{fig:box})~\cite{Geng:2008gx}. For instance, since the pole
that we have associated to the $f_0(1370)$ is placed below the two
$\rho$ meson threshold, it can decay neither to this channel nor to
those which are even heavier. Thus, the width of around $124\,{\rm
MeV}$ that can be read off from Table \ref{tab:000} accounts only for
the decay of the resonance into the open channels ($\pi\pi$,
$\bar{K}K$, $\eta\eta$). However, the resonance can decay into two
virtual $\rho$ mesons, and each of them subsequently will decay into
two pions, giving rise to four and two pion decay modes through
processes like those sketched in Fig.~\ref{fig:box}. These decays will
increase the width of the resonance~\cite{Geng:2008gx}. Obvious
modifications to these mechanisms should be considered, taking into
account the specific details of the dominant decays of the
corresponding vector mesons, for other channels. For instance, since
the $\omega$ meson decays predominantly into three pions, the coupling
of a resonance to two $\omega$ mesons will produce six or four pion
decays.

Following the findings of Ref.~\cite{Geng:2008gx}, a substantial increase of
both the $f_0(1370)$ and $f_0(1710)$ widths with respect to those deduced from
the pole position is to be expected. On the other hand, the above mechanisms
could explain a large $K\bar K$ decay mode of the $f_0(1710)$ resonance that
in our model couples strongly to the $K^*\bar K^*$ and the $\phi\phi$
channels. This also supports the picture of Ref.~\cite{Close:2005vf}, where it
is guessed that the $f_0(1710)$ is dominantly $s\bar s$. Besides, we predict a
sizeable decay of this resonance into $\eta \eta$.

The experimental $f_0(1500)$, on the other hand, has a mass of $1505 \pm
6\,{\rm MeV}$ and it is relatively narrow ($\Gamma= 109 \pm 7\,{\rm MeV}$)
with dominant decays into two and four pion channels. Owing to the above
discussion, it would be difficult to assign it to our lowest pole, and thus it
is a clear candidate to have a dominant glueball
structure~\cite{Amsler:1995td,Amsler:2004ps}. This is also in agreement with
the recent claims of Albaladejo and Oller~\cite{Albaladejo:2008qa}, though it
looks more difficult to reconcile the picture that emerges from our analysis
with this latter work in the case of the $f_0(1710)$ resonance. This is
because in Ref.~\cite{Albaladejo:2008qa}, the $f_0(1710)$ resonance is
identified as an unmixed glueball with a large $\eta^\prime\eta^\prime$
coupling, and this latter channel is not included in our scheme. In
Ref.~\cite{Geng:2008gx}, only the $f_0(1370)$ and $f_0(1710)$ resonances are
found as well, and in agreement with our findings, there the $f_0(1500)$ is
not dynamically generated either. However, there appear some differences with
our results, since in this latter reference the $f_0(1370)$ is mainly
$\rho\rho$, and the $f_0(1710)$ is mostly $K^*\bar K^*$. Such a distinction is
not so clear in our scheme, where $\omega\omega $, $\omega\phi $ and $\phi\phi
$ channels play a more significant role than in the hidden gauge unitarity
approach advocated in \cite{Geng:2008gx}.

In previous studies of pseudoscalar-pseudoscalar and vector-vector
interactions, the first three channels and the last five channels in
Table~\ref{tab:000} were considered separately (see for instance
Refs.~\cite{Oller:1997ng,Oller:1997ti} and \cite{Geng:2008gx},
respectively). It is interesting to check how the results change if they are
also considered separately. That is, if all couplings that connect the
pseudoscalar-pseudoscalar and vector-vector meson sectors are set to zero.
Comparing the results of Tables ~\ref{tab:000} and ~\ref{tab:000bis}, we
observe that, though the number of resonances is the same in both cases (four)
their positions and the relative strengths of couplings to different channels
have changed. The inclusion of the three pseudoscalar-pseudoscalar channels
has a large impact on the two poles of higher energy.
\begin{table}[b]
      \renewcommand{\arraystretch}{2} \setlength{\tabcolsep}{0.1cm}
     \caption{Pole positions and modulus of the couplings (MeV units)
       in the $(Y,I,J)=(0,0,1)$ sector. $I^G(J^{PC})=0^-(1^{+-})$
       [$h's$] and $0^+(1^{++})$ [$f's$].  We have slightly moved away
       from the choice $a=-2$ for the subtraction constants (see
       Eq.~(\ref{eq:rs})) and have used $a=-1.3$ and $a=-2.9$ for the
       $0^-(1^{+-})$ and $0^+(1^{++})$ sub-sectors,
       respectively. Possible PDG counterparts: $h_1(1170)$,
       $h_1(1380)$, $h_1(1595)$ and $f_1(1285)$.}\label{tab:001}
\vspace{0.2cm}
\begin{tabular}{c|ccccccc|c}
\hline\hline
$\sqrt{s_R}$ & $\eta\phi$ &  $\eta\omega$ & $\pi\rho$ & $(\bar{K}K^*)_A$ & $\bar{K}^* K^*$ & $\omega\phi$ &    $(\bar{K}K^*)_S$  &$G$ \\\hline 
$(1006,-85)$ & 52 & 26 & 4362 & 1192 & 1029 & & &$-$\\
$(1373,-17)$ & 2957 & 2484 & 1273 & 5322 & 1779& & & $-$\\
$(1600,-67)$ & 2414 & 3655 & 1006 & 1033 & 12869&&& $-$\\
$(1286,0)$ &  & & & & & 4089 & 6790&$+$\\
\hline\hline
\end{tabular}
\end{table}

\subsection{Hypercharge 0, isospin 0 and spin 1}

In this sector (see Table~\ref{tab:001}), there are two sets of quantum numbers: $I^G(J^{PC})$ equal
$0^-(1^{+-})$ and $0^+(1^{++})$, corresponding to those of the $h_1$ and $f_1$
resonances, respectively. Interactions turn out to be block diagonal since
strong interactions conserve $G-$parity and charge conjugation.

For $0^-(1^{+-})$ there are five coupled channels, namely, $\eta\phi$,
$\eta\omega$, $\pi\rho$, $(\bar{K}K^*)_A=\frac{1}{\sqrt{2}}(\bar{K} K^*-K
\bar{K}^*)$, $K^* \bar{K}^*$.  Three poles are found in the complex
plane,
which can be tentatively associated to the $h_1(1170)$, $h_1(1380)$, and
$h_1(1595)$. (These are the only three $0^-(1^{+-})$ resonances below 2 GeV
quoted in the PDG~\cite{Amsler:2008zz}.)  Naturally, huge couplings are found
for the $h_1(1170)$, $h_1(1380)$ and $h_1(1595)$ resonances to the $\pi\rho$,
$(\bar{K} K^*)_A$ and $\eta\omega$ and $\bar K^* K^*$ channels,
respectively. While the two latter resonances are omitted from the summary PDG
table, and even the isospin and $G-$parity of the $h_1(1380)$ is not quoted,
 the $h_1(1170)$ is firmly established experimentally. We predict for it a
width smaller than that quoted in the PDG, because within our model it comes
out lighter than the experimental one and thus the available phase space for
$\pi\rho$ decay is much smaller.

The $h_1(1170)$ and $h_1(1380)$ obtained here are placed at similar positions
and have similar couplings and widths as those obtained in
Ref.~\cite{Roca:2005nm}. This is not surprising since, as already noted, at
LO in the chiral expansion our coupled-channel interaction for $PV\to PV$
scattering coincides with that used in Ref.~\cite{Roca:2005nm}.

Within our scheme, the vector-vector channel $K^* \bar{K}^*$ plays an
important role in the dynamics of the pole placed at $(1600,-67)$
MeV. Presumably, this is the reason why a third $h_1$ pole was not found in
Ref.~\cite{Roca:2005nm}, which misses the $K^* \bar{K}^*$ channel.  On the
other hand, a $K^* \bar{K}^*$ resonance is found in Ref.~\cite{Geng:2008gx}
located almost at threshold [$\sqrt{s_R}=(1802,-39)$]. This pole was not
identified in Ref.~\cite{Geng:2008gx} with the $h_1(1595)$, because of the
very different mass. It can be conjectured that this pole corresponds to the
one found in our approach at $(1600,-67)$.  The latter is strongly modified by
the inclusion of the $\omega\eta$ channel in the dynamics.  If the picture
presented by our model is correct, with the pole at $(1600,-67)$ assigned to
$h_1(1595)$, this resonance cannot be generated just by
$\omega\eta$, as attempted in Ref.~\cite{Roca:2005nm}. Indeed, the diagonal
$\omega\eta$ potential is zero in this sector. And also it cannot be described
using only $K^* \bar{K}^*$, as in ~\cite{Geng:2008gx}, since the mass turns
out to be too high and furthermore its dominant decay mode, $\omega \eta$,
is ignored.

In the $0^+(1^{++})$ subsector only one pole is found, at $(1286,0)$, quite
similar to that reported in Ref.~\cite{Roca:2005nm}. The PDG quotes three
$f_1$ resonances below 2 GeV: $f_1(1285)$, $f_1(1420)$, and $f_1(1510)$. The
$f_1(1285)$ has a mass of $1281.8\pm0.6$ MeV and a width of $24.3\pm1.1$ MeV;
the $f_1(1420)$ has a mass of $1426.4\pm0.9$ MeV and a width of $54.9\pm2.6$
MeV; the $f_1(1510)$ has a mass of $1518\pm5$ MeV and a width of $73\pm25$
MeV.  The decay modes of $f_1(1420)$ and $f_1(1510)$ are dominated by the
$(K\bar{K}^*)_S$ mode. On the other hand, because the $f_1(1285)$ is below the
$\bar{K}K^*$ threshold, it cannot decay through this channel, though the
branching fraction into $\bar K K\pi$ is about 10\% and it might hint at a
non-negligible $\bar K K^*$ component in its wave function.

Because of the position of the pole at $(1286,0)$, it makes sense to assign
this pole to the $f_1(1285)$ resonance, as it was done in
Ref.~\cite{Roca:2005nm}. The reason why no width is found for this resonance,
while the PDG quotes 24 MeV for it, is that there are other decay channels
different to $VP$ that are obviously not considered in our scheme [like $4\pi$
  (33\%), $\eta \pi\pi$ (52\%), or $K\bar K \pi$ (10\%)\footnote{This latter
    decay mode can be easily understood from the decay of the resonance to a
    virtual $\bar K K^*$ pair.}]. Nevertheless, the assignment of the pole to
the $f_1(1420)$ resonance, whose dominant decay is $(K{\bar K}^*)_S$ and it is
also placed close to threshold, cannot be completely discarded either.

\begin{table}[htpb]
      \renewcommand{\arraystretch}{2}
     \setlength{\tabcolsep}{0.4cm}
     \caption{Pole positions and modulus of the couplings (MeV units) in the
       $(Y,I,J)=(0,0,2)$ sector. $I^G(J^{PC})=0^+(2^{++})$. In this subsector
       the value of $-2.77$ for the subtraction constant in the $\rho\rho$
       channel and $-2.5$ for the other ones have been used.  Possible PDG
       counterparts: $f_2(1270)$ and $f_2(1640)$.  }\label{tab:002}
\vspace{0.2cm}
\begin{tabular}{c|ccccc}
\hline\hline
 $\sqrt{s_R}$& $\rho\rho$ & $\omega\omega$& $\omega\phi$ &  $\bar{K}^* K^*$ &  $\phi\phi$ \\\hline 
$( 1289, 0)$ & 22138 & 17793 & 3642 & 18385 & 5865\\
 $(1783,-19)$ &    2235 &1541 &  2846 &  5201 &    2427\\
\hline\hline
\end{tabular}
\end{table}

\subsection{Hypercharge 0, isospin 0 and spin 2}

In this sector, there are five coupled channels: $\rho\rho$, $\omega\omega$,
$\omega\phi$, $K^*\bar{K}^*$, and $\phi\phi$, and we find two poles
(see Table~\ref{tab:002}) in the
SRS/FRS of our amplitudes. Experimentally, many $f_2$ resonances below
$2\,{\rm GeV}$ have been reported, including $f_2(1270)$, $f_2(1430)$,
$f_2^\prime(1525)$, $f_2(1565)$, $f_2(1640)$, $f_2(1810)$, $f_2(1910)$, and
$f_2(1950)$. Most of them [$f_2(1430)$, $f_2(1565)$, $f_2(1640)$, $f_2(1810)$,
  and $f_2(1910)$] have not been confirmed yet. It is tempting to associate
the two $f_2$ poles found within our approach to the two lowest lying
confirmed resonances $f_2(1270)$ and the $f_2^\prime(1525)$. In the first
case, the mass agrees well with that quoted in the PDG, however experimentally
the $f_2(1270)$ resonance is quite broad ($\Gamma\sim 185$ MeV) while in our
case, it appears as a bound state (pole in the FRS) of zero
width. Contributions as those depicted in Fig.~\ref{fig:box} might provide a
sizeable width to this pole. Besides there exist other mechanisms like
$d-$wave $\pi\pi$ decays, which could also be important in this case because
the large available phase space. Those associated to the left diagram
of Fig.~\ref{fig:box} are considered in
Refs.~\cite{Molina:2008jw,Geng:2008gx}. Regarding our identification of the
$f_2^\prime(1525)$ resonance, we find dominant couplings to the $K^*\bar K^*$
and $\phi\phi$ channels, which will naturally account for the experimental
dominant decay mode into $K \bar K$ of this resonance~\cite{Amsler:2008zz}
through loop mechanisms (Fig.~\ref{fig:box}). However, the mass position
disagrees much more in this case, while its sizeable coupling in our approach
to $\rho\rho$ seems difficult to reconcile with its experimental small
branching fractions into $\pi\pi$ and $\pi\pi\pi\pi$. Thus, we have some
reservation with this identification, and perhaps it could also be possible to
identify the pole with the resonances $f_2(1565)$ or $f_2(1640)$, which are
placed closer to the pole and have decay modes involving an even number of
pions, or an $\omega\omega$ pair.  Possibly further ingredients, like $d-$wave
$K\bar K$ pairs, would also be needed to correctly describe the dynamics of
the $f_2^\prime(1525)$ resonance.

In Ref.~\cite{Geng:2008gx} two states are also generated in this channel, and
are associated with the $f_2(1270)$ and $f_2^\prime(1525)$.  The real part of
both poles agree remarkably well with the masses of these two resonances.
This was achieved by a suitable fine-tuning of the subtraction constants.  A
similar good agreement could not be achieved within our scheme by fine-tuning
of the subtraction constants. In ~\cite{Geng:2008gx}, these two resonances
appear mostly as $\rho\rho$ and $\bar K^* K^*$ bound states, respectively. In
our case, these channels are still dominant but with a substantial
contribution from the subdominant channels. The hidden gauge interaction for
vector mesons model used in~\cite{Geng:2008gx} and our approach are related
for $PV\to PV$ scattering, thanks to chiral symmetry, but they are completely
unrelated in the $VV$ sector, where we believe that the nature of the contact
terms can only be unraveled by requiring consistency with the QCD asymptotic
behavior at high energies~\cite{Ecker:1989yg}. Besides, the $VV$ interactions
of our model are weaker than those deduced in Ref.~\cite{Geng:2008gx} due to
the use of $f_V$ instead of the pion decay constant.

\begin{table}[htpb]
      \renewcommand{\arraystretch}{2}
     \setlength{\tabcolsep}{0.4cm}
     \caption{Pole positions and modulus of the couplings in the
       $(Y,I,J)=(0,1,0)$ sector (MeV units).  The subtraction constants
       $a=-3.5$ for the $PP$ channels and $a=-1$ for the three $VV$ ones have
       been used. $I^G(J^{PC})=1^-(0^{++})$. Possible PDG counterparts:
       $a_0(980)$, $a_0(1450)$ and $a_0(2020)$.}\label{tab:010}
\vspace{0.3cm}
\begin{tabular}{c|ccccc}
\hline\hline
 $\sqrt{s_R}$ & $\eta\pi$ & $\bar{K}K$ & $\omega\rho$ & $\phi\rho$& $\bar{K}^*K^*$  \\\hline 
 $(991, -46)$ & 2906 & 3831& 775 & 4185& 5541\\
 $(1442,-5)$ &907 & 285 & 10898 & 677 & 3117\\
 $(1760,-12)$ &  790 & 1241 & 667 & 5962 & 5753\\
\hline\hline
\end{tabular}
\end{table}

\subsection{Hypercharge 0, isospin 1 and spin 0}

There are five coupled channels in this sector: $\pi\eta$, $K\bar{K}$,
$\rho\omega$, $\rho\phi$ and $K^*\bar{K}^*$ and our model produces three poles
in the SRS of the amplitudes. These are compiled in
Table~\ref{tab:010}. The lowest pole should correspond to the
$a_0(980)$, which has been obtained in all previous studies considering
only pseudoscalar-pseudoscalar coupled-channel. In our model, its couplings
to the $\pi\eta$ and $K\bar{K}$ are large, in agreement with the results of
earlier studies and with the data, but it also presents large couplings to the
heaviest channels, $\phi\rho$ and ${\bar K}^* K^*$.

The pole at $\sqrt{s_R}=(1442,-5)$ can be associated to the
$a_0(1450)$. Within our scheme, it can decay to $\pi\eta$ and
$K\bar{K}$, which is in agreement with the data. Its huge coupling
to $\omega \rho$ will give rise to a significant $\omega \pi\pi$ decay
mode and to an important enhancement of its width, thanks to the
broad spectral function of the $\rho$ resonance. 

On the other hand, the PDG only reports two $a_0$ resonances below $2\,{\rm
  GeV}$. Therefore, the third pole in this sector at $\sqrt{s_R}=(1760,-12)$
cannot be associated, in principle, to any known state. Nevertheless, it is
interesting to note that in Ref.~\cite{Geng:2008gx} an $a_0$-like pole was
found located close to the $K^*\bar{K}^*$ threshold and with large couplings
to $K^*\bar{K}^*$ and $\phi\rho$. On the other hand, a resonance $a_0(2020)$
has been reported in \cite{Anisovich:1999jv}, with these quantum numbers
around $2\,{\rm GeV}$ ($2025\pm 30$ MeV), but extremely wide ($330\pm 75$
MeV).  The large width of this state makes less meaningful the difference
between its mass and that of our pole, which might be then associated to this
resonance. Still, it should be noted that the $a_0(2020)$ resonance is not yet
firmly established and needs further confirmation~\cite{Amsler:2008zz}.

\begin{table}[htpb]
      \renewcommand{\arraystretch}{2}
     \setlength{\tabcolsep}{0.2cm}
     \caption{Pole positions and modulus of the 
     couplings (MeV units )  in the $(Y,I,J)=(0,1,1)$
     sector. $I^G(J^{PC})=1^+(1^{+-})$ [$b's$] and
       $1^-(1^{++})$ [$a's$].  A
     subtraction constant of $a=-1.57$ has been used for positive $G-$parity
     states. Possible PDG counterparts:
     $b_1(1235)$, $b_1(1960)$, $a_1(1260)$,
     $a_1(1640)$. }\label{tab:011}
\vspace{0.2cm}
\begin{tabular}{c|cccccccccc|c}
\hline\hline
$\sqrt{s_R}$ &  $\bar{K}^* K^*$ & $\pi\phi$ & $\pi\omega$ & $\eta\rho$ &
 $\rho\rho$ & $(\bar{K}K^*)_S$ &$\pi\rho$ & $\omega\rho$& $( \bar{K} K^*)_A$  &
$\phi\rho$ &$G$ \\ \hline 
 $(1234,-57)$ & 4516 & 438 & 3398 & 900 & 9025 & 3165 & & & & & $+$\\
 $(1642,-139)$ & 10433 & 4214 & 912 & 3321 & 965 & 523 & & & & & $+$\\
$(1021,-251)$ &  & & & && &7988 & 7580 & 5284 & 929 &$-$\\
 $(1568,-145)$ & & & & & & &   679 &    1423 & 6314 &  9973 &$-$\\
\hline\hline
\end{tabular}
\end{table}

\subsection{Hypercharge 0, isospin 1 and spin 1}

There are two sets of quantum numbers in this sector: $1^+(1^{+-})$
and $1^-(1^{++})$, corresponding to those of $b_1$ and $a_1$
resonances. Our results for this channel are compiled in Table~\ref{tab:011}.

In the $1^+(1^{+-})$ subsector, two poles are found in the SRS: the lower one
can be associated to the $b_1(1235)$. The predicted mass, width and decay
modes agree well with the data [$\omega\pi$ (dominant), $\rho\eta$ (seen),
  4$\pi$ ($<$ 50\%) and $K\bar K \pi$ (14\%)]. This state has also been found
in Refs.~\cite{Lutz:2003fm,Roca:2005nm}, and this indicates that it mainly
originates from the pseudoscalar-vector interaction. The second $b_1$ state is
found at $\sqrt{s_R}=(1642,-139)$. It couples strongly to $K^*\bar{K}^*$ and
lacks a clear PDG counterpart yet. A $b_1$ state is also found in
Ref.~\cite{Geng:2008gx} at $\sim$1700 MeV. It is tempting to associate our
second pole with the resonance $b_1(1960)$, though this state is not firmly
established yet~\cite{Anisovich:1999jv}.  The $b_1(1960)$ turns out to be also
quite wide ($\Gamma = 230\pm 50$), as it was the case of the $a_0(2020)$
resonance above, which makes less important the large difference existing
between the masses. Moreover, the data suggest that $b_1(1960)$ has non-zero
overlaps with the $\pi \omega$ and $\eta \pi \omega$
channels~\cite{Anisovich:2002su}. This is compatible with the features of our
pole. (Note that the $\eta \rho$ coupling could lead to a non zero contribution
to the $\eta \pi \omega$ decay mode.)

In the $1^-(1^{++})$ subsector, also two poles are found in the SRS. It is
tempting to associate them to the $a_1(1260)$ and $a_1(1640)$, the only two
$a_1$ resonances below $2\,{\rm GeV}$ reported in the
PDG~\cite{Amsler:2008zz}.

The mass and width of $a_1(1260)$ suffer from large uncertainties, being
quoted in the PDG values of $1230 \pm 40$ MeV and $250-600$ MeV,
respectively. Its dominant decay modes are $3\pi$ and $({\bar K}K^*)_A$. This
is in total agreement with the largest couplings of our lightest pole in this
sector. In addition, the main properties of this pole are similar to those of
the pole found in the approach of Ref.~\cite{Roca:2005nm}.

The resonance $a_1(1640)$ is much worse established experimentally and it is
not reported in the approach of Ref.~\cite{Geng:2008gx}. Nevertheless, our
second pole couples strongly to $VV$ channels and its features fit well with
those known for the $a_1(1640)$ resonance.

\begin{table}[htpb]
      \renewcommand{\arraystretch}{2}
     \setlength{\tabcolsep}{0.2cm}
     \caption{Pole positions and modulus of the couplings (MeV units ) in the
       $(Y,I,J)=(0,1,2)$ sector. $I^G(J^{PC})=1^-(2^{++})$. The subtraction
       constant has been set to $a=-3.4$.  Possible PDG counterparts:
       $a_2(1320)$ and $a_2(1700)$.}\label{tab:012}

\vspace{0.2cm}
\begin{tabular}{c|ccc}
\hline\hline
 $\sqrt{s_R}$ & $\omega\rho$ & $\phi\rho$  & $\bar{K}^* K^*$\\\hline
$(1228,0)$ &  11287 & 2637 & 6281\\
$(1775,-6)$ & 1454 & 3167 &4362\\
\hline\hline
\end{tabular}
\end{table}

\subsection{Hypercharge 0, isospin 1 and spin 2}

There are three coupled channels in this sector: $\bar{K}^* K^*$,
$\omega\rho$, and $\phi \rho$, and we find two poles, one in the FRS
and a second one in the SRS of the amplitudes (see
Table~\ref{tab:012}), which might be associated to the $a_2(1320)$ and
$a_2(1700)$ resonances. In our model, the bound state strongly couples
to the $\omega\rho$ channel, which would give rise to the observed
$3\pi$ and $\omega\pi\pi$ decay modes of the $a_2(1320)$ thanks to the
width of the virtual $\rho$ meson. Furthermore, if the pole position
were closer to the experimental mass, the width would also
increase. Fine-tuning of the subtraction constant did not work to
achieve a better agreement in the mass position.

Little is known about the $a_2(1700)$, but the assignment of our second pole
with it might get supported by the decays of this resonance into $\omega \rho$
and $K\bar{K}$ pairs. Indeed, this latter decay mode can be obtained from the
decays of the resonance to virtual $\phi\rho$ or $K^*\bar{K}^*$ pairs, through
loop mechanisms as those depicted in Fig.~\ref{fig:box}. The hidden gauge
interaction for vector mesons model used in~\cite{Geng:2008gx} gives rise only
to one pole, whose features corresponds to the heaviest of the poles found
here. As it is the case here, though its mass is close to that quoted in the
PDG for the $a_2(1700)$ resonance, it turns out to be much narrower than this
resonance.  This  could be an indication of the fact that either the
identification of this pole with the $a_2(1700)$ resonance is incorrect or
that other mechanisms, such as coupled-channel $d-$wave dynamics, might play
an important role in this case.

\begin{table}[htpb]
      \renewcommand{\arraystretch}{2}
     \setlength{\tabcolsep}{0.2cm}
     \caption{Pole positions and modulus of the couplings (MeV units) in the
       $(Y,I,J)=(1,1/2,0)$ sector. $I(J^{P})=\frac{1}{2}(0^{+})$.  The
       subtraction constant has been set to $a=-1.5$ for the $VV$
       channels. Possible PDG counterparts: $K^*_0(800)$, $K^*_0(1430)$ and
       $K^*_0(1950)$.}
\begin{tabular}{c|ccccc}
\hline\hline
 $\sqrt{s_R}$ & $K\pi$ & $\eta K$ & $K^*\rho$ & $K^*\omega$ & $K^*\phi$\\\hline
$(830,-170)$ &   4446 &   1879 &  5868 &  2029 & 1805\\
$(1428,-24)$ &  1805 &    892 & 8007 &    10803 &      5556\\
$(1787,-37)$ &  45 &   2662 &   657 &  1107 & 12181\\
\hline\hline
\end{tabular}\label{tab:1-12-0}
\end{table}

\subsection{Hypercharge 1, isospin 1/2 and spin 0}

In this sector there are five coupled channels: $\pi K$, $\eta K$,
$\rho K^*$, $\omega K^*$, and $\phi K^*$, and three poles are found in
the SRS of the amplitudes. The first one at $\sqrt{s_R}=(830,-170)$
can be associated to the $K^*_0(800)$. There is still a controversy
about the existence and the origin~\cite{Cherry:2000ut} of this broad
resonance ($\Gamma \sim 550$ MeV), being $K\pi$ its dominant decay
mode. It is very similar to the $f_0(600)$, and hence it cannot be
interpreted as a Breit-Wigner narrow resonance.

We identify the second pole at $\sqrt{s_R}=(1428,-24)$ with the
$K^*_0(1430)$ resonance, despite being the latter one much wider than
the pole found in our scheme. The $K\pi$ branching fraction for this
resonance is $93\%\pm 10\%$ \cite{Amsler:2008zz}. The pole generated
in our scheme couples more than twice stronger to the $K\pi$ channel
than to the $\eta K$ one, which is also open. However, the coupling to
the $K^*\rho$ channel is four times bigger and does not contribute to
the width of 48 MeV quoted in Table~\ref{tab:1-12-0} because it is not
open. Nevertheless, the resonance can decay into a virtual  $K^*\rho$ pair
which will significantly enhance the $K\pi$ decay probability, thanks
to the broad $\rho$ and $K^*$ widths and the fact that the pole is not
placed too far from threshold (see left panel of Fig.~\ref{fig:box}).

In Ref.~\cite{Geng:2008gx}, where only $VV-$channels are considered, only one
pole at $(1643, -24)$ with a strong $\rho K^*$ coupling was found. The authors
of \cite{Geng:2008gx} argue, although with reservations, that it might
correspond to the $K(1630)$ resonance. We conjecture that with an adequate
subtraction constant the pole found in that reference might be similar to our
second pole and thus it would rather correspond to the $K^*_0(1430)$
resonance.

The situation of the third pole is less clear. It would be tempting to
associate this third pole at $\sqrt{s_R}=(1787,-37)$ to the $K(1630)$ (with
yet undetermined $J^P$ \cite{Amsler:2008zz}). Our pole is wider than the
$K(1630)$ resonance, whose reported width is compatible with zero.  This could
be explained because our pole is located above the $K^*\omega$ threshold and
experimentally it is below this threshold although close to it. Note that this
channel gives rise to a decay mode $K\pi\pi$, as reported in the
PDG. Nevertheless, we believe that such identification would probably be
incorrect, since the biggest couplings of the pole found here are those
corresponding to the $\eta K$ and $K^*\phi$ channels. The first of these two
channels is open, giving rise to a sizeable width difficult to reconcile with
the narrow width quoted in the PDG for the $K(1630)$.  Besides, the huge
$K^*\phi$ coupling will lead to a $K\pi$ decay mode, through the loop
mechanisms sketched in Fig.~\ref{fig:box}, while the decay mode observed in
the PDG is $K\pi\pi$. Note that the pole at $\sqrt{s_R}=(1787,-37)$ also
couples to the $K^*\rho$ channel and that it will also contribute to the
$K\pi$ decay mode. This suggests to identify the pole found here with the wide
$K^*_0(1950)$ resonance, for which the decay mode observed in the PDG is
$K\pi$. Moreover, its large width $201\pm 90$ MeV ~\cite{Amsler:2008zz}) make
less meaningful the difference between its mass $1945\pm 22$ and that of our
pole. However, it should be pointed out that the $K^*_0(1950)$ resonance is
not firmly established yet and needs further
confirmation~\cite{Amsler:2008zz}.

\begin{table}[htpb]
      \renewcommand{\arraystretch}{2}
     \setlength{\tabcolsep}{0.1cm}
     \caption{Pole positions and modulus of the couplings (MeV units) in the
       $(Y,I,J)=(1,1/2,1)$ sector. $I(J^{P})=\frac{1}{2}(1^{+})$. The
       subtraction constants has been set to $a=-2.5$ for the $PV$ channels
       and to $a=-1.7$ for the three $VV$ ones. Possible PDG counterparts:
       $K_1(1270)$, $K_1(...)$, $K_1(1400)$, $K_1(1650)$.}
\vspace{0.3cm}
\begin{footnotesize}
\begin{tabular}{c|cccccccc}
\hline\hline
 $\sqrt{s_R}$ & $\pi K^*$ &  $ K \rho$ & $ K \omega$ & $\eta K^*$  & $ K\phi$
& $ K^* \rho$ & $ K^* \omega$ & $ K^* \phi$  \\\hline
 $(1188,-64)$ & 5616 & 3703 & 1959 & 1988 & 1860 & 4405 & 2824  & 2669\\
 $(1250,-31)$ & 3910 & 5267 & 2516 & 3612 & 1665 & 3225 & 6311  & 2023\\
 $(1414,-66)$ &  798 & 3326 & 3030 & 1169 & 1668 & 9866 & 2225  & 4373\\
 $(1665,-95)$ & 1358 & 1166 &  922 & 3650 & 3799 & 2880 & 3017  &10641\\
\hline\hline
\end{tabular}\label{tab:1-12-1}
\end{footnotesize}
\end{table}

\subsection{Hypercharge 1, isospin 1/2 and spin 1}

In this sector there are eight coupled channels: $K^* \pi$, $ K \rho$, 
 $ K \omega$, $ K^* \eta$,  $ K\phi$,  $ K^*\rho$, $ K^*\omega$, 
$K^*\phi$, and four poles are found on the SRS. 

In the PDG there appear three resonances below 2 GeV with these quantum
numbers, namely $K_1(1270)$, $K_1(1400)$, $K_1(1650)$, while here we found
four poles.

In Ref.~\cite{Roca:2005nm} two poles ($\sqrt{s_R}= (1112, -64)$ MeV and
$\sqrt{s_R}= (1216, -4)$ MeV) were reported, using only $PV\to PV$
interaction.  An additional pole was found at $\sqrt{s_R}= (1737, -82)$ MeV in
Ref.~\cite{Geng:2008gx}, using only the $VV\to VV$ sector. The work of Roca et
al.~\cite{Roca:2005nm} was revisited in Ref.~\cite{Geng:2006yb}. In this
latter reference, the double pole structure of the $K_1(1270)$, uncovered in
\cite{Roca:2005nm}, is further confirmed. Let us summarize here some of the
most relevant findings of Ref.~\cite{Geng:2006yb}. There, one pole is found at
$\sim$ 1200 MeV with a width of $\sim 250$ MeV and the other is found at $\sim
1280$ MeV with a width of $\sim 150$ MeV. The lower pole couples more to the
$K^* \pi$ channel and the higher pole couples dominantly to the $ K \rho$
channel. The peak in the $K\pi\pi$ mass distribution in the WA3
data~\cite{Daum:1981hb} on $K^-p \to K^-\pi^+\pi^-p$ is explained in
\cite{Geng:2006yb} as a superposition of two poles, but in the $K^*\pi$
channel the lower pole dominates and in the $\rho K$ channel, the higher pole
gives the biggest contribution. Finally, it is argued in \cite{Geng:2006yb}
that different reaction mechanisms may prefer different channels and this
would explain the different invariant mass distributions seen in various
experiments.

The results compiled in Table~\ref{tab:1-12-1}, show two poles around 1.2 GeV
which correspond to those reported in Ref.~\cite{Geng:2006yb}, though the
couplings turn out to be somehow different. This is partially due to the
inclusion here of the $VV$ channels. When those channels are switched off, the
agreement improves, but there still remain some differences between the
couplings obtained in both approaches, specially on the strength of the
$K^*\eta$ coupling for the lightest resonance. This can be attributed, at this
stage, to the approximation $m_P=0$ used here when computing the
potential. Our scheme implements an extra SU(6) symmetry breaking pattern
induced by the use of different pseudoscalar and vector decay constants ($f_P
\ne f_V$), however, in Ref.~\cite{Geng:2006yb} the WA3 $K^-p \to
K^-\pi^+\pi^-p$ data were successfully fitted with a value of $f^2\sim (115 \,
{\rm MeV})^2$, which numerically is rather similar to $f_P f_V$ used here (see
Table~\ref{tab:cts}). On the other hand, taking into account the finite $\rho$
and $K^*$ widths in the intermediate loops will increase the imaginary parts
of the poles, specially that of the higher pole which has a large coupling to
the $\rho K$ channel. This will then bring its width close to $\sim 150\,{\rm
  MeV}$, as found in in Ref.~\cite{Geng:2006yb}. Thus, our findings here
reinforce the double pole picture for the $K_1(1270)$ resonance predicted in
Refs.~\cite{Roca:2005nm,Geng:2006yb}. It is also noteworthy that
Ref.~\cite{Lutz:2003fm} did not find this double pole structure.

We move now to the third of the poles found here [$\sqrt{s_R}= (1414, -66)$
  MeV\,], which has large $K^*\pi$, $K\rho$, $K\omega$, $K^*\phi$ and
specially $K^*\rho$ couplings.  Given its mass and width, it can be naturally
associated to the $K_1(1400)$ resonance.  However in the PDG, branching
fractions of $(94\pm 6)\%$, $(3\pm 3)\%$ and $(1\pm 1)\%$, for the $K^*\pi$,
$K\rho$ and $K\omega$ modes, respectively, are quoted for this resonance. The
couplings shown in Table~\ref{tab:1-12-1} cannot be easily reconciled with
the above fractions. Refs.~\cite{Roca:2005nm,Geng:2006yb} did not find the
$K_1(1400)$ resonance, while in Ref.~\cite{Lutz:2003fm} a broad bump in the
speed plot was associated to it. In the $VV-$work of Ref.~\cite{Geng:2008gx},
a pole at $\sqrt{s_R}= (1737, -82)$ MeV is reported with a dominant $\rho K^*$
coupling. Indeed, when the $PV-VV$ interferences are switched off, we find the
two $K_1(1270)$ poles in the $PV$ sector and a third pole in the $VV$ sector
with a large $\rho K^*$ coupling, whose position depends strongly on the value
of the subtraction constant. Our conjecture is that it is precisely this pole,
which manifests itself as a $\rho K^*$ bound or resonant state when only $VV$
interactions are considered, the one that moves down to $\sqrt{s_R}= (1414,
-66)$ when the $PV$ channels are also included.

Here, we envisage two different possibilities:
\begin{itemize}

\item[(i)] To identify the $\sqrt{s_R}= (1414, -66)$ pole with the $K_1(1400)$
  resonance, despite the PDG branching fractions quoted above.  It is worth
  stressing here that the properties quoted in the PDG obtained from the WA3
  data analysis rely upon considering only one pole for the $K_1(1270)$.  The
  reanalysis of the WA3 data carried out in \cite{Geng:2006yb}, where the
  double $K_1(1270)$ pole structure is taken into account and the totality of
  the $PV$ channels studied here are considered, is also inconsistent with the
  PDG $K_1(1400)$ branching fractions, being the $K\rho$ mode almost
  comparable to the $K^*\pi$ one (see Fig. 7 of this reference) and certainly
  it is not around 30 times smaller. On the other hand, there exists another
  ingredient which should be considered. Our state has a huge $K^*\rho$
  coupling, which will provide a $K\pi\pi\pi$ signature and that of course
  will also contribute to the inclusive WA3 $K^-p \to K^-\pi^+\pi^-p$
  reaction. This latter mechanism was considered neither in the original
  analysis of Ref.~\cite{Daum:1981hb} nor in the better theoretical founded
  re-analysis of Ref.~\cite{Geng:2006yb}.

  Within this scenario, the fourth pole at $\sqrt{s_R}= (1665, -95)$
  shown in Table~\ref{tab:1-12-1}, could be assigned to the
  $K_1(1650)$ with a mass of $1650\pm 50$ MeV and a width of $150\pm
  50$~\cite{Amsler:2008zz}. The only decay channels observed are
  $K\pi\pi$ and $K\phi$, which could be easily associated to the large
  $K\phi$ coupling of the pole together with its sizeable $K^*\pi$ and
  $K\rho$ components (see Table~\ref{tab:1-12-1}). This pole appears
  due to the interplay between the $K^*\phi$ and $K\phi$ channels,
  similarly as it was discussed earlier in the case of the $h_1(1595)$, and
  indeed it disappears when  only the $VV$ sector is considered. In
  Ref.~\cite{Geng:2008gx}, the above mentioned $\sqrt{s_R}= (1737,
  -82)$ pole was tentatively assigned to the $K_1(1650)$ resonance
  despite the fact that its large $\rho K^*$ coupling is difficult 
  to accommodate with the $K_1(1650)$ known decays.

  Nevertheless, this is still a questionable scenario, since the couplings
  quoted in Table~\ref{tab:1-12-1}, for the pole at $(1414, -66)$, indicate
  that its $K\rho$ decay mode is much larger than the $K^*\pi$, and this is
  difficult to reconcile even with the results of the re-analysis of
  Ref.~\cite{Geng:2006yb}.

\item[(ii)] Alternatively, the subtraction constants could be fine-tuned so
  that the third pole is pushed up in energy and thus it could be associated
  to the $K_1(1650)$ (see for instance Table~\ref{tab:1-12-1-bis}). Properties
  of the poles, other than the mass and width, are not much affected by the
  fine-tuning, and one still gets the two pole structure for the $K_1(1270)$
  resonance. In this scenario no pole is assigned to the $K_1(1400)$, which
  then will not be dynamically generated, as advocated in the picture of
  Refs.~\cite{Roca:2005nm,Geng:2008gx,Geng:2006yb}. However, the assignment of
  the third pole to the $K_1(1650)$ would suffer from the problems
  mentioned above in the case of the $\sqrt{s_R}= (1737, -82)$ resonance found
  in Ref.~\cite{Geng:2008gx}. In addition, a further $K_1$ above 1.8 GeV and
  not included in the PDG, will be predicted with a large $K\phi$ decay mode.

\end{itemize}

\begin{table}[htpb]
      \renewcommand{\arraystretch}{2}
     \setlength{\tabcolsep}{0.1cm}
     \caption{Same as Table~\ref{tab:1-12-1}
     [$I(J^{P})=\frac{1}{2}(1^{+})$], but using  subtraction constants
     $a=-3.1$ for the $PV$ channels and $a=-1.0$ for the three $VV$
     ones. Possible PDG counterparts: $K_1(1270)$, $K_1(...)$,
     $K_1(1650)$, $K_1(...)$.}
\label{tab:1-12-1-bis}
\vspace{0.3cm}
\begin{footnotesize}
\begin{tabular}{c|cccccccc}
\hline\hline
 $\sqrt{s_R}$ & $\pi K^*$ &  $ K \rho$ & $ K \omega$ & $\eta K^*$  & $ K\phi$  & $ K^*\rho$ & $ K^*\omega$ & $ K^*\phi$  \\\hline
 $(1169,-46)$ & 4595 & 3643 & 1862 & 1873 & 1484 & 2468 & 2328  & 1418\\
 $(1266,-44)$ & 4539 & 5789 & 2797 & 4504 & 2870 & 1077 & 7956  & 1837\\
 $(1576,-43)$ &  888 & 1868 & 2508 &  796 &  682 & 9874 & 1364  & 3366\\
 $(1823,-61)$ &  474 &  154 &  221 & 2636 & 3105 & 770  & 1295  &11756\\
\hline\hline
\end{tabular}
\end{footnotesize}
\end{table}
\begin{table}[htpb]
      \renewcommand{\arraystretch}{2}
     \setlength{\tabcolsep}{0.2cm}
     \caption{Pole positions and modulus of the couplings (MeV units) in the
       $(Y,I,J)=(1,1/2,2)$ sector. $I(J^{P})=\frac{1}{2}(2^{+})$.  Possible
       PDG counterparts: $K_2^*(1430)$.}
\label{tab:2-1/2-2} 
\vspace{0.3cm}
\begin{tabular}{c|ccc}
\hline\hline
 $\sqrt{s_R}$  & $ K^*\rho$ & $ K^*\omega$ & $ K^*\phi$  \\\hline 
$(1708,-156)$ &    7227  & 2834 & 2299\\
\hline\hline
\end{tabular}
\end{table}

\subsection{Hypercharge 1, isospin 1/2 and spin 2}

In this sector, a pole is found in the SRS of the amplitudes. In the PDG, two
$K^*_2$ resonances below 2 GeV [$K^*_2(1430)$ and $K^*_2(1980)$] are reported,
though only the lightest one is firmly established. The $K^*_2(1430)$ has a
mass of $1429\pm 1.4 \,{\rm MeV}$ and a width of $104\pm 4 \,{\rm MeV}$; the
second resonance has a mass of $1973\pm26$ MeV and a width of $373\pm70$
MeV. It is not clear to which one to associate the state we find. The
subtraction constants cannot be fine-tuned to achieve the mass of the pole to
lie much closer to 1.43 GeV than in Table~\ref{tab:2-1/2-2}.  Nevertheless, we
believe that the pole found here might correspond to the $K^*_2(1430)$ and its
nature is somehow related to those of the $f_2(1270)$ and
$f^\prime_2(1525)$. In both cases, an important influence of $d-$wave
interactions is to be expected. Indeed in the case of the
$K^*_2(1430)$, the PDG
branching fractions are  around 50\%, 25\%, 9\% and 3\% for the $d-$wave
modes $K\pi$, $K^*\pi$, $K\rho$ and $K\omega$, respectively. In addition, the
branching fraction of the $K^*\pi\pi$ channel is only about 13\%. This latter
decay mode looks like the only one more or less related to the dynamics
included within our model, thanks to the dominant coupling $K^*\rho$ of the
pole displayed in Table~\ref{tab:2-1/2-2}.  This would explain why our model
does not describe properly the mass and the width of the $K^*_2(1430)$. From
this point of view, what is somewhat more surprising is the fact that our
scheme were able to describe the mass of the $f_2(1270)$ at all. However,
there is here a distinctive feature: the possible influence of the $d-$wave
pseudoscalar-vector meson $K^* \pi$ channel, which lies closer to the
resonance mass than the pseudoscalar-pseudoscalar channels. Notice that the
equivalent channel in the case of $f_2(1270)$ would be $\pi\rho$, but it is
not allowed by $G-$parity conservation.

The approach of Ref.~\cite{Geng:2008gx} for $VV\to VV$ scattering produces a
resonance in this sector, with mass fine-tuned to $1430\,{\rm MeV}$, even if
all type $d-$wave interactions are also ignored.

\subsection{Exotics}
Exotics refers here to meson states with quantum numbers that cannot be formed
by a $q\bar{q}$ pair. Quantum numbers with $I \ge 3/2$ or $|Y|= 2$ are
exotic. Our model produces five poles on the complex plane with the following
quantum numbers: $2^+(0^{++})$ with $Y=0$, $3/2(0^+)$ and $3/2(1^+)$ with
$Y=1$, and $0(1^+)$ and $1(0^+)$ with $Y=2$. Remarkably, no exotic state was
reported in Ref.~\cite{Geng:2008gx}. This is a direct consequence of
the different dynamics implicit in both approaches. Future experiments may be
needed to distinguish between these two schemes.

\begin{table}[htpb]
      \renewcommand{\arraystretch}{2}
     \setlength{\tabcolsep}{0.2cm}
     \caption{Poles positions and modulus of the couplings (MeV units) in the
       $(Y,I,J)=(0,2,0)$ sector.  The subtraction constant has been set to
       $a=-1.5$. $I^G(J^{PC})=2^+(0^{++})$ . Possible PDG counterparts:
       $X(1420)$.}\label{tab:X020}
\vspace{0.2cm}
\begin{tabular}{c|cc}
\hline\hline
 $\sqrt{s_R}$& $\pi\pi$ & $\rho\rho$ \\\hline
 $(1419,-54)$ & 2719 &10069\\
\hline\hline 
\end{tabular}
\end{table}

\subsubsection{Hypercharge 0, isospin 2 and spin 0}

In this sector, a pole is found that, given its mass and width, can be
naturally associated to the $X(1420)$ resonance (see
Table~\ref{tab:X020}). This resonance needs further confirmation and
its current evidence comes from a statistical
indication~\cite{Filippi:2000is} for a $\pi^+\pi^+$ resonant state in
the $\bar n p \to \pi^+ \pi^+ \pi^-$ annihilation reaction with data
collected by the OBELIX experiment. Within our scheme, the pole is
essentially a $\rho\rho$ bound state with a small coupling to the
$\pi\pi$ channel that moves the pole to the SRS. The fact that the
strength of the coupling to $\pi^+\pi^+$ is not large might explain
why the resonance distorts weakly the spectrum of the outgoing pair of
positive pions in the OBELIX data. Within our scheme, the $\rho\rho\to
\rho\rho$ amplitude is symmetric under $I \leftrightarrow J$
exchange. For ${\mathcal D}_{\rm kin}$ this comes as a result of SU(6)
symmetry. On the other hand, the interaction ${\mathcal D}_m$ is a
contact term and this ensures the invariance under $I \leftrightarrow
J$.\footnote{ Indeed, the most general contact interaction in the
$\rho\rho$ sector is of the form $ {\mathcal L}_{\rm int}=g_1
\rho_{ai}\rho_{ai}\rho_{bj}\rho_{bj} + g_2
\rho_{ai}\rho_{aj}\rho_{bi}\rho_{bj} $ which is symmetric under
exchange of spin and isospin labels.} As a consequence our $\rho\rho$
potential in this sector ($I=2, J=0$) is the same as that in the $I=0,
J=2$ one. BS amplitudes in both sectors will become different because
of coupled-channel and renormalization effects. Nevertheless, we
expect the $X(1420)$ to be the counterpart of the $f_2(1270)$, which
appeared mostly as a $\rho\rho$ $J=2$ isoscalar bound state. This
situation is distinctively different in the hidden gauge interaction
model used of Ref.~\cite{Geng:2008gx}, where near threshold, the
$\rho\rho$ interaction in the $I=2, J=0$ sector becomes repulsive and
five times smaller, in absolute value, than that in the $I=0, J=2$
sector~\cite{Molina:2008jw}. Indeed, while in the latter sector the
$\rho\rho$ interaction is attractive and gives rise to the $f_2(1270)$
resonance, the model of Ref.~\cite{Molina:2008jw, Geng:2008gx} does
not provide any ($I=2, J=0$) resonance. However, it is found a dip in
the $\rho\rho$ amplitude squared in this latter work. There, it is
suggested that such a dip in the $\rho\rho$ amplitude might lead to a
bump in $\pi^+\pi^+$ production.

The $\pi\pi$ diagonal potential is repulsive in this sector, however, the
$\pi\pi \to \rho\rho$ transition potential leads to an interaction more
attractive than that deduced from the diagonal $\rho\rho$ potential. Indeed,
from Tables~ \ref{tab:020} and~\ref{tab:020m}, one finds eigenvalues $\pm 2$
for ${\mathcal D}_{\rm kin}$ and $8/3$, $-8$ for ${\mathcal D}_m$ (the two
matrices entering in the kernel potential) while the $\rho\rho$ diagonal
matrix elements are $-1$ and $-16/3$, respectively. [Notice that our
  conventions are such that negative diagonal matrix elements, or eigenvalues,
  of ${\mathcal D}_{\rm kin}$, ${\mathcal D}_m$ and ${\mathcal D}_a$
  correspond to attractive interactions.]

\begin{table}[htpb]
      \renewcommand{\arraystretch}{2}
     \setlength{\tabcolsep}{0.2cm}
     \caption{Pole positions and modulus of the couplings (MeV units) in the
       $(Y,I,J)=(1,3/2,0)$ sector. $I^G(J^{P})=\frac32(0^{+}).$}\vspace{0.3cm}
\begin{tabular}{c|ccc}
\hline\hline
 $\sqrt{s_R}$ & $ K\pi$ & $ K^*\rho$  \\\hline 
 $(1433,-70)$ &  3242 & 10962\\
\hline\hline
\end{tabular}\label{tab:1320}
\end{table}
\begin{table}[htpb]
      \renewcommand{\arraystretch}{2}
     \setlength{\tabcolsep}{0.2cm}
     \caption{Pole positions and modulus of the couplings (MeV units) in the
       $(Y,I,J)=(2,1,0)$ sector. $I(J^{P})=1(0^{+}).$} \vspace{0.3cm}
\begin{tabular}{c|cc}
\hline\hline
 $\sqrt{s_R}$ & $K K$ & $K^* K^*$  \\\hline
$(1564,-66)$ &   3484 &  11593\\
\hline\hline
\end{tabular}\label{tab:210}
\end{table}

\subsubsection{Hypercharge 1, isospin 3/2 and spin 0  and hypercharge 2,
  isospin 1 and spin 0 }

The matrices ${\mathcal D}_{\rm kin}$ and ${\mathcal D}_m$ are the same
in both sectors, and identical to those appearing in $(Y,I,J)=(0,2,0)$. Thus,
the two resonances displayed in Tables~\ref{tab:1320} and~\ref{tab:210} belong
to the same multiplet of scalars that the resonance $X(1420)$, and masses and
widths are similar. We will come back to this point below.

\begin{table}[htpb]
      \renewcommand{\arraystretch}{2}
     \setlength{\tabcolsep}{0.2cm}
     \caption{Pole positions and modulus of the couplings (MeV units) in the
       $(Y,I,J)=(1,3/2,1)$ sector. $I(J^{P})=\frac32(1^{+}).$}\vspace{0.3cm}
\begin{tabular}{c|ccc}
\hline\hline
 $\sqrt{s_R}$  & $\pi K^*$ & $ K\rho$ & $ K^*\rho$  \\
\hline 
  $(1499,-127)$ &     3791 & 3699 &    8513\\
\hline\hline
\end{tabular}\label{tab:1321}
\end{table}

\begin{table}[htpb]
      \renewcommand{\arraystretch}{2}
     \setlength{\tabcolsep}{0.2cm}
     \caption{Pole positions and modulus of the couplings (MeV units) in the
       $(Y,I,J)=(2,0,1)$ sector. $I(J^{P})=0(1^{+}).$} 
\vspace{0.3cm}
\begin{tabular}{c|cc}
\hline\hline
 $\sqrt{s_R}$  & $ K K^*$ & $K^* K^*$  \\\hline
$(1608,-114)$ & 5614 & 9303\\
\hline\hline
\end{tabular}\label{tab:201}
\end{table}

\subsubsection{Hypercharge 1, isospin 3/2 and spin 1  and hypercharge 2, 
isospin 0 and spin 1}

We find one pole in each sector (see Tables~\ref{tab:1321} and
\ref{tab:201}). Masses and widths of these two resonances are quite similar
and we will argue below that they belong to the same axial vector multiplet.

\subsubsection{Hypercharge 0, isospin 2 and spin 1 and 2, 
hypercharge 1, isospin 3/2 and spin 2,  and hypercharge  2, isospin 1 and spin 1 and 2 }

The interaction in these five sectors is repulsive and they present no poles.

\section{Summary and conclusions}

\label{sec:concl}

Tables ~\ref{table:polesum} and ~\ref{table:polesum-2} compile the different
poles found within the present approach. It must be observed that the widths
obtained are only a first approximation and they could receive substantial
corrections in some cases. This is because of the following reasons. First,
the only decay channels considered are $PP$, $PV$ and $VV$ $s$-wave
pairs. Second, the widths of the vector mesons have been neglected in their
propagators in the loop functions. The effect might be particularly important
for the $\rho$ and the $K^*$ resonances. It is to be expected that this
mechanism will enhance the width of the resonances with a very small impact on
the masses ~\cite{Roca:2005nm}. The same mechanism should also introduce
contributions of the type displayed in Fig.~\ref{fig:box}. Such contributions
have appeared repeatedly during the discussion of our results.

\begin{table}[htpb]
      \renewcommand{\arraystretch}{2}
     \setlength{\tabcolsep}{0.4cm}
     \caption{ Poles having non exotic quantum numbers found in this work  and
       possible PDG counterparts. Units are given in MeV. Those resonances
       marked with $\dagger$ need to be confirmed. A question mark symbol
       expresses our reservations on the assignment.} 
\label{table:polesum}
\vspace{0.3cm}
\begin{tabular}{c|cc|ccc}
\hline\hline
$(Y,I,J)$   &$I^G(J^{PC})$ & This model & \multicolumn{3}{c}{PDG~\protect\cite{Amsler:2008zz}}\\
            &              &      Pole position ($\sqrt{s_R}$)         &  Name & Mass & Width\\\hline
$(0,0,0)$   &$0^+(0^{++})$ & $(635,-202)$ & $f_0(600)$ & $400\sim1200$ & $600\sim1000$\\
            &$0^+(0^{++})$ & $(969,0)$ & $f_0(980)$     & $980\pm10$ & $40\sim100 $\\
            &$0^+(0^{++})$ & $(1350,-62)$ & $f_0(1370)$     &
$1200\sim 1500$ & $200\sim 500 $\\
            &$0^+(0^{++})$ & $(1723,-52)$ & $f_0(1710)$     & $1720\pm
6 $ & $135 \pm 8 $\\\hline
$(0,0,1)$   &$0^-(1^{+-})$ & $(1006,-85)$ & $h_1(1170)$ & $1170\pm20$ &
$360\pm 40$\\
            &$0^-(1^{+-})$ & $(1373,-17)$ & $h_1(1380)^\dagger$ & $1386\pm19$ & $91\pm30$\\
            &$0^-(1^{+-})$ & $(1600,-67)$ & $h_1(1595)^\dagger$& $1594^{+18}_{-60}$ &$384^{+90}_{-120}$\\
            &$0^+(1^{++})$ & $(1286,0)$ & $f_1(1285)$ & $1281.8\pm 0.8$
& $24.3\pm 1.1$\\
\hline
$(0,0,2)$   &$0^+(2^{++})$ & $(1289,0)$ &$f_2(1270)$ & $1275.1\pm 1.2$
& $185.1^{+2.9}_{-2.4}$ \\
            &$0^+(2^{++})$ & $(1783,-19)$ & $f_2(1640)^\dagger$ or
$f_2^\prime(1525)$ & $1639\pm 6$
& $99^{+60}_{-40}$ \\
            & &   & or $f_2(1430)^\dagger$ or  $f_2(1565)^\dagger$, $\cdots$ \\\hline
$(0,1,0)$   &$1^-(0^{++})$ & $(991,-46)$ & $a_0(980)$ & $980\pm 20$ & $50\sim100$\\
            &$1^-(0^{++})$ & $(1442,-5)$ & $a_0(1450)$& $1474\pm19$ & $265\pm13$\\
            &$1^-(0^{++})$ & $(1760,-12)$ &$a_0(2020)^\dagger$ ? & $2025
\pm 30$ & $330 \pm 75$ \\\hline
$(0,1,1)$   &$1^+(1^{+-})$ & $(1234,-57)$ & $b_1(1235)$ & $1229.5\pm3.2$ & $142\pm9$\\
            &$1^+(1^{+-})$ & $(1642,-139)$ & $b_1(1960)^\dagger$ ? &
$1960 \pm 35 $ & $230 \pm 50$ \\
            &$1^-(1^{++})$ & $(1021,-251)$ & $a_1(1260)$ & $1230\pm40$
& $250 \sim 600$\\
            &$1^-(1^{++})$ & $(1568,-145)$ & $a_1(1640)^\dagger$ &
$1647\pm 22$ & $254\pm27$\\\hline
$(0,1,2)$   &$1^-(2^{++})$ & $(1228,0)$ & $a_2(1320)$ & $1318.3\pm
0.6$ & $107\pm 5$\\
   &$1^-(2^{++})$ & $(1775,-6)$ & $a_2(1700)^\dagger$ & $1732\pm16$ & $194\pm40$\\\hline
$(1,1/2,0)$   & $1/2(0^+)$   & $(830,-170)$ & $K^*_0(800)^\dagger$ & $672\pm40$ & $550\pm34$\\
              & $1/2(0^+)$   & $(1428,-24)$ & $K^*_0(1430)$ & $1425\pm50$ & $270\pm80$ \\
              & $1/2(0^+)$   & $(1787,-37)$ & $K_0^*(1950)^\dagger$ &
$1945\pm 22$ & $201\pm 90$\\\hline
$(1,1/2,1)$   & $1/2(1^+)$   & $(1188,-64)$ & $K_1(1270)$ & $1272\pm7$ & $90\pm20$\\
              & $1/2(1^+)$   & $(1250,-31)$ & $K_1(...)$ &  & \\
              & $1/2(1^+)$   & $(1414,-66)$ & $K_1(1400)$ ? & $1403\pm7$
& $174\pm 13$\\
              & $1/2(1^+)$   & $(1665,-95)$ & $K_1(1650)^\dagger$ ? & $1650\pm50$ & $150\pm50$\\\hline
$(1,1/2,2)$   & $1/2(2^+)$   & $(1708,-156)$ & $K_2^*(1430)$ ? & $1429\pm1.4$ & $104\pm4$\\\hline\hline
\end{tabular}
\end{table}

\begin{table}[htpb]
      \renewcommand{\arraystretch}{2}
     \setlength{\tabcolsep}{0.4cm}
     \caption{Poles with exotic quantum numbers found in this work  and
       possible PDG counterparts. Units are given in MeV. Those resonances
       marked with $\dagger$ need to be confirmed.} 
\label{table:polesum-2}
\vspace{0.3cm}
\begin{tabular}{c|cc|ccc}
\hline\hline
$(Y,I,J)$   &$I^G(J^{PC})$ & This model & \multicolumn{3}{c}{PDG~\protect\cite{Amsler:2008zz}}\\
            &              &      Pole position ($\sqrt{s_R}$)         &  Name & Mass & Width\\\hline
$(0,2,0)$   & $ 2^+(0^{++})$& $(1419,-54)$ & $X(1420)^\dagger$ & $1420\pm20$ & $160\pm10$\\\hline
$(1,3/2,0)$ & $3/2(0^+)$   & $(1433,-70)$ \\  
$(2,1,0)$  & $1(0^+)$   & $(1564,-66)$ \\
$(1,3/2,1)$ & $3/2(1^+)$   & $(1499,-127)$ \\
$(2,0,1)$   & $0(1^+)$    & $(1608,-114)$\\
\hline\hline
\end{tabular}
\end{table}

SU(6) symmetry of our approach has been explicitly broken to account for
physical masses and decay constants, and also when the amplitudes have been
renormalized. Nevertheless, the underlying SU(6) symmetry is still present and
serves to organize the set of even parity meson resonances found in this work,
and compiled in Tables~\ref{table:polesum} and ~\ref{table:polesum-2}.

Spin-flavor symmetry has been used to guide the construction of the $s-$wave
interactions among the members of the SU(6) {\bf 35} multiplet. The
matrix ${\mathcal D}_{\rm kin}$ that appears in the kinetic term of the
amplitudes can be expressed as
\begin{equation}
{\mathcal D}_{\rm kin} = -12 P_{\bf 1}-6 P_{{\bf 35}_s}-2P_{{\bf 189}} +
2P_{{\bf 405}}
.
\end{equation}
Therefore, this interaction is (moderately) repulsive in the ${\bf 405}$
representation and attractive in the other representations. To the extent that
${\mathcal D}_{\rm kin}$ is the dominant term, this favors the existence of up
to 225 (1+35+189) states. (This counts all states of spin and isospin as
different, not only multiplets. In terms of $J^PI^GY$ states, this
number is 45.)  The irreducible representations (irreps) of
SU(6) can be reduced in terms of irreps of SU(3)$\otimes$SU(2). In this way,
the content of the SU(6) ${\bf 1},{\bf 35}_s,{\bf 189}$ irreps is as follows
\begin{eqnarray}
{\bf 1}& =& 1_1
,
\nonumber \\ 
{\bf 35} & =& 8_1 \oplus 8_3\oplus 1_3
,
\nonumber \\ 
{\bf 189} & =& 27_1\oplus8_1\oplus1_1 \oplus 10_3\oplus 10^*_3\oplus
8_3\oplus8_3\oplus8_5\oplus1_5
,
\end{eqnarray}
where the subindex refers to $2J+1$, so e.g., $10^*_3$ stands for the
representation $10^*$ of SU(3) with $J=1$. Further, the $(Y,I)$ content of the
SU(3) irreps is as follows
\begin{eqnarray}
1 &=& (0,0)
,
 \nonumber \\
8 &=& (\pm 1,1/2), (0,1), (0,0)
, 
\nonumber \\
10\oplus10^* &=& (\pm 1,3/2), (0,1), (0,1),(\pm 1,1/2),(\pm 2,0)
,
\nonumber \\
27 &=& (\pm 2,1), (\pm 1,3/2),(\pm 1,1/2), (0,2),(0,1),(0,0)
.
\end{eqnarray}

The gross features of the states reported in Tables~\ref{table:polesum} and
~\ref{table:polesum-2} follow the above decomposition based on SU(6)
multiplets. This picture is somewhat modified by the effect of the terms added
to the kinetic contribution of the Hamiltonian (see Eq.~(\ref{eq:vsu6})),
namely, ${\mathcal D}_m$, which is mainly attractive and ${\mathcal D}_a$,
which is repulsive.  As mentioned, the use of different vector and
pseudoscalar meson masses and decay constants, and the used subtraction
constants, which in some cases have been fine-tuned to better reproduce the
experimental (PDG) resonances, produce also a deviation from the SU(6)
pattern.

In Table~\ref{tab:j012}, the poles found in this work (Tables
~\ref{table:polesum} and ~\ref{table:polesum-2}) are classified in terms of
the above SU(6) and SU(3)$\otimes$SU(2) irreps. Several comments are in order
here. First, it should be stressed that there will be mixings between
states with the same $J^PI^GY$ quantum numbers but belonging to different
SU(6) and/or SU(3) multiplets, since these symmetries are broken both within
our approach and in nature. These mixings have not been considered
when classifying the states in Table~\ref{tab:j012}. Some comments are also
pertinent regarding each spin-parity sector:
\begin{itemize}

\item[(i)] $J^P = 0^+$: As can be seen in Table~\ref{tab:j012}, the
  poles found here closely follow the pattern determined by the spin-flavor
  SU(6) symmetry, except for the absence of the singlet state associated to
  the {\bf 189} SU(6) irrep. The attractive interaction in this irrep is
  weak. We have checked that if the SU(6) symmetry breaking contact term
  (${\mathcal D}_m$) is switched off a new $f_0$ resonance (with a mass close
  to 1.9 GeV) would be generated in our calculation, corresponding to this
  $1_1$ missing state. On the other hand, the SU(6) pattern is also accurate
  when describing the PDG scalar resonances compiled in
  Table~\ref{tab:j012}. This fact has two consequences. First, it increases
  the credibility of our predictions on the existence of two exotic states in
  the region 1.4--1.6 GeV, belonging to the SU(3) $27$ irrep included in the
  SU(6) {\bf 189}, while giving further theoretical support on the reliability
  of other resonances, not yet firmly established, as for example the
  $a_0(2020)$ or the exotic isotensor $X(1420)$ state.  Second, by inspection
  of the resonances with these quantum numbers reported in the PDG and with
  masses below 2 GeV, it can be noted that there exists just one well
  established resonance that does not fit within the SU(6) classification
  pattern assumed in Table~\ref{tab:j012}.\footnote{We will omit here any
    reference to the $K(1630)$ resonance, since its $J^P$ is undetermined
    yet~\cite{Amsler:2008zz}.} This is the $f_0(1500)$ resonance, for which a
  glueball picture has been suggested by several
  authors~\cite{Amsler:1995td,Amsler:2004ps}. Our result would then be in
  support of such picture.

\item[(ii)] $J^P = 1^+$: Here the effects of the SU(6) breaking terms
  ${\mathcal D}_m$ and ${\mathcal D}_m$ terms turn out to be important. There
  are two types of channels, namely, $PV$ and $VV$ mesons coupled to total
  spin 1.

  The $PV \to PV$ amplitudes are constrained by the LO WT theorem (see
  Eq.~(\ref{eq:lowtpv})) and give rise to the states of the multiplets
  $8_3$ and $1_3$ of the SU(6)  $ {\bf 35}_s$ irrep and to those of a further
  SU(3) octet ($8_3^a$) of the SU(6) ${\bf 189}$. Note that the
  dynamics of the states of this latter multiplet is strongly
  influenced by the SU(6) breaking terms mentioned above.  Our results
  for those multiplets are in good agreement with those previously
  obtained in Ref.~\cite{Roca:2005nm}, which among others include the
  prediction of the existence of a second $K_1(1270)$
  resonance~\cite{Geng:2006yb}. 

  On the other hand the simultaneous consideration of $PV$ and $VV$ channels
  make the present approach different from that followed in
  Ref.~\cite{Roca:2005nm}, and has allowed us to dynamically generate also the
  $h_1 (1595)$ resonance.  The interference $PV \to VV$ amplitudes turn out to
  play a crucial role in producing this state, and that is presumably the
  reason why it is not generated either in the $VV\to VV$ study carried out in
  Ref.~\cite{Geng:2008gx} using the formalism of the hidden gauge interaction
  for vector mesons. Possibly, the situation is similar for the $K_1(1650)$
  state and thus we end up with a clearer SU(6) pattern, which is also
  followed to some extent in nature.

  In this sector, we also predict two exotic states belonging to the
  $10$ and $10^*$ irreps. On the other hand, we have verified that the
  missing $b_1$ pole in the symmetric octet of the SU(6) {\bf 189}
  would appear if the SU(6) symmetry were restored.

  To finish the discussion of this sector, we would like to point out that
  below 2 GeV there is only one firmly established axial vector resonance 
  that does not fit in the symmetry pattern sketched in Table~\ref{tab:j012}:
  It is the $f_1(1420)$, and similarly to the previous discussion for the
  $f_0(1500)$ resonance, this might hint at the possible existence of gluon
  components in its wave-function. Indeed arguments favoring the $f_1(1420)$
  being a hybrid $q\bar qg$ meson have been put forward by Ishida and
  collaborators~\cite{Ishida:1989xh}.

\item[(iii)] $J^P = 2^+$: In this sector is where the SU(6) pattern works
  worst. This is because the SU(6) symmetry kinetic term becomes less dominant
  when compared to $VV$ interaction contact term generated as a result of
  giving mass to the vector mesons. Moreover, we must stress here, once more,
  the little control that we have over this term. Yet, the interaction in the
  SU(6) {\bf 189} irrep associated to ${\mathcal D}_{\rm kin}$ is relatively
  weak. Thus all results displayed in Table~\ref{tab:j012} for this sector
  must be understood by actively considering the interplay between ${\mathcal
    D}_{\rm kin}$ and ${\mathcal D}_m$. The first remark is that if the
  contact term is switched off, the pole associated to the $f_2(1270)$ moves
  up in mass by more than 200 MeV and the $a_2(1320)$ resonance
  disappears. Actually, in each $YI$ subsector, ${\mathcal D}_m$ has two large
  and negative (attractive) eigenvalues, which correspond to a full nonet
  (singlet plus octet). The $a_2(1320)$ would be part of this nonet, and it
  might well be that the actual $K^*_2(1430)$ could be also a member of it. In
  that scenario, the pole at $(1708, -156)$ obtained here, and that we cannot
  move down closer to the mass of the $K^*_2(1430)$ resonance, might
  correspond to a further state, for which we do not find an easy
  correspondence with any of those reported in the PDG. On the other hand, by
  changing the subtraction constants it is possible to generate some more
  $0^+(2^{++})$ poles within our scheme, which might account for those states
  needed to fill in completely the nonet mentioned above.

  In this sector, and in contrast to the $0^+$ and $1^+$ cases, there appear
  in the PDG several even parity resonances that cannot be accommodated within
  our scheme. Some of them, might be glueballs, but we cannot be here as
  precise as we were in the previous sectors.  The hidden gauge formalism for
  vector mesons used in Ref.~\cite{Geng:2008gx} does not improve on that,
  though its choice for the contact $VV$ term might provide a more robust
  description of the $f_2(1270)$ and $K^*_2(1430)$ resonances than that
  obtained here.

\end{itemize}

 In summary, it has been shown that most of the low-lying even parity
 meson resonances, specially in the $J^P=0^+$ and $1^+$ sectors, can
 be classified according to multiplets of the spin-flavor symmetry
 group SU(6). The $f_0(1500)$, $f_1(1420)$ and some $0^+(2^{++})$
 resonances cannot be accommodated within SU(6) multiplets and thus
 they are clear candidates to be glueballs or hybrids. On the other
 hand, we predict the existence of five exotic resonances ($I \ge 3/2$
 and/or $|Y|=2$) with masses in region 1.4--1.6 GeV, which would
 complete the $27_1$ and $10_3$ and $10_3^*$ spin-flavor multiplets.

\begin{acknowledgments}
  We warmly thank E. Oset and E. Ruiz-Arriola for useful discussions.  This
  research was supported by DGI and FEDER funds, under contracts
  FIS2008-01143, FIS2006-03438, by the EU contract FLAVIAnet
  MRTN-CT-2006-035482, the Spanish Consolider-Ingenio 2010 Programme CPAN
  (CSD2007-00042), and the Junta de Andaluc{\'\i}a grants FQM225 and it is
  part of the European Community-Research Infrastructure Integrating Activity
  ``Study of Strongly Interacting Matter'' (acronym HadronPhysics2, Grant
  Agreement n. 227431) under the Seventh Framework Programme of EU.  LSG
  acknowledges support from the MICINN in the Program ``Juan de la Cierva" and
  from the Alexander von Humboldt foundation through a research fellowship.
\end{acknowledgments}

\newpage
\clearpage

\appendix
\section{Coefficients of the $s$-wave tree level amplitudes}
\label{app:tables}

This Appendix gives the  ${\mathcal D}_{\rm kin}$, ${\mathcal D}_{m}$ and ${\mathcal D}_a$
matrices of the $s$-wave
tree level meson-meson amplitudes in Eq.~(\ref{eq:vsu6}), for the
various $YIJ$ sectors (Tables~\ref{tab:initial}-\ref{tab:final}).

\subsection{Kinetic term: ${\mathcal D}_{\rm kin}$}
\begin{table}[htpb]
      \renewcommand{\arraystretch}{2}
     \setlength{\tabcolsep}{0.4cm}
     \caption{$(Y,I,J)=(0,0,0)$.}
\begin{tabular}{c|cccccccc}
\hline\hline
 & $\pi\pi$ &  $\bar{K}K$ & $\eta\eta$ & $\rho\rho$ & $\omega\omega$ & $\omega\phi$  & $\bar{K}^* K^*$ &  $\phi\phi$ \\\hline 
 & $-2$ & $\frac{\sqrt{3}}{2}$ & $0$ & $2 \sqrt{3}$ & $0$ & $0$ & $-\frac{3}{2}$ & $0$ \\
 & $\frac{\sqrt{3}}{2}$& $-\frac{3}{2}$ &$-\frac{3}{2}$& $-\frac{3}{2}$ & $\frac{\sqrt{3}}{2}$ & $\sqrt{3}$ & $\frac{3 \sqrt{3}}{2}$ & $\sqrt{3}$ \\
 & $0$ & $-\frac{3}{2}$ & $0$ & $0$ & $0$ & $0$ & $\frac{3 \sqrt{3}}{2}$ & $0$ \\
 & $2 \sqrt{3}$ & $-\frac{3}{2}$ & $0$& $-4$ & $2 \sqrt{3}$ & $0$ & $\frac{3 \sqrt{3}}{2}$ & $0$ \\
 & $0$ & $\frac{\sqrt{3}}{2}$ & $0$ & $2 \sqrt{3}$ & $-2$ & $0$ & $-\frac{3}{2}$ & $0$ \\
 & $0$ & $\sqrt{3}$ & $0$ & $0$ & $0$ & $0$ & $1$ & $0$ \\
 & $-\frac{3}{2}$ & $\frac{3 \sqrt{3}}{2}$ & $\frac{3 \sqrt{3}}{2}$ & $\frac{3 \sqrt{3}}{2}$ & $-\frac{3}{2}$ & $1$ & $-\frac{9}{2}$ & $-3$ \\
 & $0$ & $\sqrt{3}$ & $0$ & $0$ & $0$ & $0$ & $-3$ & $-4$\\
\hline\hline 
\end{tabular}\label{tab:initial}
\end{table}

\begin{table}[htpb]
      \renewcommand{\arraystretch}{2}
     \setlength{\tabcolsep}{0.4cm}
     \caption{$(Y,I,J)=(0,0,1)$.}

\begin{tabular}{c|cccccccc}
\hline\hline$G$ & $\eta\phi$ &  $\eta\omega$ & $\pi\rho$ & $(\bar{K}K^*)_A$ & $\bar{K}^* K^*$ & $\omega\phi$ &    $(\bar{K}K^*)_S$   \\\hline 
$-$&  0 & 0 & 0    & $-\sqrt{6}$ & $-\sqrt{6}$ &  &  \\
$-$&  0 & 0 & 0    & $-\sqrt{3}$ & $ \sqrt{3}$ &  &  \\
$-$&  0 & 0 & $-4$ & $\sqrt{3}$  &  $-\sqrt{3}$ &  &  \\
$-$& $-\sqrt{6}$ & $-\sqrt{3}$ & $\sqrt{3}$ & $-3$ & $-1$ &  &  \\
$-$& $-\sqrt{6}$ & $\sqrt{3}$ & $-\sqrt{3}$ & $-1$ & $-3$ &  &  \\
$+$&   &   &   &   &   & 0 & $-2$ \\
$+$&   &   &   &   &   & $-2$ & 0\\
\hline\hline
\end{tabular}
\end{table}

\begin{table}[htpb]
      \renewcommand{\arraystretch}{2}
     \setlength{\tabcolsep}{0.4cm}
     \caption{$(Y,I,J)=(0,0,2)$.}
\begin{tabular}{c|ccccc}
\hline\hline
 & $\rho\rho$ & $\omega\omega$& $\omega\phi$ &  $\bar{K}^* K^*$ &  $\phi\phi$ \\\hline 

 & $-1$ & $-\sqrt{3}$ & 0 & 0 &0 \\
 & $-\sqrt{3}$ & 1 & 0 & 0 &0 \\
 & 0 & 0 & 0 & $-2$ &0\\
 & 0 & 0 &$-2$ &0 &0\\
 & 0 & 0 & 0 & 0 &2\\
\hline\hline
\end{tabular}
\end{table}

\begin{table}[htpb]
      \renewcommand{\arraystretch}{2}
     \setlength{\tabcolsep}{0.4cm}
     \caption{$(Y,I,J)=(0,1,0)$.}
\begin{tabular}{c|ccccc}
\hline\hline
 & $\eta\pi$ & $\bar{K}K$ & $\omega\rho$ & $\phi\rho$& $\bar{K}^*K^*$  \\\hline 
 & 0 & $\sqrt{\frac{3}{2}}$ & 0 &0 & $-\frac{3}{\sqrt{2}}$ \\
 & $\sqrt{\frac{3}{2}}$ & $-\frac{1}{2}$ & $-\sqrt{\frac{3}{2}}$ & $-\sqrt{3}$ & $\frac{\sqrt{3}}{2}$ \\
 & 0 & $-\sqrt{\frac{3}{2}}$ & $-4$ & 0 & $\frac{3}{\sqrt{2}}$ \\
 & 0 & $-\sqrt{3}$ & 0 & 0 & $-1$ \\
 & $-\frac{3}{\sqrt{2}} $ & $\frac{\sqrt{3}}{2}$ & $\frac{3}{\sqrt{2}}$ & $-1$ & $-\frac{3}{2}$\\
\hline\hline
\end{tabular}
\end{table}

\begin{table}[htpb]
      \renewcommand{\arraystretch}{2}
     \setlength{\tabcolsep}{0.2cm}
     \caption{$(Y,I,J)=(0,1,1)$.}
\begin{tabular}{c|cccccccccc}
\hline\hline
 $G$ & $\pi \phi$ & $\pi \omega$ & $\eta\rho$ 
& $(\bar{K}K^*)_S$ & $\rho\rho$ &  $\bar{K}^* K^*$ 
& $\pi\rho$ & $(\bar{K}K^*)_A$ & $\omega\rho$ & $\phi\rho$ \\
\hline 
$+$ & 0 & 0 & 0 & $\sqrt{2}$ & 0 & $\sqrt{2}$ &  &  &  &  \\
$+$ & 0 & 0 & 0 & 1 & $2 \sqrt{2}$ & $-1$ &  &  &  &  \\
$+$ & 0 & 0 & 0 & $\sqrt{3}$ & 0 & $-\sqrt{3}$ &  &  &  &  \\
$+$ & $\sqrt{2}$ & 1 & $\sqrt{3}$ & $-1$ & $-\sqrt{2}$ & 1 &  &  &  &  \\
$+$ & 0 & $2 \sqrt{2}$ & 0 & $-\sqrt{2}$ & $-2$ & $\sqrt{2}$ &  &  &  &  \\
$+$ & $\sqrt{2}$ & $-1$ & $-\sqrt{3}$ & 1 & $\sqrt{2}$ & $-1$ &  &  &  &  \\
$-$ &  &  &  &  &  &  & 0 & 0 & $-2$ & 0 \\
$-$ &  &  &  &  &  &  & 0 & 0 & 0 & $-2$ \\
$-$ &  &  &  &  &  &  & $-2$ & 0 & 0 & 0 \\
$-$ &  &  &  &  &  &  & 0 & $-2$ & 0 & 0\\
\hline\hline
\end{tabular}
\end{table}

\begin{table}[htpb]
      \renewcommand{\arraystretch}{2}
     \setlength{\tabcolsep}{0.2cm}
     \caption{$(Y,I,J)=(0,1,2)$.}
\begin{tabular}{c|ccc}
\hline\hline
 & $\omega\rho$ & $\phi\rho$  & $\bar{K}^* K^*$\\\hline 
 & 2 & 0 & 0 \\
 & 0 & 0 & 2 \\
 & 0 & 2 & 0\\
\hline\hline
\end{tabular}
\end{table}

\begin{table}[htpb]
      \renewcommand{\arraystretch}{2}
     \setlength{\tabcolsep}{0.2cm}
     \caption{$(Y,I,J)=(0,2,0)$.}\label{tab:020}
\begin{tabular}{c|ccc}
\hline\hline
 & $\pi\pi$ & $\rho\rho$ \\\hline 
 & 1 & $-\sqrt{3}$ \\
 & $-\sqrt{3}$ & $-1$\\
\hline\hline
\end{tabular}
\end{table}

\begin{table}[htpb]
      \renewcommand{\arraystretch}{2}
     \setlength{\tabcolsep}{0.2cm}
     \caption{$(Y,I,J)=(0,2,1)$.}
\begin{tabular}{c|c}
\hline\hline
 & $\pi\rho$  \\\hline 
 & 2\\
\hline\hline
\end{tabular}
\end{table}

\begin{table}[htpb]
      \renewcommand{\arraystretch}{2}
     \setlength{\tabcolsep}{0.2cm}
     \caption{$(Y,I,J)=(0,2,2)$.}
\begin{tabular}{c|c}
\hline\hline
 & $\rho\rho$  \\\hline 
 & 2\\
\hline\hline
\end{tabular}
\end{table}

\begin{table}[htpb]
      \renewcommand{\arraystretch}{2}
     \setlength{\tabcolsep}{0.2cm}
     \caption{$(Y,I,J)=(1,1/2,0)$.}
\begin{tabular}{c|ccccc}
\hline\hline
& $K\pi$ & $\eta K$ & $K^*\rho$ & $K^*\omega$ & $K^*\phi$\\\hline
 & $-\frac{5}{4}$ & $\frac{3}{4}$ & $\frac{5 \sqrt{3}}{4}$ & $-\frac{3}{4}$ & $-\frac{3}{2 \sqrt{2}}$ \\
 & $\frac{3}{4}$ & $\frac{3}{4}$ & $-\frac{3 \sqrt{3}}{4}$ & $-\frac{3}{4}$ & $-\frac{3}{2 \sqrt{2}}$ \\
 & $\frac{5 \sqrt{3}}{4}$ & $-\frac{3 \sqrt{3}}{4}$ & $-\frac{7}{4}$ & $-\frac{5 \sqrt{3}}{4}$ & $\frac{3 \sqrt{\frac{3}{2}}}{2}$ \\
 & $-\frac{3}{4}$ & $-\frac{3}{4}$ & $-\frac{5 \sqrt{3}}{4}$ & $-\frac{5}{4}$ & $\frac{3}{2 \sqrt{2}}$ \\
 & $-\frac{3}{2 \sqrt{2}}$ & $-\frac{3}{2 \sqrt{2}}$ & $\frac{3 \sqrt{\frac{3}{2}}}{2}$ & $\frac{3}{2 \sqrt{2}}$ & $-\frac{5}{2}$\\
\hline\hline
\end{tabular}
\end{table}

\begin{table}[htpb]
      \renewcommand{\arraystretch}{2}
     \setlength{\tabcolsep}{0.2cm}
     \caption{$(Y,I,J)=(1,1/2,1)$.}
\begin{tabular}{c|cccccccc}
\hline\hline
 & $\pi K^*$ &  $ K \rho$ & $ K \omega$ & $\eta K^*$ & $ K\phi$  & $ K^*\rho$ & $ K^*\omega$ & $ K^*\phi$  \\\hline 
 & $-\frac{5}{4}$ & $\frac{5}{4}$ & $-\frac{\sqrt{3}}{4}$ & $-\frac{3}{4}$ & $-\frac{\sqrt{\frac{3}{2}}}{2}$ & $-\frac{5}{2 \sqrt{2}}$ & $\frac{\sqrt{\frac{3}{2}}}{2}$ & $-\frac{\sqrt{3}}{2}$ \\
 & $\frac{5}{4}$ & $-\frac{5}{4}$ & $\frac{\sqrt{3}}{4}$ & $\frac{3}{4}$ & $\frac{\sqrt{\frac{3}{2}}}{2}$ & $\frac{1}{2 \sqrt{2}}$ & $\frac{3 \sqrt{\frac{3}{2}}}{2}$ & $\frac{\sqrt{3}}{2}$ \\
 &$-\frac{\sqrt{3}}{4}$ & $\frac{\sqrt{3}}{4}$ & $\frac{1}{4}$ & $\frac{\sqrt{3}}{4}$ & $\frac{1}{2 \sqrt{2}}$ & $\frac{3 \sqrt{\frac{3}{2}}}{2}$ & $\frac{3}{2 \sqrt{2}}$ & $\frac{1}{2}$ \\
 & $-\frac{3}{4}$ & $\frac{3}{4}$ & $\frac{\sqrt{3}}{4}$ & $\frac{3}{4}$ & $\frac{\sqrt{\frac{3}{2}}}{2}$ & $-\frac{3}{2 \sqrt{2}}$ & $-\frac{\sqrt{\frac{3}{2}}}{2}$ & $\frac{\sqrt{3}}{2}$ \\
 & $-\frac{\sqrt{\frac{3}{2}}}{2}$ & $\frac{\sqrt{\frac{3}{2}}}{2}$ & $\frac{1}{2 \sqrt{2}}$ & $\frac{\sqrt{\frac{3}{2}}}{2}$ & $\frac{1}{2}$ & $-\frac{\sqrt{3}}{2}$ & $-\frac{1}{2}$ & $-\frac{3}{\sqrt{2}}$ \\
 & $-\frac{5}{2 \sqrt{2}}$ & $\frac{1}{2 \sqrt{2}}$ & $\frac{3 \sqrt{\frac{3}{2}}}{2}$ & $-\frac{3}{2 \sqrt{2}}$ & $-\frac{\sqrt{3}}{2}$ & $-\frac{3}{2}$ & $-\frac{\sqrt{3}}{2}$ & $-\sqrt{\frac{3}{2}}$ \\
 & $\frac{\sqrt{\frac{3}{2}}}{2}$ & $\frac{3 \sqrt{\frac{3}{2}}}{2}$ & $\frac{3}{2 \sqrt{2}}$& $-\frac{\sqrt{\frac{3}{2}}}{2}$ & $-\frac{1}{2}$ & $-\frac{\sqrt{3}}{2}$ & $-\frac{1}{2}$ & $-\frac{1}{\sqrt{2}}$ \\
 & $-\frac{\sqrt{3}}{2}$ & $\frac{\sqrt{3}}{2}$ &$\frac{1}{2}$ & $\frac{\sqrt{3}}{2}$& $-\frac{3}{\sqrt{2}}$ &$-\sqrt{\frac{3}{2}}$ &$-\frac{1}{\sqrt{2}}$ &$-1$\\
\hline\hline
\end{tabular}
\end{table}

\begin{table}[htpb]
      \renewcommand{\arraystretch}{2}
     \setlength{\tabcolsep}{0.2cm}
     \caption{$(Y,I,J)=(1,1/2,2)$.}
\begin{tabular}{c|ccc}
\hline\hline
 & $ K^*\rho$ & $ K^*\omega$ & $ K^*\phi$  \\\hline 
 & $-1$ & $\sqrt{3}$ &0 \\
 & $\sqrt{3}$ &1 &0 \\
 & 0 & 0& 2\\
\hline\hline
\end{tabular}
\end{table}

\begin{table}[htpb]
      \renewcommand{\arraystretch}{2}
     \setlength{\tabcolsep}{0.2cm}
     \caption{$(Y,I,J)=(1,3/2,0)$.}
\begin{tabular}{c|ccc}
\hline\hline
 & $ K\pi$ & $ K^*\rho$  \\\hline 
 & 1 & $-\sqrt{3}$ \\
 & $-\sqrt{3}$ & $-1$\\
\hline\hline
\end{tabular}
\end{table}

\begin{table}[htpb]
      \renewcommand{\arraystretch}{2}
     \setlength{\tabcolsep}{0.2cm}
     \caption{$(Y,I,J)=(1,3/2,1)$.}
\begin{tabular}{c|ccc}
\hline\hline
& $ \pi K^*$ & $ K\rho$ & $ K^*\rho$  \\\hline 
 & 1 & 1 & $-\sqrt{2}$ \\
 & 1 & 1 & $\sqrt{2}$ \\
 & $-\sqrt{2}$ & $\sqrt{2}$ & 0\\
\hline\hline
\end{tabular}
\end{table}

\begin{table}[htpb]
      \renewcommand{\arraystretch}{2}
     \setlength{\tabcolsep}{0.2cm}
     \caption{$(Y,I,J)=(1,3/2,2)$.}
\begin{tabular}{c|c}
\hline\hline
 & $K^*\rho $  \\\hline
 & 2\\ 
\hline\hline
\end{tabular}
\end{table}

\clearpage

\begin{table}[htpb]
      \renewcommand{\arraystretch}{2}
     \setlength{\tabcolsep}{0.2cm}
     \caption{$(Y,I,J)=(2,0,1)$.}
\begin{tabular}{c|cc}
\hline\hline
 & $ K K^*$ & $K^* K^*$  \\\hline
 & 0 & $-2$ \\
 & $-2$ & 0\\
\hline\hline
\end{tabular}
\end{table}

\begin{table}[htpb]
      \renewcommand{\arraystretch}{2}
     \setlength{\tabcolsep}{0.2cm}
     \caption{$(Y,I,J)=(2,1,0)$.}
\begin{tabular}{c|cc}
\hline\hline
 & $K K$ & $K^* K^*$  \\\hline
 & 1 & $-\sqrt{3}$ \\
 & $-\sqrt{3}$ & $-1$\\
\hline\hline
\end{tabular}
\end{table}

\begin{table}[htpb]
      \renewcommand{\arraystretch}{2}
     \setlength{\tabcolsep}{0.2cm}
     \caption{$(Y,I,J)=(2,1,1)$.}
\begin{tabular}{c|cc}
\hline\hline
 & $ K K^*$ \\\hline
 & 2  \\
\hline\hline
\end{tabular}
\end{table}

\begin{table}[htpb]
      \renewcommand{\arraystretch}{2}
     \setlength{\tabcolsep}{0.2cm}
     \caption{$(Y,I,J)=(2,1,2)$.}
\begin{tabular}{c|c}
\hline\hline
 &  $K^* K^*$  \\\hline
 & 2  \\
\hline\hline
\end{tabular}
\end{table}

\newpage

\subsection{Contact term: ${\mathcal D}_{m}$}
%(0,0,0)
\begin{table}[htpb]
      \renewcommand{\arraystretch}{2}
     \setlength{\tabcolsep}{0.4cm}
     \caption{$(Y,I,J)=(0,0,0)$.}
\begin{tabular}{c|cccccccc}
\hline\hline
 & $\pi\pi$ &  $\bar{K}K$ & $\eta\eta$ & $\rho\rho$ & $\omega\omega$ & $\omega\phi$  & $\bar{K}^* K^*$ &  $\phi\phi$ \\
\hline 
 & 0 & 0 & 0 & $\frac{16}{\sqrt{3}}$ & 0 & 0 & $-4$ & 0 \\
 & 0 & 0 & 0 & $-4$ & $\frac{4}{\sqrt{3}}$ & $\frac{8}{\sqrt{3}}$ & $4 \sqrt{3}$ & $\frac{8}{\sqrt{3}}$ \\
 & 0 & 0 & 0 & 0 & 0 & 0 & $4 \sqrt{3}$ & 0 \\
 & $\frac{16}{\sqrt{3}}$ & $-4$ & 0 & $-\frac{208}{3}$ & $\frac{80}{\sqrt{3}}$ & 0 & $24 \sqrt{3}$ & 0 \\
 & 0 & $\frac{4}{\sqrt{3}}$ & 0 & $\frac{80}{\sqrt{3}}$ & $-\frac{80}{3}$ & 0 & $-24$ & 0 \\
 & 0 & $\frac{8}{\sqrt{3}}$ & 0 & 0 & 0 & 0 & $\frac{16}{3}$ & 0 \\
 & $-4$ & $4 \sqrt{3}$ & $4 \sqrt{3}$ & $24 \sqrt{3}$ & $-24$ & $ \frac{16}{3}$ & $-72$ & $-48$ \\
 & 0 & $\frac{8}{\sqrt{3}}$ & 0 & 0 & 0 & 0 & $-48$ & $-\frac{160}{3}$\\
\hline\hline
\end{tabular}
\end{table}

%(0,0,1)
\begin{table}[htpb]
      \renewcommand{\arraystretch}{2}
     \setlength{\tabcolsep}{0.4cm}
     \caption{$(Y,I,J)=(0,0,1)$.}
\begin{tabular}{c|cccccccc}
\hline\hline$G$ & $\eta\phi$ &  $\eta\omega$ & $\pi\rho$ & $(\bar{K}K^*)_A$ & $\bar{K}^* K^*$ & $\omega\phi$ &    $(\bar{K}K^*)_S$   \\\hline 
$-$ & 0 & 0 & 0 & $4 \sqrt{\frac{2}{3}}$ & 0 &  &  \\
$-$ & 0 & 0 & 0 & $\frac{4}{\sqrt{3}}$ & 0 &  &  \\
$-$ & 0 & 0 & $\frac{16}{3}$ & $-\frac{4}{\sqrt{3}}$ & 0 &  &  \\
$-$ & $4 \sqrt{\frac{2}{3}}$ & $\frac{4}{\sqrt{3}}$ & $-\frac{4}{\sqrt{3}}$ & $4$ & 0 &  &  \\
$-$ & 0 & 0 & 0 & 0 & $-28$ &  &  \\
$+$ &  &  &  &  &  & 0 & 0 \\
$+$ &  &  &  &  &  & 0 & $-12$ \\
\hline\hline
\end{tabular}
\end{table}

%(0,0,2)
\begin{table}[htpb]
      \renewcommand{\arraystretch}{2}
     \setlength{\tabcolsep}{0.4cm}
     \caption{$(Y,I,J)=(0,0,2)$.}
\begin{tabular}{c|ccccc}
\hline\hline
 & $\rho\rho$ & $\omega\omega$& $\omega\phi$ &  $\bar{K}^* K^*$ &  $\phi\phi$ \\\hline 
 & $-\frac{16}{3}$ & $\frac{32}{\sqrt{3}}$ & 0 & $4 \sqrt{3}$ & 0 \\
 & $\frac{32}{\sqrt{3}}$ & $-\frac{32}{3}$ & 0 & $-4$ & 0 \\
 & 0 & 0 & 0 & $\frac{40}{3}$ & 0 \\
 & $4 \sqrt{3}$ & $-4$ & $\frac{40}{3}$ & $-12$ & $-8$ \\
 & 0 & 0 & 0 & $-8$ & $-\frac{64}{3}$ \\
\hline\hline
\end{tabular}
\end{table}

%(0,1,0)
\begin{table}[htpb]
      \renewcommand{\arraystretch}{2}
     \setlength{\tabcolsep}{0.4cm}
     \caption{$(Y,I,J)=(0,1,0)$.}
\begin{tabular}{c|ccccc}
\hline\hline
 & $\eta\pi$ & $\bar{K}K$ & $\omega\rho$ & $\phi\rho$& $\bar{K}^*K^*$  \\\hline 
 & 0 & 0 & 0 & 0 & $-4 \sqrt{2}$ \\
 & 0 & 0 & $-4 \sqrt{\frac{2}{3}}$ & $-\frac{8}{\sqrt{3}}$ & $\frac{4}{\sqrt{3}}$ \\
 & 0 & $-4 \sqrt{\frac{2}{3}}$ & $-\frac{160}{3}$ & 0 & $24 \sqrt{2}$ \\
 & 0 & $-\frac{8}{\sqrt{3}}$ & 0 & 0 & $-\frac{16}{3}$ \\
 & $-4 \sqrt{2}$ & $\frac{4}{\sqrt{3}}$ & $24 \sqrt{2}$ & $-\frac{16}{3}$ & $-24$ \\
\hline\hline
\end{tabular}
\end{table}

%(0,1,1)
\begin{table}[htpb]
      \renewcommand{\arraystretch}{2}
     \setlength{\tabcolsep}{0.2cm}
     \caption{$(Y,I,J)=(0,1,1)$.}
\begin{tabular}{c|cccccccccc}
\hline\hline
 $G$ & $\pi \phi$ & $\pi \omega$ & $\eta\rho$ 
& $(\bar{K}K^*)_S$ & $\rho\rho$  &  $\bar{K}^* K^*$ 
& $\pi\rho$ & $(\bar{K}K^*)_A$ & $\omega\rho$ & $\phi\rho$ \\
\hline 
 $+$& 0 & 0 & 0 & $-\frac{4 \sqrt{2}}{3}$ & 0 & 0 &  &  &  &  \\
 $+$& 0 & 0 & 0 & $-\frac{4}{3}$ & 0 & 0 &  &  &  &  \\
 $+$& 0 & 0 & 0 & $-\frac{4}{\sqrt{3}}$ & 0 & 0 &  &  &  &  \\
 $+$& $-\frac{4 \sqrt{2}}{3}$ & $-\frac{4}{3}$ & $-\frac{4}{\sqrt{3}}$ & $\frac{4}{3}$ & 0 & 0 &  &  &  &  \\
 $+$& 0 & 0 & 0 & 0 & $-\frac{56}{3}$ & $\frac{28 \sqrt{2}}{3}$ &  &  &  &  \\
 $+$& 0 & 0 & 0 & 0 & $\frac{28 \sqrt{2}}{3}$ & $-\frac{28}{3}$ &  &  &  &  \\
 $-$&  &  &  &  &  &  & $-8$ & $4 \sqrt{2}$ & 0 & 0 \\
 $-$&  &  &  &  &  &  & $4 \sqrt{2}$ & $-4$ & 0 & 0 \\
 $-$&  &  &  &  &  &  & 0 & 0 & 0 & 0 \\
 $-$&  &  &  &  &  &  & 0 & 0 & 0 & 0 \\
\hline\hline
\end{tabular}
\end{table}

%(0,1,2)
\begin{table}[htpb]
      \renewcommand{\arraystretch}{2}
     \setlength{\tabcolsep}{0.2cm}
     \caption{$(Y,I,J)=(0,1,2)$.}
\begin{tabular}{c|ccc}
\hline\hline
 & $\omega\rho$ & $\phi\rho$  & $\bar{K}^* K^*$\\\hline 
 & $-\frac{64}{3}$  & 0 & $4 \sqrt{2}$ \\
 & 0 & 0 & $-\frac{40}{3}$ \\
 & $4 \sqrt{2}$ & $-\frac{40}{3}$ & $-4$ \\
\hline\hline
\end{tabular}
\end{table}

%(0,2,0)
\begin{table}[htpb]
      \renewcommand{\arraystretch}{2}
     \setlength{\tabcolsep}{0.2cm}
     \caption{$(Y,I,J)=(0,2,0)$.} \label{tab:020m}
\begin{tabular}{c|cc}
\hline\hline
 & $\pi\pi$ & $\rho\rho$ \\\hline 
 & 0 & $-\frac{8}{\sqrt{3}}$ \\
 & $-\frac{8}{\sqrt{3}}$ & $-\frac{16}{3}$\\
\hline\hline
\end{tabular}
\end{table}

%(0,2,1)
\begin{table}[htpb]
      \renewcommand{\arraystretch}{2}
     \setlength{\tabcolsep}{0.2cm}
     \caption{$(Y,I,J)=(0,2,1)$.}
\begin{tabular}{c|c}
\hline\hline
 & $\pi\rho$  \\\hline 
 & $-\frac{8}{3}$\\
\hline\hline
\end{tabular}
\end{table}

%(0,2,2)
\begin{table}[htpb]
      \renewcommand{\arraystretch}{2}
     \setlength{\tabcolsep}{0.2cm}
     \caption{$(Y,I,J)=(0,2,2)$.}
\begin{tabular}{c|c}
\hline\hline
 & $\rho\rho$  \\\hline 
 & $-\frac{40}{3}$\\
\hline\hline
\end{tabular}
\end{table}

%(0,1/2,0)
\begin{table}[htpb]
      \renewcommand{\arraystretch}{2}
     \setlength{\tabcolsep}{0.2cm}
     \caption{$(Y,I,J)=(1,1/2,0)$.}
\begin{tabular}{c|ccccc}
\hline\hline
& $K\pi$ & $\eta K$ & $K^*\rho$ & $K^*\omega$ & $K^*\phi$\\\hline
& 0 & 0 & $\frac{10}{\sqrt{3}}$ & $-2$ & $-2 \sqrt{2}$ \\
& 0 & 0 & $-2 \sqrt{3}$ & $-2$ & $-2 \sqrt{2}$ \\
& $\frac{10}{\sqrt{3}}$ & $-2 \sqrt{3}$ & $-\frac{100}{3}$ & $-\frac{44}{\sqrt{3}}$ & $12 \sqrt{6}$ \\
& $-2$ & $-2$ & $-\frac{44}{\sqrt{3}}$ & $-\frac{44}{3}$ & $12 \sqrt{2}$ \\
& $-2 \sqrt{2}$ & $-2 \sqrt{2}$ & $12 \sqrt{6}$ & $12 \sqrt{2}$ & $-\frac{88}{3}$\\
\hline\hline
\end{tabular}
\end{table}

%(1,1/2,1)
\begin{table}[htpb]
      \renewcommand{\arraystretch}{2}
     \setlength{\tabcolsep}{0.2cm}
     \caption{$(Y,I,J)=(1,1/2,1)$.}
\begin{tabular}{c|cccccccc}
\hline\hline
 & $\pi K^*$ &  $ K \rho$ & $ K \omega$ & $\eta K^*$ & $ K\phi$  & $ K^*\rho$ & $ K^*\omega$ & $ K^*\phi$  \\\hline 
 & $-\frac{4}{3}$ & $-\frac{14}{3}$ & $-\frac{2}{\sqrt{3}}$ & $4$ & $-2 \sqrt{\frac{2}{3}}$ & 0 & 0 & 0 \\
 & $-\frac{14}{3}$ & $-\frac{4}{3}$ & $-\frac{4}{\sqrt{3}}$ & $2$ & $-4 \sqrt{\frac{2}{3}}$ & 0 & 0 & 0 \\
 & $-\frac{2}{\sqrt{3}}$ & $-\frac{4}{\sqrt{3}}$ & $-\frac{4}{3}$ & $\frac{2}{\sqrt{3}}$ & $-\frac{4 \sqrt{2}}{3}$ & 0 & 0 & 0 \\
 & $4$ & $2$ & $\frac{2}{\sqrt{3}}$ & $-4$ & $2 \sqrt{\frac{2}{3}}$ & 0 & 0 & 0 \\
 & $-2 \sqrt{\frac{2}{3}}$ & $-4 \sqrt{\frac{2}{3}}$ & $-\frac{4 \sqrt{2}}{3}$ & $2 \sqrt{\frac{2}{3}}$ & $-\frac{8}{3}$ & 0 & 0 & 0 \\
 & 0 & 0 & 0 & 0 & 0 & $-14$ & $-\frac{14}{\sqrt{3}}$ & $-14 \sqrt{\frac{2}{3}}$ \\
 & 0 & 0 & 0 & 0 & 0 & $-\frac{14}{\sqrt{3}}$ & $-\frac{14}{3}$ & $-\frac{14 \sqrt{2}}{3}$ \\
 & 0 & 0 & 0 & 0 & 0 & $-14 \sqrt{\frac{2}{3}}$ & $-\frac{14 \sqrt{2}}{3}$ & $-\frac{28}{3}$ \\
\hline\hline
\end{tabular}
\end{table}

%(1,1/2,2)
\begin{table}[htpb]
      \renewcommand{\arraystretch}{2}
     \setlength{\tabcolsep}{0.2cm}
     \caption{$(Y,I,J)=(1,1/2,2)$.}
\begin{tabular}{c|ccc}
\hline\hline
 & $ K^*\rho$ & $ K^*\omega$ & $ K^*\phi$  \\\hline 
 & $\frac{2}{3}$ & $-\frac{26}{\sqrt{3}}$ & $2 \sqrt{6}$ \\
 & $-\frac{26}{\sqrt{3}}$ & $-\frac{26}{3}$ & $2 \sqrt{2}$ \\
 & $2 \sqrt{6}$ & $2 \sqrt{2}$ & $-\frac{52}{3}$\\
\hline\hline
\end{tabular}
\end{table}

%(1,3/2,0)
\begin{table}[htpb]
      \renewcommand{\arraystretch}{2}
     \setlength{\tabcolsep}{0.2cm}
     \caption{$(Y,I,J)=(1,3/2,0)$.}
\begin{tabular}{c|ccc}
\hline\hline
 & $ K\pi$ & $ K^*\rho$  \\\hline 
 & $0$ & $-\frac{8}{\sqrt{3}}$ \\
 & $-\frac{8}{\sqrt{3}}$ & $-\frac{16}{3}$\\
\hline\hline
\end{tabular}
\end{table}

%(1,3/2,1)
\begin{table}[htpb]
      \renewcommand{\arraystretch}{2}
     \setlength{\tabcolsep}{0.2cm}
     \caption{$(Y,I,J)=(1,3/2,1)$.}
\begin{tabular}{c|ccc}
\hline\hline
& $\pi K^*$ & $ K\rho$ & $ K^*\rho$  \\\hline 
& $-\frac{4}{3}$ & $-\frac{4}{3}$ & 0 \\
& $-\frac{4}{3}$ & $-\frac{4}{3}$ &0\\
& 0 & 0 & 0 \\
\hline\hline
\end{tabular}
\end{table}

%(1,3/2,2)
\begin{table}[htpb]
      \renewcommand{\arraystretch}{2}
     \setlength{\tabcolsep}{0.2cm}
     \caption{$(Y,I,J)=(1,3/2,2)$.}
\begin{tabular}{c|c}
\hline\hline
 & $K^*\rho $  \\\hline
 & $-\frac{40}{3}$\\ 
\hline\hline
\end{tabular}
\end{table}

\clearpage

%(2,0,1)
\begin{table}[htpb]
      \renewcommand{\arraystretch}{2}
     \setlength{\tabcolsep}{0.2cm}
     \caption{$(Y,I,J)=(2,0,1)$.}
\begin{tabular}{c|cc}
\hline\hline
 & $ K K^*$ & $K^* K^*$  \\\hline
 & 0 & 0 \\
 & 0 & 0\\
\hline\hline
\end{tabular}
\end{table}

%(2,1,0)
\begin{table}[htpb]
      \renewcommand{\arraystretch}{2}
     \setlength{\tabcolsep}{0.2cm}
     \caption{$(Y,I,J)=(2,1,0)$.}
\begin{tabular}{c|cc}
\hline\hline
 & $K K$ & $K^* K^*$  \\\hline
 & $0$ & $-\frac{8}{\sqrt{3}}$ \\
 & $-\frac{8}{\sqrt{3}}$ & $-\frac{16}{3}$\\
\hline\hline
\end{tabular}
\end{table}

%(2,1,1)
\begin{table}[htpb]
      \renewcommand{\arraystretch}{2}
     \setlength{\tabcolsep}{0.2cm}
     \caption{$(Y,I,J)=(2,1,1)$.}
\begin{tabular}{c|cc}
\hline\hline
 & $ K K^*$ \\\hline
 & $-\frac{8}{3}$  \\
\hline\hline
\end{tabular}
\end{table}

%(2,1,2)
\begin{table}[htpb]
      \renewcommand{\arraystretch}{2}
     \setlength{\tabcolsep}{0.2cm}
     \caption{$(Y,I,J)=(2,1,2)$.}
\begin{tabular}{c|c}
\hline\hline
 &  $K^* K^*$  \\\hline
 &  $-\frac{40}{3}$  \\
\hline\hline
\end{tabular}
\end{table}

%-------------------------------------------------

\newpage
\subsection{$(u-t)$ term:  ${\mathcal D}_a$}

The matrix elements corresponding to ${\mathcal D}_a$ are displayed.  The
$(Y,I,J)$ or $(Y,I,J,G)$ sectors with identically zero matrices are omitted.
$PP$ and $VV$ channels (for which the matrix vanishes) are also omitted.

\begin{table}[htpb]
      \renewcommand{\arraystretch}{2}
     \setlength{\tabcolsep}{0.4cm}
     \caption{$(Y,I,J,G)=(0,0,1,+)$.}
\begin{tabular}{c|c}
\hline\hline
 &    $(\bar{K}K^*)_S$   
\\\hline 
 & 3\\
\hline\hline
\end{tabular}
\end{table}

\begin{table}[htpb]
      \renewcommand{\arraystretch}{2}
     \setlength{\tabcolsep}{0.2cm}
     \caption{$(Y,I,J,G)=(0,1,1,-)$.}
\begin{tabular}{c|cc}
\hline\hline
& $\pi\rho$ & $(\bar{K}K^*)_A$  \\
\hline 
& 2  & $-\sqrt{2}$  \\
& $-\sqrt{2}$ & 1 \\
\hline\hline
\end{tabular}
\end{table}

\begin{table}[htpb]
      \renewcommand{\arraystretch}{2}
     \setlength{\tabcolsep}{0.2cm}
     \caption{$(Y,I,J)=(1,1/2,1)$.}
\begin{tabular}{c|ccccc}
\hline\hline
 & $\pi K^*$ &  $ K \rho$ & $ K \omega$ & $\eta K^*$ & $ K\phi$ \\
\hline 
 &  $\frac{3}{4}$ & $\frac{3}{4}$ & $\frac{\sqrt{3}}{4}$ & $-\frac{3}{4}$ & $\frac{\sqrt{\frac{3}{2}}}{2}$ \\
 & $\frac{3}{4}$ & $\frac{3}{4}$ & $\frac{\sqrt{3}}{4}$ & $-\frac{3}{4}$ & $\frac{\sqrt{\frac{3}{2}}}{2}$ \\
 &$\frac{\sqrt{3}}{4}$ & $\frac{\sqrt{3}}{4}$ & $\frac{1}{4}$ & $-\frac{\sqrt{3}}{4}$ & $\frac{1}{2 \sqrt{2}}$ \\
 & $-\frac{3}{4}$ & $-\frac{3}{4}$ & $-\frac{\sqrt{3}}{4}$ & $\frac{3}{4}$ & $-\frac{\sqrt{\frac{3}{2}}}{2}$  \\
 & $\frac{\sqrt{\frac{3}{2}}}{2}$ & $\frac{\sqrt{\frac{3}{2}}}{2}$ & $\frac{1}{2 \sqrt{2}}$ & $-\frac{\sqrt{\frac{3}{2}}}{2}$ & $\frac{1}{2}$ \\ 
\hline\hline 
\end{tabular}\label{tab:final}
\end{table}

\clearpage

\section{Computation of the eigenvalues}
\label{app:eigenvalues}

To get the eigenvalues associated to each of the SU(6) projectors in
Eqs.~(\ref{eq:hsu6}) and (\ref{eq:hsu6s}), we just let act ${\mathcal H}^{\rm
  SU(6)}_{\pm}$ on one of the states of the vector space associated to each
representation. For instance, the vector space associated to the singlet
representation is generated by the state
\begin{equation}
|1\rangle\, = M^{\dagger a}_b M^{\dagger b}_a |0\rangle, \,
\quad
 a,b=1,\cdots 2 N_F
.
\end{equation}
This gives
\begin{eqnarray}
\hat {\mathcal G}_+ |1\rangle\, &=& :{\rm
       Tr}\left((MM^\dagger)^2-M^2M^{\dagger 2} \right): M^{\dagger
       a}_b M^{\dagger b}_a|0\rangle
       \nonumber \\
&=&2 \left( \underwick{21}{M^{\dagger i}_j
<1M^j_k M^{\dagger k}_l <2M^l_i >2M^{\dagger a}_b >1M^{\dagger b}_a} -
\underwick{21}{M^{\dagger i}_j
M^{\dagger j}_k <1M^k_l <2M^l_i >2M^{\dagger a}_b >1M^{\dagger b}_a}
\right) |0\rangle = -4N_F |1\rangle
\end{eqnarray}
and similarly for $\hat {\mathcal G}_{\mathcal M}$. For the remaining symmetric
representations, a convenient choice of states is
\begin{equation}
|2\rangle\, = M^{\dagger 1}_b M^{\dagger b}_2 |0\rangle, \qquad
|3\rangle\, = \left (M^{\dagger 1}_2 M^{\dagger 3}_4- M^{\dagger
3}_2 M^{\dagger 1}_4 \right) |0\rangle,  \qquad 
|4\rangle\, = M^{\dagger 1}_2 M^{\dagger 1}_2 |0\rangle
\end{equation}
for the ${\bf 35}_s$, {\bf 189} and {\bf 405} representations,
respectively. 

The $p-$wave part ${\mathcal G}_d-{\mathcal G}_c$ can be represented as
the matrix element of an operator in a fermion space
\begin{equation}
{\mathcal G}_d-{\mathcal
  G}_c = - \langle 0 |
M^{i^\prime}_{j^\prime}M^{k^\prime}_{l^\prime} {\rm Tr}(M^{\dagger 2}M^2) M^{\dagger i}_{j} M^{\dagger
  k}_{l} | 0 \rangle
\end{equation}
where now the $M^{i}_{\ j}$ operators satisfy an anticommutation relation
\begin{equation}
\{ M^i_j, M^{\dagger k}_l \} = \delta^k_j\delta^i_l-\frac{1}{2N_F}
\delta^i_j\delta^k_l
\end{equation}
and Eq.~(\ref{eq:wick}) still holds. One can follow the same techniques as
those outlined above for the $\hat {\mathcal G}_{+}$ operators, using e.g. the
states
\begin{equation}
|5\rangle\, = M^{\dagger 1}_b M^{\dagger b}_2 |0\rangle, \qquad 
|6\rangle\, = M^{\dagger 1}_3 M^{\dagger 2}_3 |0\rangle, \qquad
|7\rangle\, = M^{\dagger 3}_1 M^{\dagger 3}_2 |0\rangle
\end{equation}
for the ${\bf 35}_a$, ${\bf 280}$ and ${\bf 280}^*$ antisymmetric
representations.


\begin{thebibliography}{99}

%1\cite{Weinberg:1978kz}
\bibitem{Weinberg:1978kz}
  S.~Weinberg,
  %``Phenomenological Lagrangians,''
  Physica A {\bf 96}, 327 (1979).
  %%CITATION = PHYSA,A96,327;%%

%2\cite{Gasser:1983yg}
\bibitem{Gasser:1983yg}
  J.~Gasser and H.~Leutwyler,
  %``Chiral Perturbation Theory To One Loop,''
  Annals Phys.\  {\bf 158}, 142 (1984).
  %%CITATION = APNYA,158,142;%%

%3\cite{Gasser:1984gg}
\bibitem{Gasser:1984gg}
  J.~Gasser and H.~Leutwyler,
  %``Chiral Perturbation Theory: Expansions In The Mass Of The Strange Quark,''
  Nucl.\ Phys.\ B {\bf 250}, 465 (1985).
  %%CITATION = NUPHA,B250,465;%%


%4\cite{Meissner:1993ah}
\bibitem{Meissner:1993ah}
  U.~G.~Meissner,
  %``Recent developments in chiral perturbation theory,''
  Rept.\ Prog.\ Phys.\  {\bf 56}, 903 (1993).
  %[arXiv:hep-ph/9302247].
  %%CITATION = HEP-PH 9302247;%%

%5\cite{Bernard:1995dp}
\bibitem{Bernard:1995dp}
  V.~Bernard, N.~Kaiser and U.~G.~Meissner,
  %``Chiral Dynamics In Nucleons And Nuclei,''
  Int.\ J.\ Mod.\ Phys.\ E {\bf 4}, 193 (1995).
  %[arXiv:hep-ph/9501384].
  %%CITATION = HEP-PH 9501384;%%

%6\cite{Pich:1995bw}
\bibitem{Pich:1995bw}
  A.~Pich,
  %``Chiral perturbation theory,''
  Rept.\ Prog.\ Phys.\  {\bf 58}, 563 (1995).
  %[arXiv:hep-ph/9502366].
  %%CITATION = HEP-PH 9502366;%%

%7\cite{Ecker:1994gg}
\bibitem{Ecker:1994gg}
  G.~Ecker,
  %``Chiral Perturbation Theory,''
  Prog.\ Part.\ Nucl.\ Phys.\  {\bf 35}, 1 (1995).
  %[arXiv:hep-ph/9501357].
  %%CITATION = HEP-PH 9501357;%%

%8\cite{Scherer:2002tk}
\bibitem{Scherer:2002tk}
  S.~Scherer,
  %``Introduction to chiral perturbation theory,''
  Adv.\ Nucl.\ Phys.\  {\bf 27}, 277 (2003).
  %[arXiv:hep-ph/0210398].
  %%CITATION = ANUPB,27,277;%%

%9\cite{Bernard:2007zu}
\bibitem{Bernard:2007zu}
  V.~Bernard,
  %``Chiral Perturbation Theory and Baryon Properties,''
  Prog.\ Part.\ Nucl.\ Phys.\  {\bf 60}, 82 (2008).
  %[arXiv:0706.0312 [hep-ph]].
  %%CITATION = PPNPD,60,82;%%

%10\cite{Scherer:2009bt}
\bibitem{Scherer:2009bt}
  S.~Scherer,
  %``Chiral Perturbation Theory: Introduction and Recent Results in the
  %One-Nucleon Sector,''
  Prog.\ Part.\ Nucl.\ Phys.\  {\bf 64}, 1 (2010)
  %[arXiv:0908.3425 [hep-ph]].
  %%CITATION = PPNPD,64,1;%%

%\cite{Truong:1988zp}
\bibitem{Truong:1988zp}
  T.~N.~Truong,
  %``Chiral Perturbation Theory and Final State Theorem,''
  Phys.\ Rev.\ Lett.\  {\bf 61}, 2526  (1988).
  %%CITATION = PRLTA,61,2526;%%

%\cite{Dobado:1989qm}
\bibitem{Dobado:1989qm}
  A.~Dobado, M.~J.~Herrero and T.~N.~Truong,
  %``Unitarized Chiral Perturbation Theory for Elastic Pion-Pion Scattering,''
  Phys.\ Lett.\  B {\bf 235}, 134 (1990).
  %%CITATION = PHLTA,B235,134;%%

%\cite{Dobado:1992ha}
\bibitem{Dobado:1992ha}
  A.~Dobado and J.~R.~Pelaez,
  %``A Global fit of pi pi and pi K elastic scattering in ChPT with dispersion
  %relations,''
  Phys.\ Rev.\  D {\bf 47}, 4883 (1993).
 % [arXiv:hep-ph/9301276].
  %%CITATION = PHRVA,D47,4883;%%

%\cite{Kaiser:1995eg}
\bibitem{Kaiser:1995eg}
  N.~Kaiser, P.~B.~Siegel and W.~Weise,
  %``Chiral Dynamics And The Low-Energy Kaon - Nucleon Interaction,''
  Nucl.\ Phys.\  A {\bf 594}, 325 (1995).
  %[arXiv:nucl-th/9505043].
  %%CITATION = NUPHA,A594,325;%%

 %\cite{Dobado:1996ps}
\bibitem{Dobado:1996ps}
  A.~Dobado and J.~R.~Pelaez,
  %``The inverse amplitude method in Chiral Perturbation Theory,''
  Phys.\ Rev.\  D {\bf 56}, 3057 (1997).
  %[arXiv:hep-ph/9604416].
  %%CITATION = PHRVA,D56,3057;%%


%\cite{Hannah:1997ux}
\bibitem{Hannah:1997ux}
  T.~Hannah,
  %``The inverse amplitude method and chiral perturbation theory to two
  %loops,''
  Phys.\ Rev.\  D {\bf 55}, 5613 (1997).
  %[arXiv:hep-ph/9701389].
  %%CITATION = PHRVA,D55,5613;%%


%\cite{Oset:1997it}
\bibitem{Oset:1997it}
  E.~Oset and A.~Ramos,
  %``Non perturbative chiral approach to s-wave anti-K N interactions,''
  Nucl.\ Phys.\  A {\bf 635}, 99 (1998).
  %[arXiv:nucl-th/9711022].
  %%CITATION = NUPHA,A635,99;%%

%\cite{Oller:1997ng}
\bibitem{Oller:1997ng}
  J.~A.~Oller, E.~Oset and J.~R.~Pelaez,
  %``Non-perturbative Approach to effective chiral Lagrangians and Meson
  %Interactions,''
  Phys.\ Rev.\ Lett.\  {\bf 80}, 3452 (1998).
  %[arXiv:hep-ph/9803242].
  %%CITATION = PRLTA,80,3452;%%

%\cite{Oller:1997ti}
\bibitem{Oller:1997ti}
  J.~A.~Oller and E.~Oset,
  %``Chiral symmetry amplitudes in the S-wave isoscalar and isovector  channels
  %and the sigma, f0(980), a0(980) scalar mesons,''
  Nucl.\ Phys.\  A {\bf 620}, 438 (1997)
  [Erratum-ibid.\  A {\bf 652}, 407 (1999)].
  %[arXiv:hep-ph/9702314].
  %%CITATION = NUPHA,A620,438;%%


%\cite{Nieves:1998hp}
\bibitem{Nieves:1998hp}
  J.~Nieves and E.~Ruiz Arriola,
  %``Bethe-Salpeter approach for meson meson scattering in chiral  perturbation
  %theory,''
  Phys.\ Lett.\  B {\bf 455}, 30 (1999).
  %[arXiv:nucl-th/9807035].
  %%CITATION = PHLTA,B455,30;%%

%\cite{Kaiser:1998fi}
\bibitem{Kaiser:1998fi}
  N.~Kaiser,
  %``pi pi S-wave phase shifts and non-perturbative chiral approach,''
  Eur.\ Phys.\ J.\  A {\bf 3}, 307 (1998).
  %%CITATION = EPHJA,A3,307;%%


%\cite{Oller:1998hw}
\bibitem{Oller:1998hw}
  J.~A.~Oller, E.~Oset and J.~R.~Pelaez,
  %``Meson meson and meson baryon interactions in a chiral non-perturbative
  %approach,''
  Phys.\ Rev.\  D {\bf 59}, 074001 (1999)
  [Erratum-ibid.\  D {\bf 60}, 099906 (1999);  
  Erratum-ibid. D75, 099903 (2007)].
  %[arXiv:hep-ph/9804209].
  %%CITATION = PHRVA,D59,074001;%%

%\cite{Oller:1998zr}
\bibitem{Oller:1998zr}
  J.~A.~Oller and E.~Oset,
  %``N/D Description of Two Meson Amplitudes and Chiral Symmetry,''
  Phys.\ Rev.\  D {\bf 60} (1999) 074023
  [arXiv:hep-ph/9809337].
  %%CITATION = PHRVA,D60,074023;%%


%54\cite{Nieves:1999bx}
\bibitem{Nieves:1999bx}
  J.~Nieves and E.~Ruiz Arriola,
  %``Bethe-Salpeter approach for unitarized chiral perturbation theory,''
  Nucl.\ Phys.\  A {\bf 679}, 57 (2000).
  %[arXiv:hep-ph/9907469].
  %%CITATION = NUPHA,A679,57;%%

\bibitem{GomezNicola:2000wk}
  A.~Gomez Nicola, J.~Nieves, J.~R.~Pelaez and E.~Ruiz Arriola,
  %``Improved unitarized heavy baryon chiral perturbation theory for pi N
  %scattering,''
  Phys.\ Lett.\  B {\bf 486}, 77 (2000).
  %[arXiv:hep-ph/0006043].
  %%CITATION = PHLTA,B486,77;%%


%\cite{Nieves:2000km}
\bibitem{Nieves:2000km}
  J.~Nieves and E.~Ruiz Arriola,
  %``Bethe-Salpeter approach for the P(33) elastic pion nucleon scattering  in
  %heavy baryon chiral perturbation theory,''
  Phys.\ Rev.\  D {\bf 63}, 076001 (2001).
  %[arXiv:hep-ph/0008034].
  %%CITATION = PHRVA,D63,076001;%%


%\cite{Markushin:2000fa}
\bibitem{Markushin:2000fa}
  V.~E.~Markushin,
  %``The radiative decay phi - gamma pi pi in a coupled channel model and the
  %structure of f0(980),''
  Eur.\ Phys.\ J.\  A {\bf 8}, 389 (2000).
  %[arXiv:hep-ph/0005164].
  %%CITATION = EPHJA,A8,389;%%

%\cite{Oller:2000fj}
\bibitem{Oller:2000fj}
  J.~A.~Oller and U.~G.~Meissner,
  %``Chiral dynamics in the presence of bound states: Kaon nucleon  interactions  %revisited,''
  Phys.\ Lett.\  B {\bf 500}, 263 (2001).
  %[arXiv:hep-ph/0011146].
  %%CITATION = PHLTA,B500,263;%%


%\cite{Nieves:2001de}
\bibitem{Nieves:2001de}
  J.~Nieves, M.~Pavon Valderrama and E.~Ruiz Arriola,
  %``The Inverse Amplitude Method in $\pi\pi$ Scattering in Chiral Perturbation
  %Theory to Two Loops,''
  Phys.\ Rev.\  D {\bf 65}, 036002 (2002).
  %[arXiv:hep-ph/0109077].
  %%CITATION = PHRVA,D65,036002;%%

%\cite{GomezNicola:2001as}
\bibitem{GomezNicola:2001as}
  A.~Gomez Nicola and J.~R.~Pelaez,
  %``Meson meson scattering within one loop chiral perturbation theory and  its
  %unitarization,''
  Phys.\ Rev.\  D {\bf 65}, 054009 (2002).
  %[arXiv:hep-ph/0109056].
  %%CITATION = PHRVA,D65,054009;%%

%\cite{Lutz:2001yb}
\bibitem{Lutz:2001yb}
  M.~F.~M.~Lutz and E.~E.~Kolomeitsev,
  %``Relativistic chiral SU(3) symmetry, large N(c) sum rules and meson  baryon
  %scattering,''
  Nucl.\ Phys.\  A {\bf 700}, 193 (2002).
  %[arXiv:nucl-th/0105042].
  %%CITATION = NUPHA,A700,193;%%

%\cite{Nieves:2001wt}
\bibitem{Nieves:2001wt}
  J.~Nieves and E.~Ruiz Arriola,
  %``The S(11) N(1535) and N(1650) resonances in meson baryon unitarized
  %coupled channel chiral perturbation theory,''
  Phys.\ Rev.\  D {\bf 64}, 116008 (2001).
  %[arXiv:hep-ph/0104307].
  %%CITATION = PHRVA,D64,116008;%%


\bibitem{Hyodo:2002pk}
  T.~Hyodo, S.~I.~Nam, D.~Jido and A.~Hosaka,
  %``Flavor SU(3) breaking effects in the chiral unitary model for meson baryon
  %scatterings,''
  Phys.\ Rev.\  C {\bf 68}, 018201 (2003).
  %[arXiv:nucl-th/0212026].
  %%CITATION = PHRVA,C68,018201;%%


%\cite{GarciaRecio:2002td}
\bibitem{GarciaRecio:2002td}
  C.~Garcia-Recio, J.~Nieves, E.~Ruiz Arriola and M.~J.~Vicente Vacas,
  %``$S=-1$ Meson-Baryon Unitarized Coupled Channel Chiral Perturbation Theory
  %and the $S_{01}-$ $\Lambda$(1405) and $- \Lambda$(1670) Resonances,''
  Phys.\ Rev.\  D {\bf 67}, 076009 (2003).
  %[arXiv:hep-ph/0210311].
  %%CITATION = PHRVA,D67,076009;%%

%\cite{Kolomeitsev:2003kt}
\bibitem{Kolomeitsev:2003kt}
  E.~E.~Kolomeitsev and M.~F.~M.~Lutz,
  %``On baryon resonances and chiral symmetry,''
  Phys.\ Lett.\  B {\bf 585}, 243 (2004).
  %[arXiv:nucl-th/0305101].
  %%CITATION = PHLTA,B585,243;%%

%\cite{GarciaRecio:2003ks}
\bibitem{GarciaRecio:2003ks}
  C.~Garcia-Recio, M.~F.~M.~Lutz and J.~Nieves,
  %``Quark mass dependence of s-wave baryon resonances,''
  Phys.\ Lett.\  B {\bf 582}, 49 (2004).
  %[arXiv:nucl-th/0305100].
  %%CITATION = PHLTA,B582,49;%%

%\cite{Lutz:2003fm}
\bibitem{Lutz:2003fm}
  M.~F.~M.~Lutz and E.~E.~Kolomeitsev,
  %``On meson resonances and chiral symmetry,''
  Nucl.\ Phys.\  A {\bf 730}, 392 (2004).
  %[arXiv:nucl-th/0307039].
  %%CITATION = NUPHA,A730,392;%%


%\cite{Jido:2003cb}
\bibitem{Jido:2003cb}
  D.~Jido, J.~A.~Oller, E.~Oset, A.~Ramos and U.~G.~Meissner,
  %``Chiral dynamics of the two Lambda(1405) states,''
  Nucl.\ Phys.\  A {\bf 725}, 181 (2003).
  %[arXiv:nucl-th/0303062].
  %%CITATION = NUPHA,A725,181;%%


%\cite{Nicola:2003zi}
\bibitem{Nicola:2003zi}
  A.~G.~Nicola, J.~Nieves, J.~R.~Pelaez and E.~R.~Arriola,
  %``Improved unitarized heavy baryon chiral perturbation theory for pi N
  %scattering to fourth order,''
  Phys.\ Rev.\  D {\bf 69}, 076007 (2004).
  %[arXiv:nucl-th/0312034].
  %%CITATION = PHRVA,D69,076007;%%

%\cite{Sarkar:2004jh}
\bibitem{Sarkar:2004jh}
  S.~Sarkar, E.~Oset and M.~J.~Vicente Vacas,
  %``Baryonic resonances from baryon decuplet-meson octet interaction,''
  Nucl.\ Phys.\  A {\bf 750}, 294 (2005)
  [Erratum-ibid.\  A {\bf 780}, 78 (2006)].
  %[arXiv:nucl-th/0407025].
  %%CITATION = NUPHA,A750,294;%%


%\cite{GarciaRecio:2005hy}
\bibitem{GarciaRecio:2005hy}
  C.~Garcia-Recio, J.~Nieves and L.~L.~Salcedo,
  %``SU(6) extension of the Weinberg-Tomozawa meson-baryon Lagrangian,''
  Phys.\ Rev.\  D {\bf 74}, 034025 (2006).
  %[arXiv:hep-ph/0505233].
  %%CITATION = PHRVA,D74,034025;%%

%\cite{Borasoy:2005ie}
\bibitem{Borasoy:2005ie}
  B.~Borasoy, R.~Nissler and W.~Weise,
  %``Chiral dynamics of kaon nucleon interactions, revisited,''
  Eur.\ Phys.\ J.\  A {\bf 25}, 79 (2005).
  %[arXiv:hep-ph/0505239].
  %%CITATION = EPHJA,A25,79;%%


%\cite{Roca:2005nm}
\bibitem{Roca:2005nm}
  L.~Roca, E.~Oset and J.~Singh,
  %``Low lying axial-vector mesons as dynamically generated resonances,''
  Phys.\ Rev.\  D {\bf 72}, 014002 (2005).
  %[arXiv:hep-ph/0503273].
  %%CITATION = PHRVA,D72,014002;%%


%\cite{GarciaRecio:2006vx}
\bibitem{GarciaRecio:2006vx}
  C.~Garcia-Recio, J.~Nieves and L.~L.~Salcedo,
  %``Resonances and the Weinberg-Tomozawa 56-baryon --35-meson interaction,''
  Eur.\ Phys.\ J.\  A {\bf 31}, 499 (2007).
  %[arXiv:hep-ph/0610127].
  %%CITATION = EPHJA,A31,499;%%

%\cite{GomezNicola:2007qj}
\bibitem{GomezNicola:2007qj}
  A.~Gomez Nicola, J.~R.~Pelaez and G.~Rios,
  %``The Inverse Amplitude Method and Adler Zeros,''
  Phys.\ Rev.\  D {\bf 77}, 056006 (2008).
  %[arXiv:0712.2763 [hep-ph]].
  %%CITATION = PHRVA,D77,056006;%%

%\cite{Toki:2007ab}
\bibitem{Toki:2007ab}
  H.~Toki, C.~Garcia-Recio and J.~Nieves,
  %``Photon induced Lambda(1520) production and the role of the K^* exchange,''
  Phys.\ Rev.\  D {\bf 77}, 034001 (2008).
  %[arXiv:0711.3536 [hep-ph]].
  %%CITATION = PHRVA,D77,034001;%%

%\cite{Pelaez:2003dy}
\bibitem{Pelaez:2003dy}
  J.~R.~Pelaez,
  %``On the nature of light scalar mesons from their large $N_c$ behavior,''
  Phys.\ Rev.\ Lett.\  {\bf 92}, 102001 (2004).
  %[arXiv:hep-ph/0309292].
  %%CITATION = PRLTA,92,102001;%%

%\cite{Pelaez:2006nj}
\bibitem{Pelaez:2006nj}
  J.~R.~Pelaez and G.~Rios,
  %``Nature of the f_0(600) from its N_c dependence at two loops in unitarized
  %Chiral Perturbation Theory,''
  Phys.\ Rev.\ Lett.\  {\bf 97}, 242002 (2006).
  %[arXiv:hep-ph/0610397].
  %%CITATION = PRLTA,97,242002;%%

%\cite{GarciaRecio:2006wb}
\bibitem{GarciaRecio:2006wb}
  C.~Garcia-Recio, J.~Nieves and L.~L.~Salcedo,
  %``Large N(c) Weinberg-Tomozawa interaction and negative parity s-wave  baryon
  %resonances,''
  Phys.\ Rev.\  D {\bf 74}, 036004 (2006).
  %[arXiv:hep-ph/0605059].
  %%CITATION = PHRVA,D74,036004;%%

%\cite{Hyodo:2007np}
\bibitem{Hyodo:2007np}
  T.~Hyodo, D.~Jido and L.~Roca,
  %``Structure of the (1405) baryon resonance from its large N(c) behavior,''
  Phys.\ Rev.\  D {\bf 77}, 056010 (2008).
  %[arXiv:0712.3347 [hep-ph]].
  %%CITATION = PHRVA,D77,056010;%%

%\cite{Roca:2008kr}
\bibitem{Roca:2008kr}
  L.~Roca, T.~Hyodo and D.~Jido,
  %``On the nature of the $\Lambda(1405)$ and $\Lambda(1670)$ from their $N_c$
  %behavior in chiral dynamics,''
  Nucl.\ Phys.\  A {\bf 809}, 65 (2008).
  %[arXiv:0804.1210 [hep-ph]].
  %%CITATION = NUPHA,A809,65;%%

%\cite{Geng:2008ag}
\bibitem{Geng:2008ag}
  L.~S.~Geng, E.~Oset, J.~R.~Pelaez and L.~Roca,
  %``Nature of the axial-vector mesons from their Nc behavior within the chiral
  %unitary approach,''
  Eur.\ Phys.\ J.\  A {\bf 39}, 81 (2009).
  %[arXiv:0811.1941 [hep-ph]].
  %%CITATION = EPHJA,A39,81;%%

%\cite{Nieves:2009ez}
\bibitem{Nieves:2009ez}
  J.~Nieves and E.~Ruiz Arriola,
  %``Properties of the rho and sigma Mesons from Unitary Chiral Dynamics,''
  Phys.\ Rev.\  D {\bf 80}, 045023 (2009).
  %[arXiv:0904.4344 [hep-ph]].
  %%CITATION = PHRVA,D80,045023;%%



%\cite{Nieves:2009kh}
\bibitem{Nieves:2009kh}
  J.~Nieves and E.~Ruiz Arriola,
  %``Meson Resonances at large Nc: Complex Poles vs Breit-Wigner Masses,''
  Phys.\ Lett.\  B {\bf 679}, 449 (2009).
  %[arXiv:0904.4590 [hep-ph]].
  %%CITATION = PHLTA,B679,449;%%

 %\cite{MartinezTorres:2007sr}
\bibitem{MartinezTorres:2007sr}
  A.~Martinez Torres, K.~P.~Khemchandani and E.~Oset,
  %``Three body resonances in two meson-one baryon systems,''
  Phys.\ Rev.\  C {\bf 77}, 042203(R) (2008).
  %[arXiv:0706.2330 [nucl-th]].
  %%CITATION = PHRVA,C77,042203;%%


%\cite{MartinezTorres:2008gy}
\bibitem{MartinezTorres:2008gy}
  A.~Martinez Torres, K.~P.~Khemchandani, L.~S.~Geng, M.~Napsuciale and E.~Oset,
  %``The X(2175) as a resonant state of the $\phi K \bar{K}$ system,''
  Phys.\ Rev.\  D {\bf 78}, 074031 (2008).
  %[arXiv:0801.3635 [nucl-th]].
  %%CITATION = PHRVA,D78,074031;%%


%\cite{Kolomeitsev:2003ac}
\bibitem{Kolomeitsev:2003ac}
  E.~E.~Kolomeitsev and M.~F.~M.~Lutz,
  %``On heavy-light meson resonances and chiral symmetry,''
  Phys.\ Lett.\  B {\bf 582}, 39 (2004).
  %[arXiv:hep-ph/0307133].
  %%CITATION = PHLTA,B582,39;%%

  %\cite{Hofmann:2003je}
\bibitem{Hofmann:2003je}
  J.~Hofmann and M.~F.~M.~Lutz,
  %``Open-charm meson resonances with negative strangeness,''
  Nucl.\ Phys.\  A {\bf 733}, 142 (2004).
  %[arXiv:hep-ph/0308263].
  %%CITATION = NUPHA,A733,142;%%

%\cite{Guo:2006fu}
\bibitem{Guo:2006fu}
  F.~K.~Guo, P.~N.~Shen, H.~C.~Chiang and R.~G.~Ping,
  %``Dynamically generated 0+ heavy mesons in a heavy chiral unitary
  %approach,''
  Phys.\ Lett.\  B {\bf 641}, 278 (2006).
  %[arXiv:hep-ph/0603072].
  %%CITATION = PHLTA,B641,278;%%

  %\cite{Gamermann:2006nm}
\bibitem{Gamermann:2006nm}
  D.~Gamermann, E.~Oset, D.~Strottman and M.~J.~Vicente Vacas,
  %``Dynamically Generated Open and Hidden Charm Meson Systems,''
  Phys.\ Rev.\  D {\bf 76}, 074016 (2007).
  %[arXiv:hep-ph/0612179].
  %%CITATION = PHRVA,D76,074016;%%


%\cite{Lutz:2003jw}
\bibitem{Lutz:2003jw}
  M.~F.~M.~Lutz and E.~E.~Kolomeitsev,
  %``On charm baryon resonances and chiral symmetry,''
  Nucl.\ Phys.\  A {\bf 730}, 110 (2004).
  %[arXiv:hep-ph/0307233].
  %%CITATION = NUPHA,A730,110;%%


%\cite{Hofmann:2005sw}
\bibitem{Hofmann:2005sw}
  J.~Hofmann and M.~F.~M.~Lutz,
  %``Coupled-channel study of crypto-exotic baryons with charm,''
  Nucl.\ Phys.\  A {\bf 763}, 90 (2005).
  %[arXiv:hep-ph/0507071].
  %%CITATION = NUPHA,A763,90;%%
%\cite{Hofmann:2006qx}

\bibitem{Hofmann:2006qx}
  J.~Hofmann and M.~F.~M.~Lutz,
  %``D-wave baryon resonances with charm from coupled-channel dynamics,''
  Nucl.\ Phys.\  A {\bf 776}, 17 (2006).
  %[arXiv:hep-ph/0601249].
  %%CITATION = NUPHA,A776,17;%%

%\cite{Mizutani:2006vq}
\bibitem{Mizutani:2006vq}
  T.~Mizutani and A.~Ramos,
  %``D mesons in nuclear matter: A D N coupled-channel equations approach,''
  Phys.\ Rev.\  C {\bf 74}, 065201 (2006).
  %[arXiv:hep-ph/0607257].
  %%CITATION = PHRVA,C74,065201;%%


%\cite{GarciaRecio:2008dp}
\bibitem{GarciaRecio:2008dp}
  C.~Garcia-Recio, V.~K.~Magas, T.~Mizutani, J.~Nieves, A.~Ramos, L.~L.~Salcedo and L.~Tolos,
  %``The s-wave charmed baryon resonances from a coupled-channel approach with
  %heavy quark symmetry,''
  Phys.\ Rev.\  D {\bf 79}, 054004 (2009).
  %[arXiv:0807.2969 [hep-ph]].
  %%CITATION = PHRVA,D79,054004;%%

%\cite{Gamermann:2010zz}
\bibitem{Gamermann:2010zz}
  D.~Gamermann, C.~Garcia-Recio, J.~Nieves, L.~L.~Salcedo and L.~Tolos,
  %``Exotic dynamically generated baryons with negative charm quantum number,''
  Phys.\ Rev.\  D in press and arXiv:1002.2763.
  %%CITATION = ARXIV:1002.2763;%%


%\cite{Weinberg:1966kf}
\bibitem{Weinberg:1966kf}
  S.~Weinberg,
  %``Pion scattering lengths,''
  Phys.\ Rev.\ Lett.\  {\bf 17}, 616 (1966).
  %%CITATION = PRLTA,17,616;%%

%68\cite{Tomozawa:1966jm}
\bibitem{Tomozawa:1966jm}
  Y.~Tomozawa,
  %``Axial vector coupling renormalization and the meson baryon scattering
  %lengths,''
  Nuovo Cim.\  {\bf 46A}, 707 (1966).
  %%CITATION = NUCIA,46A,707;%%

%\cite{Molina:2008jw}
\bibitem{Molina:2008jw}
  R.~Molina, D.~Nicmorus and E.~Oset,
  %``The \rho\rho interaction in the hidden gauge formalism and the f_0(1370)
  %and f_2(1270) resonances,''
  Phys.\ Rev.\  D {\bf 78}, 114018 (2008).
  %%CITATION = PHRVA,D78,114018;%%


%\cite{Geng:2008gx}
\bibitem{Geng:2008gx}
  L.~S.~Geng and E.~Oset, 
%``Vector meson-vector meson interaction in a hidden gauge unitary approach,''
  Phys.\ Rev.\ D {\bf 79}, 074009 (2009); L.~S.~Geng, E.~Oset, R.~Molina and
  D.~Nicmorus, arXiv:0905.0419 [hep-ph].
%``Dynamically generated resonances from vector meson-vector meson interaction 
%based on a hidden-gauge unitary approach,'' 


%\cite{Oset:2009vf}
\bibitem{Oset:2009vf}
  E.~Oset and A.~Ramos,
  %``Dynamically generated resonances from the vector octet-baryon octet
  %interaction,''
  arXiv:0905.0973 [hep-ph].
  %%CITATION = ARXIV:0905.0973;%%


%\cite{Gonzalez:2008pv}
\bibitem{Gonzalez:2008pv}
  P.~Gonzalez, E.~Oset and J.~Vijande,
  %``An explanation of the $\Delta_{5/2^{-}}(1930)$ as a $\rho\Delta$ bound
  %state,''
  Phys.\ Rev.\  C {\bf 79}, 025209 (2009).
%  [arXiv:0812.3368 [hep-ph]].
  %%CITATION = PHRVA,C79,025209;%%


%\cite{Sarkar:2009kx}
\bibitem{Sarkar:2009kx}
  S.~Sarkar, B.~X.~Sun, E.~Oset and M.~J.~V.~Vacas,
  %``Dynamically generated resonances from the vector octet-baryon decuplet
  %interaction,''
  arXiv:0902.3150 [hep-ph].
  %%CITATION = ARXIV:0902.3150;%%

%\cite{Bando:1984ej}
\bibitem{Bando:1984ej}
  M.~Bando, T.~Kugo, S.~Uehara, K.~Yamawaki and T.~Yanagida,
  %``Is Rho Meson A Dynamical Gauge Boson Of Hidden Local Symmetry?,''
  Phys.\ Rev.\ Lett.\  {\bf 54}  1215 (1985).
  %%CITATION = PRLTA,54,1215;%%

%\cite{Bando:1987br}
\bibitem{Bando:1987br}
  M.~Bando, T.~Kugo and K.~Yamawaki,
  %``Nonlinear Realization and Hidden Local Symmetries,''
  Phys.\ Rept.\  {\bf 164}  217 (1988).
  %%CITATION = PRPLC,164,217;%%

%\cite{MartinezTorres:2009uk}
\bibitem{MartinezTorres:2009uk}
  A.~Martinez Torres, L.~S.~Geng, L.~R.~Dai, B.~X.~Sun, E.~Oset and B.~S.~Zou, 
  %``Study of the $J/\psi \to \phi (\omega) f_2(1270)$, $J/\psi \to \phi 
  %(\omega) f'_2(1525)$ and $J/\psi \to K^{*0}(892) \bar{K}^{* 0}_2(1430)$ 
  %decays,''
  Phys.\ Lett.\ B {\bf 680}, 310 (2009).  
  %%CITATION= PHLTA,B680,310;%%

%\cite{Branz:2009cv}
\bibitem{Branz:2009cv}
  T.~Branz, L.~S.~Geng and E.~Oset, 
  %``Two-photon and one photon-one vector meson decay widths of the 
  %$f_0(1370)$, $f_2(1270)$, $f_0(1710)$, $f'_2(1525)$, and $K^*_2(1430)$,'' 
  Phys.\ Rev.\ D {\bf 81}, 054037 (2010).
  %%CITATION =ARXIV:0911.0206;%%

%\cite{Geng:2009iw}
\bibitem{Geng:2009iw}
  L.~S.~Geng, F.~K.~Guo, C.~Hanhart, R.~Molina, E.~Oset and B.~S.~Zou,
  %``Study of the $f_2(1270)$, $f_2'(1525)$, $f_0(1370)$ and $f_0(1710)$ in the
  %$J/\psi$ radiative decays,''
  Eur.\ Phys.\ J.\  A {\bf 44}, 305 (2009).
  %arXiv:0910.5192 [hep-ph].
  %%CITATION = EPHJA,A44,305;%%


%\cite{Caldi:1975tx}
\bibitem{Caldi:1975tx}
  D.~G.~Caldi and H.~Pagels,
  %``A Solution To The Rho-Pi Puzzle: Spontaneously Broken Symmetries Of The
  %Quark Model,''
  Phys.\ Rev.\  D {\bf 14}, 809 (1976).
  %%CITATION = PHRVA,D14,809;%%

%\cite{Caldi:1976gz}
\bibitem{Caldi:1976gz}
  D.~G.~Caldi and H.~Pagels,
  %``Spontaneous Symmetry Breaking And Vector Meson Dominance,''
  Phys.\ Rev.\  D {\bf 15}, 2668 (1977).
  %%CITATION = PHRVA,D15,2668;%%

%\cite{Ibanez:1979vb}
\bibitem{Ibanez:1979vb}
  L.~E.~Ibanez,
  %``Is The Rho Meson A Dormant Goldstone Boson?,''
  Phys.\ Lett.\  B {\bf 86}, 340 (1979).
  %%CITATION = PHLTA,B86,340;%%

%\cite{Meissner:1986it}
\bibitem{Meissner:1986it}
  U.~G.~Meissner and V.~Pasquier,
  %``A CHIRAL SOLITON MODEL WITH MESONIC DEMOCRACY,''
  Phys.\ Lett.\  B {\bf 235}, 153 (1990).
  %%CITATION = PHLTA,B235,153;%%

%\cite{Smit:1980nf}
\bibitem{Smit:1980nf}
  J.~Smit,
  %``Chiral Symmetry Breaking In QCD: Mesons As Spin Waves,''
  Nucl.\ Phys.\  B {\bf 175}, 307 (1980).
  %%CITATION = NUPHA,B175,307;%%

%\cite{Amsler:2008zz}
\bibitem{Amsler:2008zz}
  C.~Amsler {\it et al.}  [Particle Data Group],
  %``Review of particle physics,''
  Phys.\ Lett.\  B {\bf 667}, 1 (2008).
  %%CITATION = PHLTA,B667,1;%%


%\cite{Birse:1996hd}
\bibitem{Birse:1996hd}
  M.~C.~Birse,
  %``Effective chiral Lagrangians for spin-1 mesons,''
  Z.\ Phys.\  A {\bf 355}, 231 (1996).
%  [arXiv:hep-ph/9603251].
  %%CITATION = ZEPYA,A355,231;%%

%
\bibitem{Hyodo:2006kg}
  T.~Hyodo, D.~Jido and A.~Hosaka,
  %``Study of exotic hadrons in s-wave scatterings induced by chiral interaction
  %in the flavor symmetric limit,''
  Phys.\ Rev.\  D {\bf 75}, 034002 (2007)
  [arXiv:hep-ph/0611004].
  %%CITATION = PHRVA,D75,034002;%%

%\cite{Gursey:1992dc}
\bibitem{Gursey:1992dc}
  F.~Gursey and L.~A.~Radicati,
  %``Spin and unitary spin independence of strong interactions,''
  Phys.\ Rev.\ Lett.\  {\bf 13}, 173 (1964).
  %%CITATION = PRLTA,13,173;%%

%%\cite{Pais:1964}
\bibitem{Pais:1964}
A.~Pais,
Phys.\ Rev.\ Lett.\  {\bf 13}, 175 (1964).
%%CITATION = PRLTA,13,175;%%

%\cite{Sakita:1964qq}
\bibitem{Sakita:1964qq}
  B.~Sakita,
  %``Supermultiplets of elementary particles,''
  Phys.\ Rev.\  {\bf 136}, B1756 (1964).
  %%CITATION = PHRVA,136,B1756;%%


%\cite{de Swart:1963gc}
\bibitem{deSwart:1963gc}
  J.~J.~de Swart,
  %``The Octet Model And Its Clebsch-Gordan Coefficients,''
  Rev.\ Mod.\ Phys.\  {\bf 35}  916 (1963)
  [Erratum-ibid.\  {\bf 37}  326  (1965)].
  %%CITATION = RMPHA,35,916;%%


\bibitem{Cook65}
C. L. Cook and G. Murtaza, Nuovo Cim. 39  531 (1965).

%
\bibitem{Feynman:1964fk}
  R.~P.~Feynman, M.~Gell-Mann and G.~Zweig,
  %``Group U(6) x U(6) generated by current components,''
  Phys.\ Rev.\ Lett.\  {\bf 13}, 678 (1964).
  %%CITATION = PRLTA,13,678;%%

\bibitem{Coleman:1967ad}
  S.~R.~Coleman and J.~Mandula,
  %``ALL POSSIBLE SYMMETRIES OF THE S MATRIX,''
  Phys.\ Rev.\  {\bf 159}, 1251 (1967).
  %%CITATION = PHRVA,159,1251;%%

\bibitem{Ecker:1988te}
  G.~Ecker, J.~Gasser, A.~Pich and E.~de Rafael,
  %``The Role Of Resonances In Chiral Perturbation Theory,''
  Nucl.\ Phys.\  B {\bf 321}, 311 (1989).
  %%CITATION = NUPHA,B321,311;%%

%\cite{Ecker:1989yg}
\bibitem{Ecker:1989yg}
  G.~Ecker, J.~Gasser, H.~Leutwyler, A.~Pich and E.~de Rafael,
  %``Chiral Lagrangians for Massive Spin 1 Fields,''
  Phys.\ Lett.\  B {\bf 223}, 425 (1989).
  %%CITATION = PHLTA,B223,425;%%


\bibitem{Jaminon:1998de}
  M.~Jaminon and E.~Ruiz Arriola,
  %``Vector mesons from tensor couplings in the NJL model,''
  Phys.\ Lett.\  B {\bf 443}, 33 (1998).
  %%CITATION = PHLTA,B443,33;%%


%\cite{Weinberg:1968de}
\bibitem{Weinberg:1968de}
  S.~Weinberg,
  %``Nonlinear realizations of chiral symmetry,''
  Phys.\ Rev.\  {\bf 166}, 1568 (1968).
  %%CITATION = PHRVA,166,1568;%%

%\cite{Caldi:1984xr}
\bibitem{Caldi:1984xr}
  D.~G.~Caldi,
  %``Skyrmions And Vector Mesons: A Symmetric Approach,''
  %%CITATION = C84/06/02;%%

%\cite{Manohar:2000dt}
\bibitem{Manohar:2000dt}
  A.~V.~Manohar and M.~B.~Wise,
  %``Heavy quark physics,''
  Camb.\ Monogr.\ Part.\ Phys.\ Nucl.\ Phys.\ Cosmol.\  {\bf 10}, 1 (2000).
  %%CITATION = CMPCE,10,1;%%

%\cite{Caldi-Skyrmion}
\bibitem{Caldi-Skyrmion}


%\cite{Nagahiro:2008cv}
\bibitem{Nagahiro:2008cv}
  H.~Nagahiro, L.~Roca, A.~Hosaka and E.~Oset,
  %``Hidden gauge formalism for the radiative decays of axial-vector mesons,''
  Phys.\ Rev.\  D {\bf 79}, 014015 (2009).
 % [arXiv:0809.0943 [hep-ph]].
  %%CITATION = PHRVA,D79,014015;%%


%\cite{Close:2005vf}
\bibitem{Close:2005vf}
  F.~E.~Close and Q.~Zhao,
  %``Production of $f_0(1710)$, $f_0(1500)$, and $f_0(1370)$ in $J/\psi$
  %hadronic decays,''
  Phys.\ Rev.\  D {\bf 71}, 094022 (2005).
  %[arXiv:hep-ph/0504043].
  %%CITATION = PHRVA,D71,094022;%%


%\cite{Amsler:1995td}
\bibitem{Amsler:1995td}
  C.~Amsler and F.~E.~Close,
  %``Is $f_0(1500)$ a Scalar Glueball?,''
  Phys.\ Rev.\  D {\bf 53}, 295 (1996).
%  [arXiv:hep-ph/9507326].
  %%CITATION = PHRVA,D53,295;%%
%\cite{Amsler:2004ps}

\bibitem{Amsler:2004ps}
  C.~Amsler and N.~A.~Tornqvist,
  %``Mesons beyond the naive quark model,''
  Phys.\ Rept.\  {\bf 389}, 61 (2004).
  %%CITATION = PRPLC,389,61;%%


%\cite{Albaladejo:2008qa}
\bibitem{Albaladejo:2008qa}
  M.~Albaladejo and J.~A.~Oller,
  %``Identification of a Scalar Glueball,''
  Phys.\ Rev.\ Lett.\  {\bf 101}, 252002 (2008).
%  [arXiv:0801.4929 [hep-ph]].
  %%CITATION = PRLTA,101,252002;%%

%\cite{Anisovich:1999jv}
\bibitem{Anisovich:1999jv}
  A.~V.~Anisovich {\it et al.}  [Crystal Barrel Collaboration],
  %``Anti-p p --> pi0 eta and pi0 eta' from 600-MeV/c to 1940-MeV/c,''
  Phys.\ Lett.\  B {\bf 452}, 173 (1999).
  %%CITATION = PHLTA,B452,173;%%

%\cite{Anisovich:2002su}
\bibitem{Anisovich:2002su}
  A.~V.~Anisovich {\it et al.},
  %``Combined analysis of meson channels with I = 1, C = -1 from 1940-MeV to
  %2410-MeV,''
  Phys.\ Lett.\  B {\bf 542}, 8 (2002).
  %%CITATION = PHLTA,B542,8;%%


\bibitem{Cherry:2000ut}
  S.~N.~Cherry and M.~R.~Pennington,
  %``There is no $\kappa(900)$,''
  Nucl.\ Phys.\  A {\bf 688}, 823 (2001).
  %[arXiv:hep-ph/0005208].
  %%CITATION = NUPHA,A688,823;%%

%\cite{Geng:2006yb}
\bibitem{Geng:2006yb}
  L.~S.~Geng, E.~Oset, L.~Roca and J.~A.~Oller,
  %``Clues for the existence of two K(1)(1270) resonances,''
  Phys.\ Rev.\  D {\bf 75}, 014017 (2007).
  %[arXiv:hep-ph/0610217].
  %%CITATION = PHRVA,D75,014017;%%

%\cite{Daum:1981hb}
\bibitem{Daum:1981hb}
  C.~Daum {\it et al.}  [ACCMOR Collaboration],
  %``Diffractive Production Of Strange Mesons At 63-Gev,''
  Nucl.\ Phys.\  B {\bf 187}, 1 (1981).
  %%CITATION = NUPHA,B187,1;%%


%\cite{Filippi:2000is}
\bibitem{Filippi:2000is}
  A.~Filippi {\it et al.}  [OBELIX Collaboration],
  %``An Analysis Of The Contribution Of Isospin Two Pi Pi Resonant States In The%Anti-N P $\to$ Pi+ Pi+ Pi- Annihilation Reaction,''
  Phys.\ Lett.\  B {\bf 495}, 284 (2000).
  %%CITATION = PHLTA,B495,284;%%


%\cite{Ishida:1989xh}
\bibitem{Ishida:1989xh}
  S.~Ishida, M.~Oda, H.~Sawazaki and K.~Yamada,
  %``IS THE f1 (1420) OUR FIRST HYBRID MESON?,''
  Prog.\ Theor.\ Phys.\  {\bf 82}, 119 (1989).
  %%CITATION = PTPKA,82,119;%%

\end{thebibliography}
\end{document}